\documentclass[12pt]{article}

\usepackage[margin=1.25in]{geometry}

\usepackage{subcaption}
\RequirePackage[colorlinks,citecolor=blue,linkcolor=blue,urlcolor=blue]{hyperref}

\usepackage{amsmath,amsthm,amssymb}
\usepackage[backend=biber,style=authoryear,doi=false,url=false,eprint=false]{biblatex}
\usepackage{xparse}
\NewDocumentCommand{\citet}{O{} O{} m}{
  \textcite[#1][#2]{#3}%
}
\NewDocumentCommand{\citep}{O{} O{} m}{
  \parencite[#1][#2]{#3}%
}

\theoremstyle{plain}
\newtheorem{assumption}{Assumption}
\newtheorem{theorem}{Theorem}

\theoremstyle{definition}
\newtheorem{remark}{Remark}

\usepackage{dsfont}
\DeclareMathOperator{\var}{var}
\DeclareMathOperator{\cov}{cov}
\newcommand{\ind}{\mathds{1}}
\newcommand{\N}{\mathcal{N}}
\renewcommand{\S}{\mathcal{S}}
\newcommand{\T}{\mathcal{T}}
\newcommand{\M}{\mathcal{M}}
\newcommand{\Y}{\mathcal{Y}}
\newcommand{\Z}{\mathcal{Z}}
\newcommand{\R}{\mathcal{R}}
\newcommand{\dnoeffect}{d_{0}}

\makeatletter
\newcommand\niton{\mathrel{\m@th\mathpalette\canc@l\owns}}
\newcommand\canc@l[2]{{\ooalign{$\hfil#1/\mkern1mu\hfil$\crcr$#1#2$}}}
\makeatother

\usepackage{mathtools}
\DeclarePairedDelimiter\floor{\lfloor}{\rfloor}
\DeclarePairedDelimiter\abs{\lvert}{\rvert}
\DeclarePairedDelimiter\norm{\lVert}{\rVert}

\usepackage{enumitem}
\usepackage{rotating}

\usepackage{array}
\newcolumntype{K}[1]{>{\centering\arraybackslash}p{#1}}

\usepackage{setspace}
\usepackage{placeins}

\usepackage{titling}
\makeatletter
\g@addto@macro\@floatboxreset\centering
\makeatother

\title{Causal Inference for Spatial Treatments}
\author{Michael~Pollmann\thanks{Duke~University. Email: \href{mailto:michael.pollmann@duke.edu}{michael.pollmann@duke.edu}. \\
I am grateful to my advisor, Guido Imbens, for invaluable encouragement and guidance.
I am thankful to Luis Armona, Paul Goldsmith-Pinkham, Christian Hansen, Caroline Hoxby, Joshua Kim, Matt Masten, Áureo de Paula, Daniel Pollmann, Fredrik Sävje, Jann Spiess, Melanie Wallskog, as well as numerous seminar participants and three anonymous referees for many comments and insightful discussions.
This research was supported generously by the B.F. Haley and E.S. Shaw Fellowship for Economics through a grant to the Stanford Institute for Economic Policy Research.
This paper uses data from \citet{safegraph2021}.
SafeGraph is a data company that aggregates anonymized location data from numerous applications in order to provide insights about physical places.
}}

\begin{document}

\begin{refsection}

\maketitle

\begin{abstract}
Many events and policies (treatments) occur at specific spatial locations, with researchers interested in their effects on nearby units.
I approach the \emph{spatial treatment} setting from an experimental perspective:
What ideal experiment would we design to estimate the causal effects of spatial treatments?
This perspective motivates a comparison between units near realized treatment locations and units near counterfactual (unrealized) candidate locations, which differs from current empirical practice.
I derive design-based standard errors that are straightforward to compute.
For observational data, I propose machine learning methods to find counterfactual candidate locations when observable characteristics, rather than potential outcomes, determine treatment probabilities.
To accommodate methods for high-dimensional data in the theory, I extend a double machine learning result to the design-based framework with spatial correlations.
I apply the proposed methods to study the causal effects of grocery stores on foot traffic to nearby businesses during COVID-19 shelter-in-place policies, finding a large positive effect at very short distances, with no effect at larger distances.
\end{abstract}

\paragraph{Keywords:}
Causal inference,
spatial treatments,
design-based,
ideal experiment,
machine learning,
convolutional neural networks

\clearpage

\section{Introduction}

Many actions, events, and policies studied by economists occur at locations in space and affect (geographically) nearby units or individuals.\footnote{Examples include the effects of: businesses' location decisions on local competitors, workers, or consumers; schools, hospitals, or sources of pollution on education, income, and health of nearby residents; low-income housing, local public goods, or crime risk on property values; centrally administered treatments such as deworming in schools or COVID-19 vaccination centers on treatment uptake and effectiveness. See Online~Appendix Table~OA1 for examples of papers studying these and other spatial treatments.}
I refer to the setting of such studies as the ``spatial treatment'' setting because these ``treatments'' vary at the level of locations in space.
The researcher studies the effects of such treatments on individuals who are located in the vicinity of these treatments but who are conceptually distinct units.
In contrast, in most of the theoretical literature on causal inference, each individual is thought to, in principle, be associated with a distinct treatment that generates potential outcomes, and some work considers ``spillovers'' and ``clustered assignment'' of individual-level treatments.
Such a framework sufficed when treatment and outcome information was only available at aggregated levels, such as the county level.
However, more recently, precise (geocoded) location data for treatments and individuals have become more readily available, allowing more informative analyses of the disaggregated effects of spatial treatments by distance from treatment.

This paper makes three contributions.
First, I develop a framework that allows me to formalize ideal experiments and analyze questions of causal inference in spatial settings from a design-based perspective.
Second, I show that this design-based perspective is tractable and useful by deriving (approximately) unbiased inverse probability weighting (IPW) estimators and new expressions for their variances, which differ from commonly used existing estimators and sampling-based variances.
Third, I provide theoretical results for observational data and propose using convolutional neural networks, previously used for image and satellite data, to parsimoniously yet flexibly condition on the distribution of covariates across space.

The ideal experiment for studying spatial treatments randomizes the location of the treatment among a set of plausible candidate locations.
Such an ideal experiment is a formalization of settings where the location of the treatment is quasi-random, for instance, due to the exogenous (un-) availability of candidate locations at the time the treatment is implemented.\footnote{
For instance, \citet[p. 1110]{linden2008estimates} state that ``the nature of the search for housing is also a largely random process at the local level. Individuals may choose neighborhoods with specific characteristics, but, within a fraction of a mile, the exact locations available at the time individuals seek to move into a neighborhood are arguably exogenous.'' In their empirical analyses, they estimate causal effects based on a parallel trends assumption.}
In the ideal experiment, individuals who are located near candidate treatment locations that -- by random chance alone -- did not receive treatment constitute a valid control group.
The ideal experiment does not generally justify comparisons of individuals near realized treatment (on an ``inner ring'') to individuals farther away but centered around the same treatment locations (on an ``outer ring'') even when embedded in a difference-in-differences analysis as is common in the empirical literature.
The approach developed in this paper therefore presents an alternative identification strategy for spatial treatment settings, leveraging design-based assumptions.

I propose IPW estimators, derive their (approximate) finite population design-based unbiasedness and variance under the ideal experiment, and show asymptotic normality in a sequence of finite but growing populations.
In the thought experiment underlying inference, only the realized locations of the treatment vary among a pre-defined set of plausible candidate locations.
Other locations and potential outcomes are fixed.
The design-based analysis of the variance, following \citet{neyman1923}, has both conceptual and practical advantages over sampling-based alternatives \citep[for instance,][]{conley1999gmm}:
The design-based variance reflects the variation that the researcher exploits when claiming causality of estimated effects by appealing to ``quasi-random'' variation.
In the design-based analysis, researchers do not need to distinguish between sample and population, which may be a difficult distinction to justify in spatial settings \citep{Pinkse2007}.
Estimating the design-based standard error is straightforward and does not require the correct modeling of the correlation of potential outcomes across space.
Importantly, my approach and formulas generalize to settings where individuals are exposed to multiple treatments.
In these settings, off-the-shelf alternatives for design-based inference, such as clustering at the level of the assignment \citep{Abadie2023}, are not applicable.
In Online~Appendix~2, I show that different implementations of the variance estimator of \citet{conley1999gmm} can have undesirable properties in a design-based framework.

I then study an observational data setting where, possibly unknown, treatment probabilities are determined by observable characteristics.
Suppose locations in two neighborhoods look identical in terms of their observable pre-treatment characteristics.
Then the assumption requires that the treatment is equally likely to be realized in the two locations, irrespective of potential outcomes, similar to an unconfoundedness assumption in sampling-based analyses.
I derive a novel design-based ``double machine learning''-type result for spatial treatments that allows estimation of nuisance parameters using arbitrary (for instance, machine learning) methods as long as (weak) rate conditions are satisfied.

To implement flexible estimation based on this assumption on assignment, I propose using convolutional neural networks (CNNs) in a way that may be of independent interest for settings with spatial data.
Researchers can plot many economic data, such as locations of businesses, property prices, school district quality, and the average income by census tract, on maps.
The distribution of spatial covariates across space often encodes otherwise latent information that is lost in coarse summary statistics.
However, controlling for the distribution of units or covariates across space relative to the location of estimation intrinsically is an extremely high-dimensional problem.
I propose CNNs that parsimoniously condition on the distribution of covariates across space, incorporating the economic logic that typically only relative, not absolute, locations matter.
I use such networks to find plausible counterfactual locations of the treatment that are observationally similar to realized treatment locations.

I apply the proposed methods to study whether grocery stores caused an increase in the number of visitors to nearby restaurants during COVID-19 shelter-in-place policies.
Consumers may find it convenient to grab a coffee or meal while waiting to enter the store or before returning home.
Using CNNs, I identify counterfactual grocery store locations that are in neighborhoods with business compositions and relative locations similar to the neighborhoods of real grocery stores.
I find that restaurants within a couple of minutes walking from real grocery stores had about twice as many visitors as restaurants at the same distance from counterfactual locations.
There is no such difference in visitors at longer distances.

A nascent methodological literature studies causal inference in spatial treatment and related settings.
\citet{zigler2021bipartite} define potential outcomes and estimands.
Most closely related, \citet{Wang2025} in contemporaneous work explore a similar experimental setting and show design-based unbiasedness, variance, and asymptotic normality for a special case of the experimental design (constant treatment probabilities) and class of estimators (simplifying weights) described in this paper under distinct structural assumptions (restricting treatment effect heterogeneity and ruling out any effect of the treatment at longer distances).
In addition to arguably weaker assumptions and more general estimators and designs, I address challenges arising in observational settings.
\citet{Borusyak2023} take a design-based perspective similar to this paper but focus on regression estimators, accommodating multivalued treatments.
However, the estimands of their unweighted regressions differ under treatment effect heterogeneity.
These papers do not explicitly estimate design-based standard errors in their applications;
instead, they report \citet{conley1999gmm} standard errors.
\citet{Borusyak2023} additionally propose randomization inference with confidence intervals based on the sharp null hypothesis of constant treatment effects.

The present paper contributes to the literature by showing that design-based inference, beyond identification, is conceptually attractive, analytically tractable, and computationally straightforward.
Furthermore, I propose a data-driven method for inferring a plausible counterfactual distribution of the treatment under characteristics-determined treatment probabilities using CNNs, while prior work requires the researcher to specify it based on institutional knowledge.
Similar to \citet{Borusyak2023}, the methodological contributions of the present paper are not restricted to spatial settings.
The theoretical results for observational data extend existing results on double machine learning \citep{chernozhukov2018double} from the sampling-based framework to the design-based framework, resolving challenges that arise when observations are neither independent nor identically distributed in finite populations.

The remainder of this paper proceeds as follows.
Section~\ref{sec:setup} describes the framework and notation of this paper.
Section~\ref{sec:experimental} contains estimation and inference results under the ideal experiment.
Section~\ref{sec:observational} discusses observational data and describes the use of CNNs for finding counterfactual locations.\footnote{A documented code tutorial implementing the approach using CNNs is available at \url{https://github.com/michaelpollmann/spatialTreat-example}, in addition to the replication code accompanying this paper.}
Section~\ref{sec:applications} applies the methods of this paper to study the effects of grocery stores on foot traffic to nearby restaurants.
Section~\ref{sec:conclusion} concludes.

\section{Setup and notation}\label{sec:setup}

Both individuals (outcome units) and treatments are located in a shared (geographic) space.
Individuals, indexed by \(i \in \mathbb{I}\), have fixed location, or residence, \(r_{i} \in \mathbb{R}^{2}\) such as latitude and longitude.\footnote{It is not essential that locations are in two-dimensional space.}
In contrast to the standard setting of causal inference, treatments do not share the same index \(i\) with individuals.
Instead, the treatment takes values \(S \subset \mathbb{R}^{2}\) corresponding to locations in the same space as the individuals.

Each individual has a potential outcome \(Y_{i}(S)\) for each \(S\), and treatment effects are contrasts between different potential outcomes.
The individual-level treatment effect compares the outcome of \(i\) when there is treatment at location \(s\) vs. no treatment at \(s\), holding fixed treatments at other locations: \(\tau_{i}(s\mid S) \equiv Y_{i}(S \cup \{s\}) - Y_{i}(S \setminus \{s\})\).
\(\tau_{i}(s\mid S)\) is a marginal effect with background exposure \(S \setminus\{s\}\).
Of particular interest is the treatment effect of \(s\) when there is no other (relevant) treatment: \(\tau_{i}(s) \equiv \tau_{i}(s\mid \{s\}) = Y_{i}(\{s\}) - Y_{i}(\emptyset)\).
For ease of notation, define \(Y_{i}(s) \equiv Y_{i}(\{s\})\) and \(Y_{i}(0) \equiv Y_{i}(\emptyset)\).
I state the notation and results in this paper in terms of cross-sectional data only.
With panel data and staggered treatment adoption, all results remain unchanged under the same ideal experiment after subtracting the corresponding pre-treatment outcome from each (potential) outcome.

The experimental design generates randomness in where the treatment is realized.
I use calligraphic letters or hats to denote random variables in contrast to regular and Greek letters used for fixed values.
The realized treatment locations are \(\S \subset \mathbb{R}^{2}\), such that the observed outcome for individual \(i\) is \(\Y_{i} \equiv Y_{i}(\S)\).
Let \(\pi_{s} \equiv \Pr(\S \ni s)\) be the experimental probability of treatment at location \(s\).
The term \emph{candidate treatment locations} (\(\mathbb{S}\)) refers to locations \(s \in \mathbb{S} = \{s\in\mathbb{R}^{2}:\;\pi_{s}>0\}\), such that \(\S \subset \mathbb{S}\) with probability one.

The researcher is interested in the average effects of treatments on individuals who are a specific \emph{distance} away.
I denote the distance between \(s\) and \(r_{i}\) by \(d(s,r_{i})\).
The researcher chooses the distance function which is meaningful in their application such as ``straight line distance'' or driving time during rush hour (which may be asymmetric).
Importantly, like ``pre-treatment characteristic,'' the distance used must not depend on treatment assignment.

The estimand of interest is the expected (over the design distribution) average effect of the treatment on the treated (ATT) at a distance of approximately \(d\),
\begin{equation}\label{eq:tau-ATT}
\tau(d) \equiv \frac{\sum_{s\in\mathbb{S}} \Pr(\S \ni s) \sum_{i \in \mathbb{I}} w_{i}(s,d) \tau_{i}(s)}{\sum_{s \in \mathbb{S}} \Pr(\S \ni s) \sum_{i \in \mathbb{I}} w_{i}(s,d)},
\end{equation}
where the weights \(w_{i}(s,d)\) collect individuals at distance approximately \(d\).\footnote{In this paper, the choice of weights, such as binning, corresponds to the desired estimand rather than a kernel used to estimate a function at a point.}
In practice, researchers often bin individuals within a bandwidth \(h\) around \(d\) together when distance is a continuous variable using weights \(w_{i}(s,d) \equiv \ind\{\abs{d(s,r_{i})-d}\leq h\}\).
For simplicity of the results, throughout I assume that the weights are known (rather than estimated), \(w_{i}(s,d)\geq 0\), and \(w_{i}(s,d) = 0\) if \(d(s,r_{i}) > D\) for some \(D \in \mathbb{R}\) large enough that may depend on \(d\).
Let \(\mathbb{I}_{s} = \{i\in\mathbb{I}: w_{i}(s,d) \neq 0\}\).

Estimating the effect of one treatment compared to no treatment is impractical in some settings because multiple treatments are observed even in small areas.
Instead, the researcher may focus on an average \emph{marginal} effect of the treatment on the treated at \(d\):
\begin{equation}\label{eq:tau-ATT-marginal}
\tau_{\text{marginal}}(d) \equiv \frac{\sum_{S \in 2^{\mathbb{S}}} \Pr(\S = S) \sum_{s \in S} \sum_{i \in \mathbb{I}} w_{i}(s,d) \tau_{i}(s\mid S)}{\sum_{S \in 2^{\mathbb{S}}} \Pr(\S = S) \sum_{s \in S} \sum_{i \in \mathbb{I}} w_{i}(s,d)}
.
\end{equation}
This effect aggregates the marginal effects of location \(s\) given all possible background exposures \(S\setminus\{s\}\) \citep[cf.][]{savje2021average}.
The weights again resemble the ATT, placing more weight on assignments that are more likely to be realized.

Asymptotic results in this paper refer to sequences of finite but growing populations as in, for instance, \citet{abadie2020sampling}.
All randomness is due to treatment assignment.
Along the sequence, populations, indexed by \(k\), grow in the sense that \(\abs{\mathbb{S}_{k}} \to \infty\) and \(\abs{\mathbb{I}_{k}} \to \infty\) as \(k \to \infty\).
Additionally, for all asymptotic results, I assume outcomes and weights are bounded and weak bounds on the spatial concentration of observations hold as summarized in Assumption~\ref{as:regularity} in the appendix, with all bounds uniform over \(k\).
Hence, the asymptotic results most plausibly approximate the finite sample behavior of the estimators below when the number of treatment locations is large and spread out in space.
For readability, I suppress the dependence on the sequence index in the notation of the main part of the paper.

\section{Experimental data: estimation and inference}\label{sec:experimental}

I discuss estimators of average treatment effects on the treated (ATT) for a setting where all data are for a single large region with multiple candidate treatment locations across which treatment is randomized independently.

\begin{assumption}[Independent Assignment]\label{as:assign-independent}
    Treatment is assigned to candidate locations independently, with marginal probability \(\pi_{s} \equiv \Pr(\S \ni s)\) for location \(s\).
    For \(S \subset \mathbb{S}\):
    \begin{equation*}
    \Pr(\S = S) = \prod_{s \in S} \pi_{s} \prod_{s \in \mathbb{S} \setminus S} (1-\pi_{s}),
    \end{equation*}
    and the probability of treatment is bounded away from \(1\), \(\pi_{s} < 1-c\) for some \(c>0\).
\end{assumption}

The key idea of this section is that one can use assumptions motivated by the spatial nature of the treatments to derive estimators for treatment effects, as well as their standard errors.
Even without these assumptions, the estimators estimate meaningful \emph{marginal} effects (as defined in Equation~\ref{eq:tau-ATT-marginal}), see Theorem~\ref{thm:single-additive}(iii), and only some of the structure is needed for the approximate variance of the estimator to remain valid, see Theorem~\ref{thm:single-additive}(iv).

I first focus on an assumption of additive separability, also studied by \citet{Wang2025}, before also discussing identification and estimation under an alternative assumption.
\begin{assumption}[Additively Separable Effects]\label{as:additive-separability} The effects of the treatment are additively separable. For all \(i \in \mathbb{I}\), \(S \subset \mathbb{S}\) and \(s \in S\):
\begin{equation*}
Y_{i}(S) - Y_{i}(S \setminus \{s\}) = Y_{i}(\{s\}) - Y_{i}(\emptyset) \equiv \tau_{i}(s).
\end{equation*}
\end{assumption}
Intuitively, the assumption requires that returns to additional realized treatment locations are neither increasing nor decreasing in the number of realized treatment locations nearby.
Additively separable treatment effects are an appropriate specification if the effect of each treatment is independent of the realization of other treatments.
Additive separability implies that one can write \(Y_{i}(S) - Y_{i}(\emptyset) = \sum_{s \in S} \tau_{i}(s)\).
For instance, if, to a first approximation, the air pollution due to a power plant \citep{zigler2021bipartite} adds pollutants into the air without affecting pollutants added by other power plants, the effects of the plants on exposure to pollution are likely approximately additive.
The assumption does not impose homogeneity of treatment effects: It neither requires different treatment locations to have the same effect nor does it require a treatment location to have the same effect on two distinct individuals.

The estimator based on additive separability compares individuals at the distance of interest from realized treatment locations to (properly weighted) individuals at the distance of interest from unrealized treatment locations:
\begin{equation}\label{eq:tau-hat-single-additive}
\hat{\tau}(d) \equiv 
\frac{\sum_{s\in\mathbb{S}} \ind\{\S \ni s\} \sum_{i\in\mathbb{I}} w_{i}(s,d) \Y_{i}}{\sum_{s\in\mathbb{S}} \ind\{\S \ni s\} \sum_{i\in\mathbb{I}} w_{i}(s,d)}
-
\frac{\sum_{s\in\mathbb{S}} \frac{\ind\{\S \niton s\}}{1-\pi_{s}}\pi_{s} \sum_{i\in\mathbb{I}} w_{i}(s,d) \Y_{i}}{\sum_{s\in\mathbb{S}} \frac{\ind\{\S \niton s\}}{1-\pi_{s}} \pi_{s} \sum_{i\in\mathbb{I}} w_{i}(s,d)}
.
\end{equation}

To derive standard errors and show asymptotic normality, I assume that treatment effects are local: Treatments have no or relatively small effects on individuals far away from them.
Let \(\dnoeffect{}\) be the known distance denoting ``far away,'' specified appropriately by the researcher depending on the treatment and outcome of interest.
For individual \(i\), define the approximate exposure mapping \citep{Saevje2023} based on the set \(\mathbb{M}_{i} \equiv 2^{\{s \in \mathbb{S}: \; d(s,r_{i}) \leq \dnoeffect{}\}}\).
Let the random variable \(\M_{i} \in \mathbb{M}_{i}\) be the realized approximate exposure, determining the realized treatment state of all candidate treatment locations within \(\dnoeffect{}\) of \(i\).
Denote \(i\)'s ``potential outcome'' under exposure \(m \in \mathbb{M}_{i}\) by \(Y_{i}(m) \equiv E(\Y_{i} \mid \M_{i}=m)\).

\begin{assumption}[Limit on Effects After Distance \(\dnoeffect{}\)]\label{as:no-effect-dist} Either
\begin{enumerate}
    \item[(a)] the treatment has no effect at distances larger than \(\dnoeffect{}\) such that for all \(i \in \mathbb{I}\), \(S \subset \mathbb{S}\) with \(s \in S\): if \(d(s,r_{i}) > \dnoeffect{}\), then \(Y_{i}(S) = Y_{i}(S \setminus \{s\})\); or
    \item[(b)] the effect of treatments at distances larger than \(\dnoeffect{}\) are small relative to the variation in outcomes due to treatments at shorter distances: \(\frac{\epsilon}{\sigma} = o_{p}(1)\), where \\
\(\epsilon \equiv 
\frac{1}{\abs{\mathbb{S}}}\sum_{s\in\mathbb{S}} \sum_{i\in\mathbb{I}} \frac{w_{i}(s,d)}{\bar{n}(d)}
\Bigl(\ind\{\S\ni s\} - \frac{\ind\{\S\niton s\} \pi_{s}}{1-\pi_{s}}\Bigr) (\Y_{i} - Y_{i}(\M_{i}))\) \\
with \(\bar{n}(d) \equiv \frac{1}{\mathbb{S}} \sum_{s\in\mathbb{S}} \pi_{s} \sum_{i\in\mathbb{I}} w_{i}(s,d)\) the average (per location) expected number of treated individuals, and \(\sigma^{2}\) is the variance under (a) of an infeasible estimator \(\tilde{\tau}(d) \approx \hat{\tau}(d)\), defined in Theorem~\ref{thm:single-additive}.
\end{enumerate}
\end{assumption}

Assumption~\ref{as:no-effect-dist}(b), which is implied by \ref{as:no-effect-dist}(a), allows treatments to affect outcomes of faraway individuals.
The term \(\epsilon\) captures the realized contribution to outcomes of treatment locations farther than \(\dnoeffect{}\) away.
By construction, \(E(\epsilon)=0\).
In contrast, \(\sigma\) captures the variation due to treatment locations closer than \(\dnoeffect{}\).
For instance, individuals not affected by treatments beyond a distance of \(\dnoeffect{}\) have \(\Y_{i} = Y_{i}(\M_{i})\) with probability 1 and therefore contribute only to \(\sigma\) but not to \(\epsilon\).
Hence, Assumption~\ref{as:no-effect-dist}(b) only requires that there is a distance \(\dnoeffect{}\) capturing \emph{most} of the effects.

The following theorem describes the (approximate) finite population properties of \(\hat{\tau}(d)\).
It uses an approximation to \(\hat{\tau}(d)\) that replaces its stochastic denominators by their expectations and recenters its numerators appropriately.
Specifically, let
\begin{equation*}
\tilde{\tau}(d) \equiv \tau_{\text{marginal}}(d)
+
\frac{1}{\abs{\mathbb{S}}}\sum_{s\in\mathbb{S}} \sum_{i\in\mathbb{I}} \frac{w_{i}(s,d)}{\bar{n}(d)}
\Bigl(\ind\{s\ni\S\} (\Y_{i} - \mu_{t}(d))
- \frac{\ind\{s\niton\S\} \pi_{s}}{1-\pi_{s}} (\Y_{i} - \mu_{c}(d))\Bigr)
\end{equation*}
where \(\mu_{t}(d)\) and \(\mu_{c}(d)\) are defined with the same weights as \(\tau_{\text{marginal}}(d)\) (Equation~\ref{eq:tau-ATT-marginal}) but with \(Y_{i}(S)\) and \(Y_{i}(S\setminus \{s\})\), respectively, replacing \(\tau_{i}(s\mid S)\).

\begin{theorem}\label{thm:single-additive}
The estimator \(\hat{\tau}(d)\) is similar to the infeasible estimator \(\tilde{\tau}(d)\), which has analytically tractable design-based properties:
\begin{enumerate}[label={(\roman*)}]
    \item Under regularity conditions (Assumption~\ref{as:regularity} in the appendix): \(\hat{\tau}(d) - \tilde{\tau}(d) \to_{p} 0\).
    \item Under Assumptions~\ref{as:assign-independent}~and~\ref{as:additive-separability}: \(E(\tilde{\tau}(d)) = \tau(d)\) (unbiasedness for ATT).
    \item Under Assumption~\ref{as:assign-independent}: \(E(\tilde{\tau}(d)) = \tau_{\text{marginal}}(d)\) (unbiasedness for marginal ATT).
    \item Under Assumptions~\ref{as:assign-independent}~and~\ref{as:no-effect-dist}(a): The variance of \(\tilde{\tau}(d)\) is\\
    \(
    \var(\tilde{\tau}(d))
    = \sigma^{2} \equiv
    \bigl(\tilde{V}_{t}(d) + \tilde{V}_{c}(d) + \tilde{V}_{\times}(d) - \tilde{V}_{tt}(d) - \tilde{V}_{cc}(d) - \tilde{V}_{ct}(d)\bigr) / \abs{\mathbb{S}}
    \)\\
    with the notation defined in Appendix~\ref{app:proof-thm-single-additive}.
    \item Under Assumptions~\ref{as:assign-independent},~\ref{as:no-effect-dist}(b), and regularity conditions (Assumption~\ref{as:regularity} in the appendix): \\
    \(\frac{\tilde{\tau}(d)-\tau_{\text{marginal}}(d)}{\sigma} \stackrel{d}{\to} \mathcal{N}(0,1)\).
\end{enumerate}
\end{theorem}

\begin{remark}
The variance expression contains conceptually similar terms to the finite population design-based variance of the difference-in-means estimator in standard randomized experiments with individual-level treatments.
The first two terms, \(\tilde{V}_{t}\) and \(\tilde{V}_{c}\), resemble variances of individual level potential outcomes corresponding to treatment and control of a candidate location at distance \(d\).
The third term, \(\tilde{V}_{\times}(d)\) takes observable cross-products between distinct individuals, candidate treatment locations, or treatment states.
The third term (jointly with the weighting inside \(\tilde{V}_{t}\) and \(\tilde{V}_{c}\)) adjusts for the correlation in the exposure to treatment of individuals who are close to one another, as well as for individuals with positive weight \(w_{i}(s,d)\) for multiple different locations \(s\).
The final three terms, \(\tilde{V}_{tt}(d)\), \(\tilde{V}_{cc}(d)\), and \(\tilde{V}_{ct}(d)\), are averages of squares of differences in potential outcomes that cannot be observed simultaneously, and are therefore unobservable similar to the variance of treatment effects in the finite population design-based variance of the difference-in-means estimator in standard experiments.
Dropping the final three terms yields a conservative estimator of the variance because these terms are non-negative by construction.
\end{remark}

\begin{remark}
As a general guideline, to meaningfully reduce the design-based variance of the estimator, one needs to expand the sampling area, rather than the number of individuals or candidate locations within a fixed area.
Adding individuals while holding the sample area fixed does not generally reduce the variance of the estimator.
The effect of increasing the number of (candidate) treatment locations within a fixed sample area on the variance of the estimator is more nuanced.
However, the \emph{estimator} of that variance will generally become more conservative because additional candidate locations increase the number of unobservable and hence inestimable treatment configurations (captured by \(\tilde{V}_{cc}(d)\), \(\tilde{V}_{tt}(d)\), and \(\tilde{V}_{ct}(d)\)) exponentially.
\end{remark}

\begin{remark}
The theorem shows attractive properties of an estimator for the effects of spatial treatments under a \emph{design-based} framework leveraging the randomization Assumption~\ref{as:assign-independent} for causal identification.
A popular alternative estimator, comparing individuals on an inner ring who are near the treatment to individuals on an outer ring who are farther away from treatment, typically combined with a before vs. after comparison in a difference-in-differences setup, is not justified by the same design-based assumption.
This lack of a natural design-based interpretation of a particular difference-in-differences estimator in spatial settings is in contrast to the standard difference-in-differences setting and estimator, where a natural form of randomization can ensure parallel trends by design.
\end{remark}

\begin{remark}
In contemporaneous work, \citet{Wang2025} provide a result similar to Theorem~\ref{thm:single-additive} but focus on a simple difference-in-means estimator that arises after averaging outcomes by candidate treatment location and assuming the probability of treatment is constant across candidate locations.
The setup of the present paper nests these choices but allows me to derive more general results, covering estimands targeted by existing empirical studies of spatial treatments and accommodating heterogeneous treatment probabilities that are commonly found in the non-experimental settings considered in the following section.
Expressions of the design-based variance admitting simple (conservative) estimators without restrictions on treatment effect heterogeneity were first circulated in early versions of the present paper.
The asymptotic normality result of Theorem~\ref{thm:single-additive}(v) allows treatments to have (small) effects even at large distances (Assumption~\ref{as:no-effect-dist}(b)), while the earlier result of \citet{Wang2025} relies on an analog of the stronger Assumption~\ref{as:no-effect-dist}(a) of no effects past some distance.
Due to the aggregation of outcomes by candidate treatment location in the analysis of \citet{Wang2025}, their assumptions are not expressed in terms of the individual-level potential outcomes and locations that are part of the primitive conceptual objects of the present paper.
Hence, while closely related, the assumptions are not nested and the formulas do not coincide exactly even when specialized to the estimand they consider.
\end{remark}

\begin{remark}
When the candidate treatment locations are sufficiently far apart that individuals receiving non-zero weight are affected by at most one location, the setting corresponds to clustered assignment in randomized experiments.
Intuitively, Assumption~\ref{as:no-effect-dist} allows for ``overlapping clusters'' where any individual may be part of an arbitrary number of clusters.
\end{remark}

\begin{remark}
The variance in Theorem~\ref{thm:single-additive}(iv) is for \(\tilde{\tau}(d)\) as an estimator for the in-sample ATT defined in Equation~\ref{eq:tau-ATT-marginal}.
It relies solely on randomness due to treatment assignment, not sampling.
The underlying thought experiment (repeated samples re-assign the treatment to the candidate locations) is easy to articulate and corresponds to the variation required for interpretation as a causal effect.
Hence, the researcher does not need to additionally specify a hypothetical super-population and how the sample arose from it.
\end{remark}

\begin{figure}
    \centering
    \begin{subfigure}[t]{0.49\textwidth}
    \includegraphics{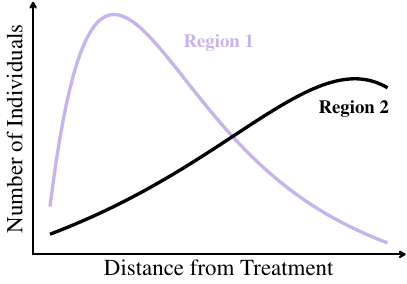}
    \caption{Individuals per distance in each region}\label{fig:att-vs-att-eq-population}
    \end{subfigure}
    \hfill
    \begin{subfigure}[t]{0.49\textwidth}
    \includegraphics{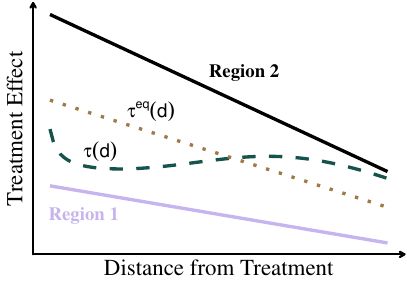}
    \caption{Estimands}\label{fig:att-vs-att-eq-effects}
    \end{subfigure}
    \caption{\label{fig:att-vs-att-eq} The estimands \(\tau(d)\) and \(\tau^{\text{eq}}(d)\) can meaningfully differ from one another.
    Suppose the sample area can be separated into regions such that outcomes in a region are only affected by the single candidate treatment location within the same region.
    Consider two types of regions, whose candidate treatment locations are equally likely to be treated and which have individuals distributed across space as given in panel~(\subref{fig:att-vs-att-eq-population}).
    Panel~(\subref{fig:att-vs-att-eq-effects}) shows the decay of ATTs over distance for each region as a solid line.
    The dashed line shows the estimand \(\tau(d)\), which weights by the relative number of individuals at distance \(d\) and is increasing in distance over some range.
    The dotted line shows the estimand \(\tau^{\text{eq}}(d)\), which weights the regions equally and decreases monotonically.
    }
\end{figure}

\begin{remark}
The estimator \(\hat{\tau}(d)\) with distance bin weights places equal weight on all individuals at distance \(d\pm h\) from a candidate treatment location (up to the ATT weights reflecting treatment probabilities).
At least two alternatives may be worthwhile.
First, one can apply equal weights to each treatment location, rather than individual, by taking \(w_{i}^{\text{eq}}(s,d) \equiv \ind\{\abs{d(s,r_{i})-d}\leq h\} / \sum_{i' \in \mathbb{I}} \ind\{\abs{d(s,r_{i'})-d}\leq h\}\).
These weights facilitate interpretation of the estimand across distance if there is substantial heterogeneity in population counts over distance and treatment effects by treatment location; see Figure~\ref{fig:att-vs-att-eq} for an illustration.
Second, researchers may deviate from the distance bin weight by choosing a kernel that is continuous in distance, such as triangular weights \(w_{i}^{\text{tri}}(s,d) \equiv (1 - \abs{d - d(s,r_{i})}/h) \ind\{\abs{d(s,r_{i})-d}\leq h\}\).
In the framework of this paper, changes to the weights change both the estimator and estimand.
In practice, the resulting estimator may be more robust to small errors in locations and may have more attractive properties if one wishes to estimate the \emph{function} \(\tau(\cdot)\) in a framework where asymptotically there are individuals arbitrarily close to any distance of interest \(d\) for this estimand to be well-defined.
\end{remark}

While the results above describe an estimator motivated by additive separability, the ideas in this paper can be used to motivate estimators and derive their properties under alternative assumptions such as:
\begin{assumption}[Only Nearest Realized Location Matters]\label{as:only-nearest} For all \(i \in \mathbb{I}\), \(S \subset \mathbb{S}\) with \(s \in S\), \(s' \in \mathbb{S}\): if \(d(s,r_{i}) \leq d(s',r_{i})\), then \(Y_{i}(S) = Y_{i}(S \cup \{s'\}) = Y_{i}(S \setminus \{s'\})\).
\end{assumption}

Typically, only the nearest realized treatment location matters if individuals only access, or visit, a single realized treatment location.
For instance, if a developing country quasi-randomly chooses locations to construct new schools \citep[as studied by][using a difference-in-differences design]{duflo2001schooling}, it may be plausible to assume that only the nearest built school matters to an individual.
For the effects of infrastructure projects, such as additional bus or subway stops, on commute times and real estate prices \citep{gupta2022take}, the appropriate assumption may depend on the type of transit stop.
An additive effects specification for bus or subway stops may be a good approximation if each stop gives access to a different transit line.
A specification where only the nearest stop matters may be more appropriate for stops of the same line.

\begin{theorem}\label{thm:only-nearest-identification}
The average effect of the treatment on the treated, \(\tau(d)\), is nonparametrically identified if Assumptions~\ref{as:assign-independent},~\ref{as:no-effect-dist}(a),~and~\ref{as:only-nearest} are satisfied.\\
\end{theorem}

The proof of Theorem~\ref{thm:only-nearest-identification} is constructive in that it suggests an estimator that exploits the combination of assumptions: 
\begin{equation*}
\hat{\tau}_{\text{nearest}}(d)
\equiv
\frac{\sum_{s \in \S} \sum_{i \in \mathbb{I}} \frac{\N_{i}(s) w_{i}(s,d)}{\Pr(\N_{i}(s)=1 \mid \S \ni s)} \Y_{i}}{\sum_{s \in \S} \sum_{i \in \mathbb{I}} \frac{\N_{i}(s) w_{i}(s,d)}{\Pr(\N_{i}(s)=1 \mid \S \ni s)}}
-
\frac{\sum_{s \in \mathbb{S}\setminus\S} \frac{\pi_{s}}{1-\pi_{s}} \sum_{i \in \mathbb{I}} \frac{\N_{i}(0) w_{i}(s,d)}{\Pr(\N_{i}(0)=1 \mid \S \niton s)} \Y_{i}}{\sum_{s \in \mathbb{S}\setminus\S} \frac{\pi_{s}}{1-\pi_{s}}  \sum_{i \in \mathbb{I}} \frac{\N_{i}(0) w_{i}(s,d)}{\Pr(\N_{i}(0)=1 \mid \S \niton s)}}
\end{equation*}
where \(\N_{i}(s)\) is an indicator for \(s\) being the nearest realized treatment location to \(i\), and \(\N_{i}(0)\) is an indicator for no treatment location within \(\dnoeffect{}\) of \(i\) being realized.
See Online~Appendix~6 for additional discussion.

\section{Observational data: theory and implementation}\label{sec:observational}

Often, researchers can only study spatial treatments in observational, rather than experimental, data.
I first lay out assumptions and theory connecting ``quasi-experimental'' analysis using observational data to the experimental analysis outlined above.
Then, I discuss the practical implementation of estimation under such assumptions with spatial data.

\subsection{Assumptions and Theory}

In Sections~\ref{ssub:latent_experiment_perspective},~\ref{ssub:hypothetical_experiment_perspective},~and~\ref{ssub:conditional_inference_perspective}, I discuss three internally coherent perspectives on ``quasi-experimental'' analysis using observational data that guide statistical inference.
The perspectives are each founded in existing theoretical work but yield distinct prescriptions for inference and interpretation.
The researcher chooses the perspective they find attractive for their application.

For any of the three perspectives, the unknown or hypothetical experiment satisfies an assumption that the probability of treatment depends solely on observable characteristics, not on potential outcomes, aligning with the concept of unconfoundedness in sampling-based analyses.
Intuitively, among observationally similar locations, treatment assignment is as good as random.
Let \(Z_{s}\) be some observable characteristics of the spatial neighborhood of \(s\in\tilde{\mathbb{S}} \subset \mathbb{R}^{2}\), where \(\tilde{\mathbb{S}}\) are the locations under consideration, and let \(\mathbb{S} \equiv \{s\in\tilde{\mathbb{S}}:\; \pi_{s}>0\}\) be the (potentially known) set of locations with positive probability of treatment.
Define \(\mathbb{Z} \equiv \{Z_{s}:\; s \in \mathbb{S}\}\) and \(\tilde{\mathbb{Z}} \equiv \{Z_{s}:\; s \in \tilde{\mathbb{S}}\}\).

\begin{assumption}[Characteristics-Determined Probabilities]\label{as:spatial-unconfoundedness}
Among a known set of locations \(s \in \tilde{\mathbb{S}} \subset \mathbb{R}^{2}\), the treatment assignment probabilities depend solely on observed characteristics, \(\pi_{s} \equiv \Pr(\S \ni s) = p(Z_{s})\) where \(Z_{s}\) are observable characteristics of the spatial neighborhood of \(s\) and \(p: \tilde{\mathbb{Z}} \to [0,1]\) is a (possibly unknown) function.
\end{assumption}
In spatial settings, researchers may commonly wish to invoke Assumption~\ref{as:spatial-unconfoundedness} based on a relatively large set of (or ``high-dimensional'') characteristics \(Z_{s}\).
These characteristics may include not just information about the point \(s\) itself, but also information about the surrounding neighborhood.

The different perspectives on inference outlined below each describe the (asymptotic) distribution \(F_{\hat{\tau},P}\) of an estimator \(\hat{\tau}\) over the treatment assignment distribution \(P\).
The perspectives differ in their choices for \(\hat{\tau}\) and \(P\).

Each of the results in this section also invokes Assumption~\ref{as:assign-independent} (independent assignment) to limit spatial dependence in the data.
If latent treatment probabilities depend on spatially correlated characteristics, incorrectly omitting such characteristics from the conditioning set when invoking Assumption~\ref{as:spatial-unconfoundedness} can lead not only to bias but also to incorrect standard errors.
Specifically, treatment assignment may behave as if it was spatially correlated rather than independent in that case.

\subsubsection{Latent Experiment Perspective}\label{ssub:latent_experiment_perspective}
The data are the product of a latent experiment with unknown treatment probabilities \(\pi_{s}\).
However, because treatment probabilities are deterministic functions of observable characteristics by Assumption~\ref{as:spatial-unconfoundedness}, one may be able to consistently estimate the \(\pi_{s}\) under Assumption~\ref{as:assign-independent}.
The experimental estimator discussed in the previous section is then consistent by standard arguments even in observational settings.
However, inference needs to account for (design-based) uncertainty in the estimated treatment probabilities:
If the latent experiment had resulted in a different set of realized treatment locations, the estimated treatment probabilities could be different.
Because the true experiment is unknown, Assumption~\ref{as:spatial-unconfoundedness} typically needs to hold for \(\tilde{\mathbb{S}}\) the subset of \(\mathbb{R}^{2}\) corresponding to the study area, while the unknown \(\mathbb{S} \equiv \{s \in \mathbb{R}^{2}: \; \pi_{s} > 0\}\) is typically finite.

I show that for a class of estimators, the uncertainty in estimated treatment probabilities does not affect the asymptotic distribution of the estimators to first order.
Extending recent theoretical advances \citep[for instance,][]{chernozhukov2018double} to a design-based finite population framework with spatial dependence, I study the estimator \(\hat{\tau}_{\text{dml}}(d) = \hat{\tau}(d,\hat{\pi},\hat{\mu})\) where
\begin{equation*}
\hat{\tau}(d,p,m) = 
\frac{\sum_{i \in \mathbb{I}} \sum_{s \in \mathbb{S}} w_{i}(s,d) \Bigl(\ind\{\S \ni s\} - \frac{\ind\{\S \not\ni s\}}{1-p_{s}} p_{s}\Bigr) (\Y_{i} - m_{s,i}(d))}
    {\sum_{i \in \mathbb{I}} \sum_{s \in \mathbb{S}} w_{i}(s,d) \ind\{\S \ni s\}}
\end{equation*}
with a choice of first step estimates for \(p\) and \(m\), \(\hat{\pi}\) and \(\hat{\mu}\), based on cross-fitting described below.
With (infeasible) \(\pi \equiv (\pi_{s})_{s\in\mathbb{S}}\) and \(\mu\) such that, using covariates \(Z_{s,i}\) containing \(Z_{s}\),
\begin{equation*}
\mu_{s,i}(d) \equiv \frac{\sum_{s'\in\mathbb{S}} \sum_{i'\in\mathbb{I}} w_{i'}(s',d) \ind\{Z_{s',i'}=Z_{s,i}\} E(\Y_{i'} \mid \S \niton s')}{\sum_{s'\in\mathbb{S}} \sum_{i'\in\mathbb{I}} w_{i'}(s',d) \ind\{Z_{s',i'}=Z_{s,i}\}}
\end{equation*}
the estimator \(\hat{\tau}(d,\pi,\mu)\) is similar to the estimator \(\hat{\tau}(d)\) in Equation~\ref{eq:tau-hat-single-additive} with ``demeaned'' outcomes, and a result analogous to Theorem~\ref{thm:single-additive} applies.

The first step estimates \(\hat{\pi}_{s}\) and \(\hat{\mu}_{s,i}\) are based on a random split of the candidate treatment locations into \(F\) folds, with \(F\) fixed, denoted by the partition \((\mathbb{S}_{f})_{f=1}^{F}\) of \(\mathbb{S}\) with \(f(s)\) the fold of \(s\).
Create estimation samples containing observations that are design-based independent of observations in a fold \(f\) as follows.
For \(I\subset\mathbb{I}\), let \(\mathbb{S}_{I} \equiv \{s\in\mathbb{S}: d(s,r_{i}) < \dnoeffect{} \text{ for some } i\in I\}\)
be the locations that may affect individuals in \(I\) under Assumption~\ref{as:no-effect-dist}(a).
Let \(\mathbb{I}_{f} \equiv \{i\in\mathbb{I}: d(s,r_{i})<\dnoeffect \text{ for some } i\in\mathbb{S}_{f}\}\) be the individuals who may be affected by treatment at locations in fold \(f\) under Assumption~\ref{as:no-effect-dist}(a).
Let \(\tilde{\mathbb{I}}_{f} \equiv \bigcup_{s\in\mathbb{S}_{f}} \mathbb{I}_{s}\)
be the individuals who receive weight by the estimand for being near a location in fold \(f\).
The estimate \(\hat{\pi}_{s}\) is based only on candidate locations \(s' \in \mathbb{S} \setminus (\mathbb{S}_{f(s)}\cup\mathbb{S}_{\tilde{\mathbb{I}}_{f(s)}})\) that are neither in the same fold nor affect individuals receiving weight in the estimand for being near a location in the same fold.
The estimate \(\hat{\mu}_{s,i}\) is based only on individuals \(i' \in \mathbb{I} \setminus (\mathbb{I}_{f(s)}\cup\tilde{\mathbb{I}}_{f(s)})\) who are neither affected by locations in the same fold nor receive weight in the estimand for being near a location in the same fold.
The estimation samples are weighted random samples: locations and individuals far away from other locations are more likely to be included, but sample splitting is random.
Hence, it is straightforward to reweight observations inversely by their inclusion probabilities, which, intuitively, are related to the probability that all locations are not in fold \(f(s)\).
These sample splitting estimates ensure design-based independence of \((\hat{\pi}_{s},\hat{\mu}_{s,i})\) and \((\ind\{\S \ni s\},\Y_{i}\})\) under Assumptions~\ref{as:assign-independent}~and~\ref{as:no-effect-dist}(a).
The individual folds cannot differ too much from the full sample, in the sense that, for each fold \(f\), \(\frac{1}{\abs{\mathbb{S}}} \sum_{s\in\mathbb{S}_{f}}\sum_{i\in\mathbb{I}_{s}} (\tilde{\mu}_{s,i} - \mu_{s,i})^{2} = O_{p}(\abs{\mathbb{S}}^{-1/2})\) where
\(
\tilde{\mu}_{s,i} = \frac{\sum_{s'\in\mathbb{S}_{f(s)}} \sum_{i'\in\mathbb{I}} w_{i'}(s') \ind\{Z_{s',i'}=Z_{s,i}\} E(\Y_{i'} \mid \S \niton s')}{\sum_{s'\in\mathbb{S}_{f(s)}} \sum_{i'\in\mathbb{I}} w_{i'}(s') \ind\{Z_{s',i'}=Z_{s,i}\}}
\)
similar to \(\mu\) but averaging only within fold \(f(s)\).
Assumption~\ref{as:regularity}(d) in the appendix formally imposes this condition.
For random sample splits, this bound is attained if the number of distinct values \(Z_{s,i}\) takes on is \(O(\sqrt{\abs{\mathbb{S}}})\).

\begin{assumption}\label{as:nuisance-quality}
The first step estimates using sample splits described above converge sufficiently quickly such that, \\
(i) \(\sqrt{\abs{\mathbb{S}}^{-1} \sum_{s\in\mathbb{S}} \sum_{i\in\mathbb{I}_{s}} (\hat{\mu}_{s,i} - \mu_{s,i})^{2}} = o_{p}(\abs{\mathbb{S}}^{-1/4})\), \\
(ii) \(\sqrt{\abs{\mathbb{S}}^{-1} \sum_{s\in\mathbb{S}} (\hat{\pi}_{s}-\pi_{s})^{2}} = o_{p}(\abs{\mathbb{S}}^{-1/4})\), and \\
(iii) with probability one, \(\hat{\pi}_{s} \in [0,1-c]\) for some \(c>0\), and \(\hat{\mu}_{s,i}\) is bounded for all \(s\) and \(i\).
\end{assumption}

\begin{theorem}\label{thm:dml}
Suppose Assumptions~\ref{as:assign-independent},~\ref{as:no-effect-dist}(a),~\ref{as:spatial-unconfoundedness},~and~\ref{as:nuisance-quality}, as well as regularity conditions (Assumption~\ref{as:regularity} in the appendix), hold, and \((\hat{\tau}(d,\pi,\mu) - \tau_{\text{marginal}}(d))/\sigma \stackrel{d}{\to}\mathcal{N}(0,1)\) where \(\sigma^{2} \equiv \var(\tilde{\tau}(d,\pi,\mu))\), with \(\tilde{\tau}(d,p,m)\) equal to \(\hat{\tau}(d,p,m)\) except that the denominator is replaced by its expectation, is such that \(1/\sigma = O(\sqrt{\abs{\mathbb{S}}})\).
Then,
\begin{equation*}
\frac{\hat{\tau}(d,\hat{\pi},\hat{\mu})-\tau_{\text{marginal}}(d)}{\sigma} \stackrel{d}{\to} \mathcal{N}(0,1)
\end{equation*}
\end{theorem}

\begin{remark}
The theorem describes the asymptotic distribution \(F_{\hat{\tau}(d,\hat{\pi},\hat{\mu}),P}\) of the estimator \(\hat{\tau}(d,\hat{\pi},\hat{\mu})\) (including nuisance function estimation) across the treatment assignment distribution \(P\) arising from independent assignment according to the marginal probabilities \(\pi_{s}\) of the true latent experiment.
\end{remark}

\begin{remark}
The variance in the theorem does not depend on the quality of the estimators of the first step estimates \(\hat{\pi}\) and \(\hat{\mu}\) as long as they satisfy the rate conditions.
Hence, when using this estimator of the ATT, one can effectively ``ignore'' the noise due to estimated treatment probabilities at the cost of also requiring an estimate of the conditional mean.
\end{remark}

\begin{remark}
The assumption on the infeasible \(\hat{\tau}(d,\pi,\mu)\) typically holds based on an analog of Theorem~\ref{thm:single-additive}.
\end{remark}

The rate conditions in the theorem above allow for growing dimensionality of characteristics.
For notational simplicity, I derive convergence rates (Theorems~\ref{thm:prop-score-hd}~and~\ref{thm:spatial-lasso}) for estimators without sample-splitting.

For a simple example of the estimator \(\hat{\pi}\), let \(\hat{\pi}_{s} = \sum_{s'\in\tilde{\mathbb{S}}} \ind\{Z_{s'} = Z_{s}\}\ind\{\S \ni s\}/ n_{Z_{s}}\), where \(n_{z} = \sum_{s\in\tilde{\mathbb{S}}} \ind\{Z_{s} = z\}\), be the average treatment status of locations with the same characteristics as \(s\).

\begin{theorem}\label{thm:prop-score-hd} Under Assumptions~\ref{as:assign-independent}~and~\ref{as:spatial-unconfoundedness}, if \(\abs{\mathbb{Z}} = o(\sqrt{\abs{\mathbb{S}}})\), then \(\sqrt{\abs{\mathbb{S}}^{-1} \sum_{s\in\mathbb{S}} (\hat{\pi}_{s}-\pi_{s})^{2}} = o_{p}(\abs{\mathbb{S}}^{-1/4})\), and \(\hat{\pi}_{s} = 0\) with probability 1 for \(s\in\tilde{\mathbb{S}}\setminus\mathbb{S}\).
\end{theorem}

\begin{remark}
The particular estimator \(\hat{\pi}\) above is most appropriate when the characteristics \(Z_{s}\) correspond to discrete types.
Intuitively, one may think of the CNNs proposed below as defining locations to have the same type if their neighborhoods, discretized through a fine grid, are identical up to rotation and mirroring.
\end{remark}

\begin{remark}
The convergence rate does not depend on the cardinality of the set of characteristics \(\tilde{\mathbb{Z}} \setminus \mathbb{Z}\) of locations where the unknown probability of treatment is zero.
This result is useful because in spatial settings there may often be very many or even infinitely many possible locations, and hence many possible characteristics, with almost all never appearing with any treatment.
\end{remark}

For the conditional outcome mean, consider a parametric assumption such as \(\mu_{s,i} = Z_{s,i} \beta\) for \(L\)-dimensional bounded row vector \(Z_{s,i}\).
Under standard assumptions, one may estimate \(\beta\) by weighted regression of \(\Y_{i}\) on \(Z_{s,i}\) using each \((s,i) \in \mathbb{S} \times \mathbb{I}\) pair with \(w_{i}(s,d) \neq 0\) and \(s \notin \S\) as observations with weights \(w_{i}(s,d)\).
In settings where the number of covariates grows (slowly) along the sequence of finite populations, the theorem below considers the estimator including a LASSO penalty under a sparsity assumption.
To define the LASSO estimator formally, define the vector \(\boldsymbol{\Y}\) and matrix \(\boldsymbol{\Z}\) to include all pairs \((s,i) \in \mathbb{S} \times \mathbb{I}\) with \(w_{i}(s,d) \neq 0\), such that for row \(j\), \(\Y_{j} \equiv \sqrt{w_{i(j)}(s(j),d)} \ind\{\S \niton s(j)\} \Y_{i(j)}\) and \(\Z_{j} \equiv \sqrt{w_{i(j)}(s(j),d)} \ind\{\S \niton s\} Z_{i(j),s(j)}\).
Both \(\boldsymbol{\Y}\) and \(\boldsymbol{\Z}\) are random due to their dependence on \(\S\).
Let \(\dot{n}\) be the number of rows of \(\boldsymbol{\Y}\), and \(\lambda\) the penalty parameter.
Then
\begin{equation*}
\hat{\beta} \in \arg \min_{\beta \in \mathbb{R}^{L}} \frac{1}{2 \dot{n}}(\boldsymbol{\Y} - \boldsymbol{\Z}\beta)'(\boldsymbol{\Y} - \boldsymbol{\Z}\beta) + \lambda \sum_{l=1}^{L} \abs{\beta_{l}},
\qquad
\hat{\mu}_{s,i} = Z_{s,i} \hat{\beta}
.
\end{equation*}

The formal result below, bounding the estimation error, uses a restriction of spatial dependence, which, effectively, is a strengthened form of Assumption~\ref{as:no-effect-dist}(b).
The condition groups location-individual pairs \((s,i)\) into clusters \(c(s,i) \in \mathbb{C}\) such that dependence across clusters is limited.
For pair \((s,i)\), define the approximate exposure mapping based on the set \(\mathbb{M}_{s,i} \equiv 2^{\{s' \in \mathbb{S}: \; c(s,i)=c(s',i') \text{ for some } i'\}}\).
Let the random variable \(\M_{s,i} \in \mathbb{M}_{s,i}\) be the realized approximate exposure, determining the realized treatment state of all candidate treatment locations in the same cluster.
Denote the \((s,i)\) ``potential outcome'' under exposure \(m \in \mathbb{M}_{s,i}\) by \(Y_{s,i}(m) \equiv E(\Y_{i} \mid \M_{s,i}=m)\), and the vector of weighted potential outcomes by \(\boldsymbol{Y}^{\boldsymbol{\M}}\) with \(\boldsymbol{Y}^{\boldsymbol{\M}}_{j} \equiv \sqrt{w_{i(j)}(s(j),d)} \ind\{\S \niton s(j)\} Y_{s(j),i(j)}(\M_{s(j),i(j)})\).

\begin{assumption}[Many approximate clusters]\label{as:approximate-clusters}
Observations \((s,i) \in \mathbb{S} \times \mathbb{I}\) with \(w_{i}(s,d) \neq 0\) can be grouped into approximate clusters \(c(s,i) \in \mathbb{C}\) such that (i) the number of observations per cluster is bounded, (ii) \(c(s,i) = c(s,i')\) for each \(s\) and any \(i,i'\), and (iii) \(\max_{l=1,\dots,L}\bigl\{\frac{\boldsymbol{\Z}_{\cdot,l}'(\boldsymbol{\Y} - \boldsymbol{Y}^{\boldsymbol{\M}})}{\dot{n}}\bigr\} = o_{p}(\sqrt{\ln(L)/\abs{\mathbb{C}}})\) with probability approaching 1.
\end{assumption}

In Assumption~\ref{as:approximate-clusters}(iii), \(\boldsymbol{\Y} - \boldsymbol{Y}^{\boldsymbol{\M}}\) computes how much treatments outside an observation's cluster affect its outcome under the realized assignment.
These out-of-cluster effects are multiplied with the regressors and averaged over observations.
The LASSO relies on the largest covariance between regressors and residuals, \(\max_{l}\{\frac{\boldsymbol{\Z}_{\cdot,l}'(\boldsymbol{\Y} - \boldsymbol{\Z}\beta)}{\dot{n}}\}\), vanishing not too slowly, where \(\boldsymbol{\Z}_{\cdot,l}\) is the \(l\)-th column of the regressor matrix \(\boldsymbol{\Z}\).
When \(\max_{l}\{\frac{\boldsymbol{\Z}_{\cdot,l}'(\boldsymbol{Y}^{\boldsymbol{\M}} - \boldsymbol{\Z} \beta)}{\dot{n}}\} = O_{p}(\sqrt{\ln(L)/\abs{\mathbb{C}}})\), as in Theorem~\ref{thm:spatial-lasso} below, Assumption~\ref{as:approximate-clusters}(iii) effectively states that variation in the residual \(\boldsymbol{\Y} - \boldsymbol{\Z}\beta\) due to assignments outside the cluster is of smaller order of magnitude than variation due to assignments within the cluster.
Assumption~\ref{as:approximate-clusters}(ii) implies that the treatment indicator is independent across clusters if also Assumption~\ref{as:assign-independent} holds.
Assumption~\ref{as:approximate-clusters}(i) ensures that, for sequences of finite populations where the number of candidate locations and the number of individuals grow at the same rate, the number of clusters grows sufficiently fast.

\begin{theorem}\label{thm:spatial-lasso}
Suppose Assumptions~\ref{as:assign-independent},~\ref{as:spatial-unconfoundedness},~and~\ref{as:approximate-clusters}, as well as regularity conditions (Assumption~\ref{as:regularity} in the appendix), hold.
If the following two conditions hold,
\begin{enumerate}
    \item linear conditional mean: \(\mu_{s,i} = Z_{s,i} \beta\) for all \((s,i) \in \mathbb{S} \times \mathbb{I}\) with \(w_{i}(s,d) \neq 0\),
    \item restricted eigenvalues: w.p. approaching 1 and \(\kappa>0\) bounded away from zero, \(\frac{1}{\dot{n}} \norm{\boldsymbol{\Z}x}_{2}^{2} \geq \kappa \norm{x}_{2}^{2}\) for all \(x \in \mathbb{R}^{L}\) such that \(\norm{x_{\beta=0}}_{1} \leq 3 \norm{x_{\beta\neq0}}_{1}\) where \(x_{\beta=0}\) and \(x_{\beta\neq0}\) are subvectors restricted to entries where corresponding entries \(\beta\) are (not) zero and \(\norm{a}_{p}\) denotes the \(\ell_{p}\) norm of the vector \(a\),
\end{enumerate}
then, for a sequence of finite populations with \(\abs{\mathbb{S}}\), \(\abs{\mathbb{I}}\), and \(\abs{\mathbb{C}}\) growing at equal rates and \(L\) growing, 
there exist constants \(C, \tilde{C} \in \mathbb{R}\) such that the LASSO with penalty parameter chosen as \(\lambda = \tilde{C} \sqrt{\ln(L)/\abs{\mathbb{S}}}\) yields estimation error
\(\sqrt{\sum_{l=1}^{L} (\hat{\beta}_{l} - \beta_{l})^{2}} \leq C \sqrt{\norm{\beta}_{0} \ln(L)/\abs{\mathbb{S}}}\)
with probability approaching one, where \(\norm{\beta}_{0}\) is the number of non-zero elements in \(\beta\).
\end{theorem}

\begin{remark}
    Theorem~\ref{thm:spatial-lasso} shows that, under the finite population design-based framework of this paper, the LASSO achieves its usual convergence rate, allowing for a growing number of regressors assuming sparsity.
    For prediction errors \(\hat{\mu}_{s,i} - \mu_{s,i}\), if \(\sum_{l=1}^{L} Z_{s,i,l}^{2}\) remains bounded as \(L\) grows (for example, a growing number of types where, for a given observation, only the fixed subset of regressors corresponding to its own type are non-zero), then Theorem~\ref{thm:spatial-lasso} implies Assumption~\ref{as:nuisance-quality}(i) as long as \(\norm{\beta}_{0} \ln(L) = o(\sqrt{\abs{\mathbb{S}}})\).
\end{remark}

\begin{remark}
    The linear conditional mean assumption can be ensured to hold by saturating the regression in settings with discrete covariates.
    Sparsity allows the estimator to achieve low error even if the number of regressors grows along the sequence of finite populations.
    The restricted eigenvalues assumption is standard in the literature on the LASSO.
    When \(\boldsymbol{\Z}'\boldsymbol{\Z}\) is invertible but all entries of \(\beta\) are non-zero, the assumption bounds the smallest eigenvalue away from zero such that the typical rank condition (from fixed-\(L\) asymptotics) holds even in the limit; sparsity of \(\beta\) weakens the assumption.
\end{remark}

\subsubsection{Hypothetical Experiment Perspective}\label{ssub:hypothetical_experiment_perspective}
The researcher identifies a particular set of counterfactual locations, determines hypothetical treatment probabilities, and analyzes the data as if these locations and probabilities were known properties of an experiment as in Section~\ref{sec:experimental}.
The researcher uses institutional knowledge and computational techniques, such as those outlined in the following section, to identify a control group that is plausibly comparable to the observed group of treated observations.
Because the hypothetical treatment probabilities \(\pi_{s}\) are determined by the researcher, they generally satisfy Assumption~\ref{as:spatial-unconfoundedness} of being based on characteristics observable to the researcher.
The approach then treats the data as resulting from the corresponding experiment satisfying Assumptions~\ref{as:assign-independent}~and~\ref{as:no-effect-dist}.
There is no assumption that such an experiment occurred in the past; it is solely a hypothetical device for constructing the ``repeated sampling'' thought experiment underlying inference.
This perspective acknowledges that the thought experiment for design-based inference around in-sample causal effects is inherently divorced from any feasible or observable procedure.
In contrast to repeated sampling from a larger population, which at least theoretically can be feasible in some settings, it is entirely impossible to carry out the design-based \emph{repetition} in practice.
Hence, the researcher justifies inference as reflecting one particular hypothetical thought experiment.

A convenient choice for the hypothetical experiment is the ideal experiment of Section~\ref{sec:experimental} because it is feasible to study and has the attractive design-based properties discussed above.
Formally, the researcher appeals to the properties stated in Theorem~\ref{thm:single-additive} as describing the (asymptotic) distribution \(F_{\hat{\tau}(d),P}\) of the estimator \(\hat{\tau}(d)\) as in Equation~\ref{eq:tau-hat-single-additive} (with \(\pi_{s}\) fixed as determined by the researcher) across the treatment assignment distribution \(P\) arising from independent assignment according to the marginal probabilities \(\pi_{s}\) determined by the researcher.

Procedurally, inference under this perspective is similar to constructing a matched sample based on estimated treatment probabilities and then taking the sample as given \citep[for instance,][ch. 17.6]{imbens2015causal}.
By never using outcome data in identifying the counterfactual locations and determining hypothetical treatment locations, this necessary first step for observational data ``cannot intentionally introduce systematic biases in the subsequent analyses for causal effects on outcomes'' \citep[p. 374]{imbens2015causal}.

\subsubsection{Conditional Inference Perspective}\label{ssub:conditional_inference_perspective}
The researcher assumes an experiment took place, but inference is not based on the full unknown experimental assignment distribution.
Instead, the researcher makes Assumption~\ref{as:spatial-unconfoundedness} with a convenient parametric functional form but an unknown parameter.
By conditioning on a sufficient statistic for the unknown parameter, the inferential distribution no longer depends on the true parameter value.
In other words, standard errors express the amount of variation in the estimator across assignments that lead to the same value of the sufficient statistic.

Formally, the researcher appeals to the properties stated in Theorem~\ref{thm:single-additive} as describing an (asymptotic) distribution \(F_{\hat{\tau}(d),P}\).
The estimator \(\hat{\tau}(d)\) is as in Equation~\ref{eq:tau-hat-single-additive} but with estimated \(\hat{\pi}_{s}\) given the realized assignment in place of the unknown \(\pi_{s}\).
The distribution \(P\) is the \emph{conditional} distribution arising from restricting the true latent experiment to assignments for which \(\hat{\pi}_{s}\) takes on the values observed in the sample and the marginal treatment probabilities of the conditional distribution coincide with \(\hat{\pi}_{s}\).
Despite the researcher acknowledging that \(\hat{\pi}_{s}\) is estimated from the sample, the conditional inference thought experiment can hold \(\hat{\pi}_{s}\) fixed across samples, simplifying finite sample inference as in the ideal experiment of Section~\ref{sec:experimental}.
However, even if the true latent experiment features independent assignment, the conditional distribution may feature dependence.
Nevertheless, some of the results of Theorem~\ref{thm:single-additive} may apply, as illustrated with a simple example below.

For the simplest example, consider an experiment according to Assumption~\ref{as:assign-independent} with a constant probability of treatment for locations in a known set \(\mathbb{S}\), and zero probability of treatment elsewhere.
The full assignment distribution depends on the unknown probability of treatment.
However, conditional on the number of treated locations in the sample, the assignment distribution is known to be a uniform distribution over all assignments that hold the number of treated locations fixed.
Hence, conditional inference treating the estimated (constant) treatment probabilities as fixed is valid.

The results of Theorem~\ref{thm:single-additive} remain useful for conditional inference in the example:
The variance assuming independent assignment is conservative in expectation (by the law of total variance), but asymptotically the designs are equivalent \citep{Hajek1960}.
The example readily extends to settings where the researcher observes several candidate locations for each value of the characteristics in Assumption~\ref{as:spatial-unconfoundedness}.
When many locations have distinct characteristics, but the treatment probability function takes a logistic functional form, conditional inference via permutation tests of sharp null hypotheses is possible \citep{Rosenbaum1984}.

\subsection{Finding counterfactual treatment locations using convolutional neural networks}\label{sec:implementation-cnn}

I propose using CNNs to identify plausible counterfactual treatment locations that are observationally similar to realized treatment locations.
The particular implementation of CNNs I advocate for is well-suited for spatial settings because it has two key features:
First, it allows the prediction of counterfactual locations to depend flexibly on the spatial configuration of characteristics in the neighborhoods around the locations.
Second, it is computationally feasible despite a very large number of possible locations to consider and very high-dimensional covariates.

The convolution operation, together with input data augmentation \citep{simard2003best}, implements the idea that the spatial distribution of characteristics relative to a location is often important for the location's plausibility as a counterfactual treatment location.
In Assumption~\ref{as:spatial-unconfoundedness}, characteristics \(Z_{s}\) of location \(s\) may contain the local values \(v_{x,y}\) and relative locations of points \((x,y)\) around \(s\).
The convolution operation \(f\) on a grid \(\boldsymbol{v}\) of input values \(v_{x,y}\) is
\begin{equation*}
f(\boldsymbol{v})_{x,y} = \sum_{a=-k}^{k} \sum_{b=-k}^{k} \beta_{a,b} \cdot v_{x+a,y+b}
\end{equation*}
such that the value at grid cell \((x,y)\) is based on input values within \(x \pm k, y \pm k\) for a fixed \(k\).
The coefficients \(\boldsymbol{\beta}\), which are estimated by the neural network, capture the weight placed on input values at locations relative to \((x,y)\).
By using the same \(\boldsymbol{\beta}\) to compute the convolution at all points \((x,y)\), CNNs can be more parsimonious than fully connected neural networks and enforce equivariance to shift.
Using multiple layers of convolutions combined with nonlinear activation functions at each grid cell allows the network to learn nonlinear relationships.
Augmenting data by spatially shifting, rotating, and mirroring the input effectively imposes an equivariance with respect to these operations.
Equivariance formalizes the economic logic that relative locations of spatial features and characteristics matter, rather than their absolute locations and orientations.
With standard methods, such as logistic regression, it appears challenging to incorporate a similarly flexible yet equivariant relationship between the output at a point and the characteristics measured at spatial locations near it.
Furthermore, the approach using CNNs is computationally feasible because the categorical predictions compute estimates for many locations (grid cells) at once, and stochastic gradient descent, used to estimate the CNN parameters, limits memory requirements and improves speed in the presence of very high-dimensional spatial data used as predictors.
I describe and discuss the use of CNNs in more detail in Online~Appendix~3.

I train a CNN to distinguish between realized locations of the treatment and other locations based on observable characteristics.
The key insight is that spatial data can typically be ``plotted on a map,'' and hence spatial data can resemble image data, for which CNNs have enjoyed recent popularity \citep{Krizhevsky2012convolution}.
The input to the CNN is a 3D tensor:
a fine 2D discretization of space approximately centered around the real or possible counterfactual location, combined with a third dimension that enumerates the values of every characteristic measured for this grid cell.
I design the objective function of the CNN to partly resemble a generative adversarial network \citep[GAN,][]{Goodfellow2014gan} while maintaining the more easily trained structure of finite categorical prediction:
The output of the CNN is an ``activation score'' for each categorical prediction of whether (``discriminator'') and in which spatial grid cell (``generator'') there is a real treatment location.
A GAN is attractive because its generator draws from the \emph{modes} (``most plausible'') of the treatment location distribution across space rather than estimating mean locations \citep{goodfellow2016nips,lotter2016unsupervised}, and the GAN implicitly maintains an internal estimate of the distribution of the treatment across space, which here resembles the treatment probabilities.

After training, the ``false positives'' of this algorithm (counterfactual locations drawn from the generator) are grid cells without real treatment locations that the CNN could not distinguish from real treatment locations based on observable characteristics.
Under Assumption~\ref{as:spatial-unconfoundedness}, observationally identical locations are a design-based valid control group.
More precisely, I use matching on the activation score to create a sample consisting of the real treatment locations and those locations with the most similar activation scores, akin to propensity score matching for sample construction.
I then estimate the treatment probability using only the matched sample, effectively setting the estimated treatment probabilities of all other locations to zero, and analyze the observational data following one of the perspectives above.

\section{Application: foot traffic in times of COVID-19}\label{sec:applications}

In this section, I demonstrate the use of the proposed methods to study the effect of grocery stores on the number of visitors to restaurants during COVID-19 shelter-in-place policies.\footnote{See Online~Appendix~4 for the exact definitions of grocery stores and restaurants used in this analysis.}
Here, grocery store locations are the treatment locations, and restaurants are the individuals for whom I estimate average effects by distance from treatment.
For examples of existing empirical studies involving spatial treatments, see Online~Appendix Table~OA1.

\begin{figure}
\begin{subfigure}{0.49\textwidth}\centering
\includegraphics{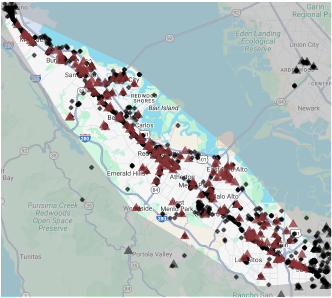}
\caption{map of sample}\label{fig:map_of_sample}
\end{subfigure}
\hfill
\begin{subfigure}{0.49\textwidth}\centering
\includegraphics{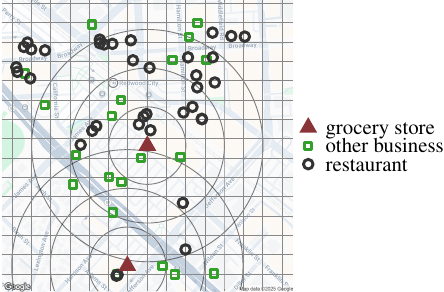}
\caption{example location}\label{fig:example_location}
\end{subfigure}
\caption{\label{fig:bay-area}The sample includes businesses in the San Francisco Bay Area between San Francisco and San Jose (panel~(\subref{fig:map_of_sample})).
Grocery stores in study area: solid red triangles (167); outside (considered fixed): black triangles.
Restaurants: black circles (1627 within 0.5 miles of real or counterfactual grocery stores).
Panel~(\subref{fig:example_location}) zooms in on a location in Redwood City, also indicating locations of other businesses (green squares), and illustrates the size of grid cells as well as circles with radii \(0.05\mathrm{mi}\), \(0.10\mathrm{mi}\), and \(0.15\mathrm{mi}\) around the two grocery store locations in the plotted area.}
\end{figure}

I use data on the location of businesses in the San Francisco Bay Area, shown in Figure~\ref{fig:bay-area}(a), and the number of visitors to them from SafeGraph, available to academic researchers.
For each business appearing in these data, SafeGraph records latitude and longitude, NAICS industry code, and the number of visitors in a given week whose smartphone location data is available to SafeGraph.
SafeGraph's proprietary algorithm defines visits based on the location of the smartphone, time spent, relative locations of other businesses, and time of day.
To cover the South Bay Area, I select grocery stores near Burlingame (within 3 miles of the city center), Belmont (5 miles), Menlo Park (5.5 miles), and Mountain View (2.95 miles).
I manually confirm that all grocery stores used in the analysis were indeed open during the study period (the week starting April 13, 2020) and set their latitude and longitude to the front door.
For restaurants, I remove duplicate observations due to typos in restaurant names and compare SafeGraph coded latitude and longitude to geocoding by Google Maps.
I restrict businesses to those with at least 7 visits, as recorded by SafeGraph, in each of the first four calendar weeks of 2020 to focus on businesses that were open before the pandemic.
Overall, I use 167 real and 162 counterfactual grocery store locations, with 1,894 and 1,612 instances of restaurants within 0.2 miles, respectively.
These instances arise from 752 unique restaurants with a median of 8 visits in the SafeGraph data for the study period.
For additional detail on sample construction, see Online~Appendix~4.1.

When consumers make only essential trips, such as getting groceries, other businesses relying on foot traffic, such as restaurants, may benefit from being located nearby.
Local governments in the San Francisco Bay Area urged residents to only make essential trips during shelter-in-place policies in April 2020.
At the same time, other businesses such as restaurants remained open for takeout business.
However, drastically reduced foot traffic and customers over time led to financial distress for many businesses \citep{Yang2020}.
In such times, a location along consumers' essential trips may benefit these businesses.

The causal interpretation of the cross-sectional estimators of this paper rests on Assumption~\ref{as:spatial-unconfoundedness} that assignment probabilities depend only on observable characteristics, not potential outcomes.
Hence, restaurants in neighborhoods differing in their number of grocery stores, but similar in terms of all other kinds of businesses and observable characteristics, would in expectation have comparable numbers of visitors if they had similar exposure to grocery stores.
The assumption may be plausible because restaurants chose their locations based on pre-pandemic potential outcomes, if at all.
Restaurants that located before the pandemic are unlikely to have (accurately) predicted and sorted based on potential foot traffic patterns during shelter-in-place policies.
Even pre-pandemic, grocery store locations may not have been the primary concern for restaurants, holding locations of all other businesses fixed.
See Online~Appendix~7 for empirical analyses assessing the identifying assumption.

For this application, I make Assumption~\ref{as:spatial-unconfoundedness} using the relative locations of businesses by industry as the observable characteristics \(Z_{s}\), which form the input into the CNN and hence treatment probability estimation.
Panel~(\subref{fig:example_location}) of Figure~\ref{fig:bay-area} illustrates these controls by plotting as green squares other businesses near a particular grocery store in the sample.
I superimpose a grid with cells of size \(0.025\mathrm{mi}\times0.025\mathrm{mi}\) that shows the discretization used by the CNN.
I divide these other businesses into seven groups by their four-digit NAICS code, as listed in Online~Appendix Table~OA3, and the count of businesses by industry for each grid cell is used as a covariate.
One could similarly control for any other variables that can be plotted on a map, such as average house price by grid cell or the fraction of individuals with college degrees in the census tract covering the grid cell if such data are available and relevant for a given application.
When training the network, I impose continuous shifts to the grid, such that the discretization becomes less relevant, as well as rotation and mirroring to build in equivariance such that only relative locations matter.
The CNN passes the spatial grid of businesses by industry through four sequential 2D convolutions, calculating 16, 36, 36, and 1 weighted averages (``channels'') of the neighborhood of each point based on the output of the previous layer, with one final fully connected layer.
Based on a matched sample of real grocery store locations and counterfactual locations predicted by the CNN, I estimate treatment probabilities based on the number of restaurants and grocery stores by distance from each location using logistic regression.
I give a complete description of the implementation of the CNN and treatment probability estimation in Online~Appendix~3~and~4.

Researchers can assess the plausibility and quality of the counterfactual grocery store locations predicted by the neural network and the estimated treatment probabilities by considering summary statistics of balance and the concentration of the difference in exposure to real grocery stores.
Researchers can also informally inspect the suitability of counterfactual locations by plotting both real and counterfactual locations on a map.
Systematic differences between real and counterfactual locations imply that estimated effects reflect not just differences in exposure to grocery stores, but also these other differences.

\begin{figure}
\begin{tabular}{@{} K{1.75in} @{} K{1.4in} @{} K{1.4in} @{} K{1.4in} @{}}
{\tiny restaurants dist. \(<0.025\)mi to grocery} & {\tiny \(0.025\)mi--\(0.05\)mi to grocery} & {\tiny \(0.05\)mi--\(0.075\)mi to grocery} & {\tiny \(0.075\)mi--\(0.1\)mi to grocery} \\
\end{tabular}
\includegraphics{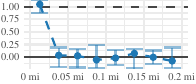} \hfill 
\includegraphics{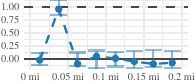} \hfill
\includegraphics{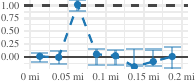} \hfill
\includegraphics{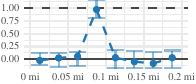} \\
\begin{tabular}{@{} K{1.5in} @{} K{1.5in} @{} K{1.5in} @{} K{1.5in} @{}}
{\tiny \(0.1\)mi--\(0.125\)mi to grocery} & {\tiny \(0.125\)mi--\(0.15\)mi to grocery} & {\tiny \(0.15\)mi--\(0.175\)mi to grocery} & {\tiny \(0.175\)mi--\(0.2\)mi to grocery} \\
\end{tabular}
\includegraphics{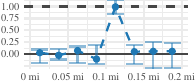} \hfill 
\includegraphics{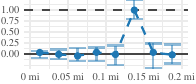} \hfill
\includegraphics{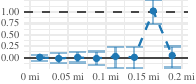} \hfill 
\includegraphics{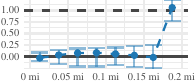}
\caption{\label{fig:exposure}Differential exposure of restaurants to grocery stores.
Each plot shows the difference in the average number of grocery stores at multiple distances (horizontal axis) between treated and untreated restaurants.
Columns show restaurants in different distance bins around candidate grocery store locations.
To give a sense of statistical uncertainty in differential exposure, I take 10,000 draws from the treatment assignment distribution (given by Assumptions~\ref{as:assign-independent}~and~\ref{as:spatial-unconfoundedness}) and display error bars covering the differential exposure realized in 95\% of these draws.}
\end{figure}

Figure \ref{fig:exposure} assesses whether restaurants near real grocery stores, compared to restaurants near counterfactual locations, are exposed to one additional grocery store at the distance of interest, with no differences in exposure at other distances.
Each panel focuses on restaurants at a different distance from (real and counterfactual) grocery store locations.
The line shows the difference in the average number of real grocery stores by distance from these restaurants.
In each panel, there is little difference in exposure for restaurants near real and counterfactual restaurants, \emph{except} at the distance for which these restaurants serve as treated and control, respectively.
Hence, the estimated effect at a particular distance indeed reflects the difference between one more/fewer grocery store at that distance.
Because balance in exposure to real grocery stores is essential for interpretation, I include covariates describing exposure directly in the treatment probability estimation.
If there were differences in exposure at other distances, one could not interpret the estimates as the effect of adding one more grocery store.
Instead, under appropriate assumptions, it may reflect the effect of shifting a grocery store from another distance to the distance of interest.

\begin{figure}
\begin{tabular}{@{} K{1.75in} @{} K{1.4in} @{} K{1.4in} @{} K{1.4in} @{}}
{\tiny restaurants dist. \(<0.025\)mi to grocery} & {\tiny \(0.025\)mi--\(0.05\)mi to grocery} & {\tiny \(0.05\)mi--\(0.075\)mi to grocery} & {\tiny \(0.075\)mi--\(0.1\)mi to grocery} \\
\end{tabular}
\rotatebox{90}{\tiny \hspace{2em} restaurant}
\includegraphics{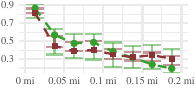} \hfill
\includegraphics{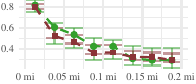} \hfill
\includegraphics{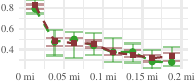} \hfill
\includegraphics{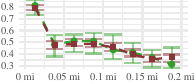} \hfill \\
\rotatebox{90}{\tiny \hspace{1em} amusement}
\includegraphics{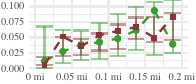} \hfill
\includegraphics{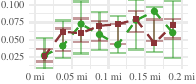} \hfill
\includegraphics{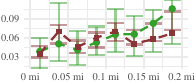} \hfill
\includegraphics{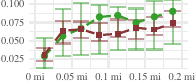} \hfill \\
\rotatebox{90}{\tiny \hspace{1em} museum}
\includegraphics{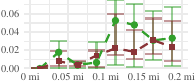} \hfill
\includegraphics{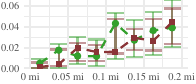} \hfill
\includegraphics{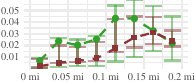} \hfill
\includegraphics{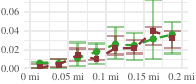} \hfill \\
\rotatebox{90}{\tiny \hspace{1em} religious org}
\includegraphics{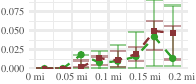} \hfill
\includegraphics{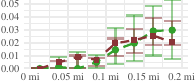} \hfill
\includegraphics{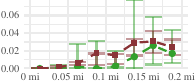} \hfill
\includegraphics{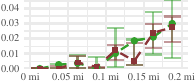} \hfill \\
\rotatebox{90}{\tiny \hspace{2em} dentist}
\includegraphics{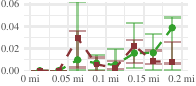} \hfill
\includegraphics{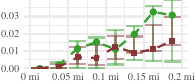} \hfill
\includegraphics{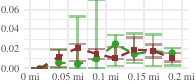} \hfill
\includegraphics{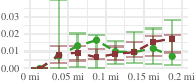} \hfill \\
\rotatebox{90}{\tiny \hspace{1em} auto repair}
\includegraphics{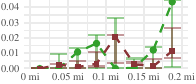} \hfill
\includegraphics{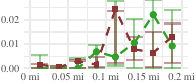} \hfill
\includegraphics{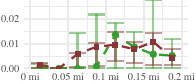} \hfill
\includegraphics{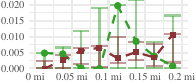} \hfill
\caption{\label{fig:balance}Balance in neighborhood characteristics.
Each plot shows the fraction of businesses with, in different rows, 4-digit NAICS 7225, 7139, 7121, 8131, 6212, or 8111 at multiple distances (horizontal axis) from treated (red squares) and untreated (green dots) restaurants.
Columns show restaurants in different distance bins around candidate grocery store locations.
For restaurants at longer distances from grocery stores, see Online~Appendix~Figure~OA1.
To give a sense of statistical uncertainty in characteristics balance, I take 10,000 draws from the treatment assignment distribution (given by Assumptions~\ref{as:assign-independent}~and~\ref{as:spatial-unconfoundedness}) and display error bars covering the balance in characteristics realized in 95\% of these draws.}
\end{figure}

Figure \ref{fig:balance} shows that other observable characteristics are balanced between the neighborhoods of treated and untreated restaurants at different distances.
The top left plot focuses on restaurants within \(0.025\) miles of real and counterfactual grocery stores.
For each treated or untreated restaurant, I compute the fraction of businesses that fall into particular industries within different distance bins from the restaurant (horizontal axis).
Red squares show the average fraction among the treated restaurants.
Green dots show the ATT-weighted average fraction among the untreated restaurants.
Columns focus on treated and untreated restaurants at different distances from grocery stores.
Rows show the neighborhood proportions of different industries.
Overall, the reasonable balance in business composition patterns across distances suggests that the neural networks succeeded in finding counterfactual locations similar to real grocery store locations.
Note that restaurants, amusement, museums, and religious locations are used as predictors in the neural network, but dentists and automotive businesses are not.\footnote{The count of dentists and automotive businesses is used by the neural network together with all ``other industries'' as a single covariate per grid cell.}
Except for the count (not fraction) of restaurants, none of these industries are used in the treatment probability estimation, such that the balance shown in the figure is not mechanical.

\begin{figure}
\includegraphics{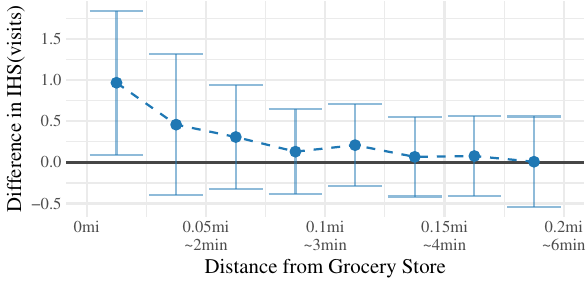}
\caption{\label{fig:effect-asinh}Estimated average effect of grocery stores on restaurants at different distances. The outcome is the inverse hyperbolic sine of the number of visits as recorded by SafeGraph. Bars indicate \(\pm 1.96\) standard errors.}
\end{figure}

Figure \ref{fig:effect-asinh} displays estimated effects and standard errors for the estimator given in Equation~\ref{eq:tau-hat-single-additive}.
Standard errors indicate the variation in the estimate expected from reassigning treatment according to the fixed estimated treatment probabilities following the ``hypothetical experiment perspective'' of Section~\ref{sec:observational}, under Assumption~\ref{as:assign-independent} of independent assignment and Assumption~\ref{as:no-effect-dist} that treatments have no effect beyond a distance of \(\dnoeffect{} \equiv 0.075\) miles.
Independent assignment may appear implausible if one believes that clustering of grocery stores close to one another is particularly likely or unlikely.
In practice, I observe real grocery stores both in isolated locations and close to other grocery stores.
However, if information on the covariances (joint location probabilities) was available one could impose it instead of independent assignment (zero covariance).
No effect beyond \(0.075\) miles appears plausible given the substantively close to zero point estimates beyond that distance in Figure \ref{fig:effect-asinh}.
Note, however, that such a figure is not proof of the sharp null hypothesis of no effect of any possible grocery store exposure beyond such distances.
Without further assumptions, the figure only suggests zero \emph{average marginal} effects.
If each grocery store brings a separate set of potential customers to nearby restaurants, treatment effects may be approximately additively separable (Assumption~\ref{as:additive-separability}), in which case average marginal effects equal average effects.

\begin{sidewaystable}
\caption{\label{tab:application-covid-19-results} Estimated effects on the number of visits to restaurants using different estimators.
The first panel uses the IPW estimator, with inference valid for observational data under the hypothetical experiment perspective described in Section~\ref{sec:observational}.
The second panel uses the double machine learning (DML) estimator that may have favorable theoretical properties under the latent experiment perspective.
For each panel, ``IHS'' refers to the effect in inverse hyperbolic sine unit and ``level'' to effects in levels.
Standard errors are given in parentheses.
Rows labeled ``percent incr.'' show the percent increase (or decrease) relative to the mean visits of the control restaurants at that distance.
}
\begin{tabular}{l ccccc ccc}
\hline
Distance: &    0.000 mi &     0.025 mi &     0.050 mi &     0.075 mi &     0.100 mi &     0.125 mi &     0.150 mi &     0.175 mi \\
          & -- 0.025 mi &  -- 0.050 mi &  -- 0.075 mi &  -- 0.100 mi &  -- 0.125 mi &  -- 0.150 mi &  -- 0.175 mi &  -- 0.200 mi \\
\hline
\multicolumn{9}{l}{\emph{IPW Estimator for Hypothetical Experiment Perspective:}} \\
IHS: & 0.97 & 0.46 & 0.31 & 0.13 & 0.21 & 0.07 & 0.08 & 0.01 \\
   & (0.45) & (0.44) & (0.32) & (0.26) & (0.26) & (0.25) & (0.25) & (0.28) \\

percent incr.: & 166 & 59 & 37 & 14 & 24 & 7 & 8 & 1 \\

\\
level: & 14.92 & -0.42 &  3.78 &  1.53 &  2.52 &  0.43 &  2.93 & -7.79 \\
   & (6.33) & (8.86) & (4.18) & (3.91) & (3.00) & (3.68) & (4.70) & (6.49) \\

percent incr.: & 158 & -3 & 36 & 12 & 23 & 3 & 28 & -37 \\

\\
\multicolumn{9}{l}{\emph{DML Estimator for Latent Experiment Perspective:}} \\
IHS: &  1.07 &  0.49 &  0.10 &  0.01 & -0.02 & -0.03 & -0.17 & -0.09 \\
   & (0.44) & (0.43) & (0.36) & (0.27) & (0.25) & (0.27) & (0.30) & (0.32) \\

percent incr.: & 193 & 63 & 10 & 1 & -2 & -3 & -16 & -9 \\

\\
level: & 16.16 &  2.44 & -0.98 & -0.40 & -0.36 & -1.07 & -1.19 & -4.45 \\
   & (7.40) & (7.35) & (5.49) & (5.41) & (3.54) & (4.75) & (5.27) & (5.43) \\

percent incr.: & 171 & 15 & -9 & -3 & -3 & -9 & -11 & -21 \\

\\
\hline
\end{tabular}
\end{sidewaystable}

I find large positive effects of being located very close to a grocery store, with no effect past a few minutes of walking.
Table~\ref{tab:application-covid-19-results} shows the point estimates corresponding to Figure~\ref{fig:effect-asinh} up to a distance of \(0.2\) miles.
For the restaurants in the closest bin of up to \(0.025\) miles, the average effect more than doubles the number of SafeGraph-recorded visitors both when estimating an approximate percentage effect using inverse hyperbolic sine units\footnote{Due to the extreme reduction in foot traffic during the COVID-19 pandemic, some businesses have zero visits recorded by SafeGraph in April 2020, such that a log transformation is not feasible. The inverse hyperbolic sine is similar to the (shifted) log at values other than zero, but the usual caveats regarding the percentage change interpretation apply \citep{bellemare2020elasticities,Mullahy2023,Chen2024a}.} and when estimating effects in levels.
In the second closest bin of restaurants between \(0.025\) and \(0.05\) miles from grocery stores, the estimated effects are smaller and not statistically significant at the 5\% level.
For any longer distance, the effects are both economically smaller and statistically insignificant.
Effects close to 0 past a couple of minutes of walking may be due to either the unwillingness of consumers to walk longer distances or the lack of a need to do so because there typically is a closer alternative restaurant or coffee shop.
In a study of grocery store openings building on the framework of this paper, \citet{Qian2023} find a very similar pattern with estimated effects that are roughly half in magnitude of the estimates here.
Larger effects in this study may be due to reduced baseline foot traffic during the COVID-19 pandemic, but overlapping confidence intervals suggest the difference may also be due to random chance.

Table~\ref{tab:application-covid-19-results} also displays estimates using the double machine learning estimator that may have favorable statistical properties under the ``latent experiment perspective'' of Section~\ref{sec:observational}.
I estimate the first step treatment probability and outcome models using cross-fitting in a post-LASSO procedure with data-driven penalty \citep{Belloni2012}.
When estimating treatment probability or outcome mean for a grocery store or restaurant, I use only grocery stores or restaurants that are at least 1 mile away, satisfying design-based independence of the sample splits under the assumption that grocery stores have no effect on outcomes for restaurants at distances farther than \(0.5\) miles.
For the treatment probability, I use the same logistic regression specification as before.
For the outcome model, I use a linear model at the restaurant level with regressors for the number of grocery stores in eight distance bins of width \(0.025\) miles up to \(0.2\) miles.
To obtain estimates of the conditional mean in the absence of a marginal grocery store, I use the outcome model prediction after removing a marginal grocery store from the regressors.
The inferential results of Theorem~\ref{thm:dml} apply if the cross-fit estimates of treatment probabilities (including the CNN) and outcome predictions satisfy Assumption~\ref{as:nuisance-quality}.
The results are qualitatively and quantitatively similar to the IPW estimators discussed above.

I also implement a cross-sectional inner vs. outer ring estimator.
For given inner outer ring distances, I use restaurant-grocery store pairs on both rings as observations in a regression of restaurant outcomes on an inner ring indicator and grocery store fixed effects.
Note that, most commonly, the inner vs. outer ring strategy is used in a difference-in-differences design.
In this paper, however, the temporal differencing is not attractive for two reasons:
First, because changes in foot traffic from pre-COVID-19 periods into COVID-19 periods are very large, a parallel trends assumption (in levels or percentage terms) may be questionable, and parallel trends before COVID-19 are likely uninformative about whether trends into the COVID-19 periods are parallel.
Second, temporal differencing would at best yield estimates of the differential effect after vs. before the COVID-19 pandemic of being near real vs. counterfactual grocery store locations.
In the absence of these two issues, one may wish to use \emph{both} the estimators proposed in this paper and the inner vs. outer ring estimator in a difference-in-differences design.

The inner vs. outer ring design and the quasi-experimental design proposed in this paper are distinct both conceptually and in the implementation choices required.
Conceptually, whether identification is most plausible based on functional form (such as the comparability of the inner and outer rings), on quasi-experimental variation in the treatment location (as in the design proposed in this paper), on both, or on neither, depends on the application.
With panel data, the inner vs. outer ring design requires parallel trends of outcomes on inner and outer rings, while the quasi-experimental design requires parallel trends of outcomes near realized and counterfactual treatment locations.
In the inner vs. outer ring design, researchers typically justify the functional form assumption for identification and describe a separate sampling scheme for inference.
In the quasi-experimental design, researchers use the same design-based variation for identification and inference.
The inner vs. outer ring design requires the specification of an outer ring distance.
If the outer ring distance is too large, the outcome individuals on the outer ring may be systematically different from individuals on the inner ring, especially if treatment locations were chosen strategically.
If the outer ring distance is too small, the outcome individuals on the outer ring may be affected by the treatment such that one at best learns about the effect of being relatively closer to the treatment.
Unfortunately, the data are generally uninformative about this trade-off.
In contrast, the quasi-experimental design requires implementation choices for counterfactual locations, as discussed in Section~\ref{sec:implementation-cnn}.
Below, I discuss an additional challenge in implementing the inner vs. outer ring design when treatment locations are not far apart.

\begin{figure}
\begin{tabular}{@{} K{1.4in} @{} K{1.4in} @{} K{1.4in} @{} K{1.4in} @{}}
{\tiny restaurants dist. \(<0.025\)mi to grocery} & {\tiny \(0.025\)mi--\(0.05\)mi to grocery} & {\tiny \(0.05\)mi--\(0.075\)mi to grocery} & {\tiny \(0.075\)mi--\(0.1\)mi to grocery} \\
\end{tabular}
\includegraphics{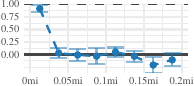} \hfill 
\includegraphics{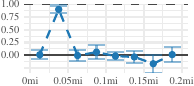} \hfill
\includegraphics{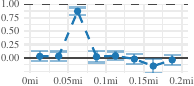} \hfill
\includegraphics{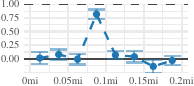} \\
\begin{tabular}{@{} K{1.4in} @{} K{1.4in} @{} K{1.4in} @{} K{1.4in} @{}}
{\tiny restaurants dist. \(0.1\)mi--\(0.125\)mi to grocery} & {\tiny \(0.125\)mi--\(0.15\)mi to grocery} & {\tiny \(0.15\)mi--\(0.175\)mi to grocery} & {\tiny \(0.175\)mi--\(0.2\)mi to grocery} \\
\end{tabular}
\includegraphics{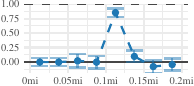} \hfill 
\includegraphics{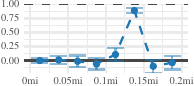} \hfill
\includegraphics{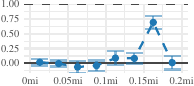} \hfill 
\includegraphics{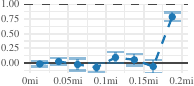}
\caption{\label{fig:exposure-outer}Differential exposure of treated and control restaurants to grocery stores at different distances when using an outer ring of distance \(0.2\)--\(0.225\).
Each panel holds fixed the inner (and outer) ring distance from a real grocery store and plots the coefficient of a regression of the number of grocery stores at a particular distance on an inner ring dummy with grocery store fixed effects.}
\end{figure}

For inner vs. outer ring comparisons using all restaurant-grocery store pairs, treated restaurants are not necessarily exposed to one more grocery store than control restaurants, on average.
Figure~\ref{fig:exposure-outer} shows the analog to Figure~\ref{fig:exposure} for the inner vs. outer ring estimator.
Each panel restricts the sample to restaurants that are either on the inner ring with a particular radius or on the outer ring with distance between \(0.2\) and \(0.225\) miles.
Within each panel, the figure plots the coefficients of regressions of the number of grocery stores at a particular distance on an inner ring dummy and grocery store fixed effects, varying the distance along the horizontal axis.
Because the outer ring restaurants, while farther away from the focal grocery store, tend to be closer to \emph{other} grocery stores, the difference in exposure (plotted coefficient) is below 1 at the distance of interest.
Hence, the inner vs. outer ring estimator does not necessarily estimate the effect of 1 additional grocery store at the distance of interest.

\begin{figure}
\includegraphics{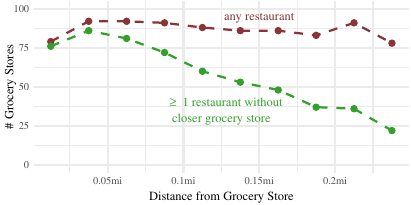}
\hfill
\includegraphics{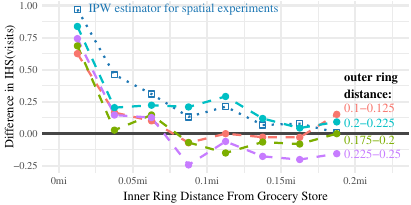}
\caption{\label{fig:isolated-cnt}Left:
Number of grocery stores with either at least one restaurant in a given distance bin (red) or at least one restaurant with no closer grocery store (green).
There are few grocery stores with restaurants unexposed to other grocery stores when considering larger outer ring distances.
Right:
Estimated average effect of grocery stores on restaurants at different distances using control groups based on different outer ring distances as labeled on the figure, not restricting to those without closer grocery stores.
The estimated effects vary noticeably depending on the choice of outer ring.
When the inner and outer ring distance bins coincide, the estimates are mechanically zero.
For comparison, the squares repeat the estimator proposed in this paper from Figure~\ref{fig:effect-asinh}.
}
\end{figure}

To avoid exposure of outer ring restaurants to other grocery stores, one may focus on isolated grocery stores where no other grocery store is closer to the outer ring.
Figure~\ref{fig:isolated-cnt} (left) shows the number of grocery stores that have at least one restaurant in a particular distance bin, as well as the number of grocery stores that have at least one restaurant without any closer grocery store in the distance bin.
If one chooses an outer ring distance, drops all restaurants with any grocery store closer than that distance, and includes grocery store fixed effects, then the estimated treatment effect is based solely on the grocery stores constituting the lower of the two curves.
Even within this subsample of grocery stores, the estimator is based on a possibly selected sample of restaurants geographically away from the more business-dense direction.
Hence, when there are multiple realized treatment locations close to one another, proper differences in exposure to the treatment can be difficult to achieve with the inner vs. outer ring strategy.

Furthermore, the neighborhoods of restaurants on inner rings and outer rings are not necessarily alike.
Figure~\ref{fig:balance} above shows that the neighborhoods of the restaurants on an outer ring are likely noticeably different from the neighborhoods of restaurants on an inner ring.
Many grocery stores are located in shopping or strip malls.
The estimators proposed in this paper, intuitively, use counterfactual treatment locations that exploit differences in the number of grocery stores in such retail-heavy areas.
In contrast, the inner vs. outer ring strategy compares restaurants in commercial areas to restaurants in more residential or industrial areas.

Inner vs. outer ring estimates are fairly sensitive to the choice of outer ring distance in this application.
To estimate the effects for given inner and outer rings, restrict the sample to restaurants on these rings.
Restaurants may appear multiple times if they are on the inner or outer ring around different grocery stores.
Estimate the treatment effect by regressing the outcome (IHS of visits) on an inner ring dummy and grocery store fixed effects.
Figure~\ref{fig:isolated-cnt} (right) plots the coefficients on the inner ring dummy for regressions using different inner ring distances (horizontal axis) and a few choices for the outer ring distance (curves).
Even for ``adjacent'' outer ring distances, for instance \(0.2\)--\(0.225\) miles and \(0.225\)--\(0.25\) miles, estimated effects can be noticeably different.
Given the differences in exposure and neighborhood characteristic balance, one may wish to choose a different outer ring for each inner ring to obtain more plausible comparisons.
Alternatively, functional form assumptions may allow for correcting imbalances in levels (or non-parallel trends if panel data and variation in the timing of the treatment were available).
The approach proposed in this paper balances by design without manual adjustments to the control group to estimate effects at different distances.
To interpret the inner vs. outer ring estimates as causal effects of a marginal grocery store at a particular distance, one needs to assume that the outer ring is unaffected (as in Assumption~\ref{as:no-effect-dist}), 
and the correct choice of this distance is required for the estimation of the effects themselves.
The approach proposed in this paper uses Assumption~\ref{as:no-effect-dist} only for standard errors.

\section{Conclusion}\label{sec:conclusion}

The causal effects of treatments occurring at locations in space on individuals located nearby are of interest across fields of economics and social sciences.
This paper presents a design-based approach for causal identification, estimation, and inference in spatial settings.
The approach differs from existing approaches in the literature to identification and estimation based on parallel trends and inference based on sampling.
I argue that identification, estimation, and inference using design-based ideas are conceptually attractive, analytically tractable, and computationally feasible.
The same design-based ideas do not validate the inner vs. outer ring empirical strategy commonly applied in empirical practice.
Instead, this ideal experiment validates the comparison of individuals near realized treatment to individuals near counterfactual locations where the treatment could have happened (but did not).
The finite population design-based variances derived in this paper express the variation due to the ideal experiment.
Design-based inference removes the need to specify a hypothetical super-population and sampling scheme.
Because counterfactual locations of treatments are typically not available in observational data, I propose a computationally feasible method using convolutional neural networks to identify locations that are observationally similar to real treatment locations.
These counterfactual locations allow the estimation of causal effects in observational data under an assumption that ensures treatment assignment is not based on potential outcomes other than through observable characteristics, similar to an unconfoundedness assumption in sampling-based analyses.

I demonstrate the use of these methods by studying the causal effects of grocery stores on the number of visitors to nearby restaurants during COVID-19 shelter-in-place policies.
The counterfactual grocery store locations proposed by the neural network are in neighborhoods that are indeed observationally similar to the neighborhoods of real grocery stores.
I estimate large effects for restaurants very close to a grocery store, on average more than doubling the number of visitors, as measured in data from SafeGraph.
The design-based standard errors take into account the complex ways in which exposure to grocery stores (the treatment) is correlated across restaurants (outcome units) \emph{by design} of the ideal experiment.
Hence, I find significant externalities between businesses.
Such externalities may lead to socially undesirable concentrations of consumers during a pandemic, as well as spatial inequities across business owners to the extent that they are unanticipated and not internalized through, for instance, differential rent.

\begin{appendix}
\section{Proofs}

\begin{assumption}\label{as:regularity} For the sequence of finite populations indexed by \(k\), there exist positive constants \(c_{1},\dots,c_{5}\) such that
\begin{enumerate}[label=(\alph*)]
    \item Bounded potential outcomes, weights, and concentration: For each \(i\in \mathbb{I}\), \(S \subseteq \mathbb{S}_{k}\), and \(s \in \mathbb{S}_{k}\): \(\abs{Y_{i}(S)} < c_{1}\), \(\abs{w_{i}(s,d)} < c_{2}\), and \(\sum_{i\in\mathbb{I}_{k}} w_{i}(s,d) \in [c_{3},c_{4}]\).
    \item Minimum observation spacing: For any \(i,i' \in \mathbb{I}_{k}\) and \(s,s' \in \mathbb{S}_{k}\), \(d(r_{k,i},r_{k,i'}) > c_{5}\) if \(i \neq i'\), and \(d(s,s') > c_{5}\) if \(s \neq s'\).
    \item Asymptotic negligibility: \(\lim \inf_{k\to\infty} \abs{\mathbb{I}_{k}} \sigma_{k}^{2} > 0\) with \(\sigma_{k}\) defined in Theorem~\ref{thm:single-additive}(iv).
    \item Sample splitting: For each fold \(f\), \(\frac{1}{\abs{\mathbb{S}}} \sum_{s\in\mathbb{S}_{f}}\sum_{i\in\mathbb{I}_{s}} (\tilde{\mu}_{s,i} - \mu_{s,i})^{2} = O_{p}(\abs{\mathbb{S}}^{-1/2})\).
\end{enumerate}
\end{assumption}

Assumption~\ref{as:regularity}(a)~and~(b) require that outcomes are bounded, and, as the population grows, the estimator places weight on a growing geographic area and hence a growing number of candidate treatment locations.
Assumption~\ref{as:regularity}(c) ensures that no single treatment location dominates the estimator.
Assumption~\ref{as:regularity}(d), needed for the double machine learning estimator, requires that the different folds each sufficiently resemble the full sample.
For random sample splits, this bound is attained if the number of distinct values \(Z_{s,i}\) takes on is \(O(\sqrt{\abs{\mathbb{S}}})\).

\subsection{Notation and Proof of Theorem~\ref{thm:single-additive}}\label{app:proof-thm-single-additive}

\paragraph*{Notation} 
Let the random variable \(\M_{i}^{m} \equiv \ind\{\M_{i}=m\}\) be the indicator for whether exposure \(m\) of individual \(i\) is realized.
Define the random variables \(\T_{s}^{t} \equiv \ind\{\S \ni s\}\) and \(\T_{s}^{c} \equiv \ind\{\S \niton s\}\) and probabilities \(\pi_{i,s}^{m,a} \equiv \Pr(\M_{i}^{m}\T_{s}^{a}=1)\) and \(\pi_{i,s,i',s'}^{m,a,m',a'} \equiv \Pr(\M_{i}^{m}\T_{s}^{a}=1 \text{ and } \M_{i'}^{m'}\T_{s'}^{a'}=1)\), which are straightforward to compute under Assumptions~\ref{as:assign-independent}~and~\ref{as:no-effect-dist}.

The variance terms used in the statement of the theorem are, for \(a \in \{c,t\}\), 
\begin{equation*}
\begin{aligned}
\tilde{V}_{a}(d) & \equiv 
\frac{1}{\abs{\mathbb{S}}} \sum_{s\in\mathbb{S}} \sum_{i\in\mathbb{I}} \sum_{m\in\mathbb{M}_{i}} \pi_{i,s}^{m,a} \frac{w_{i}(s,d)}{\bar{n}(d)} v_{i,s}^{m,a}(d) (Y_{i}(m) - \mu_{a}(d))^{2} \\
\end{aligned}
\end{equation*}
\begin{equation*}
\begin{aligned}
\tilde{V}_{\times}(d) & \equiv 
\frac{1}{\abs{\mathbb{S}}} \sum_{s\in\mathbb{S}} \sum_{s'\in\mathbb{S}} \sum_{i\in\mathbb{I}} \sum_{m\in\mathbb{M}_{i}} \sum_{i'\in\mathbb{I}} \sum_{m'\in\mathbb{M}_{i}} \sum_{a\in\{c,t\}} \sum_{a'\in\{c,t\}} \Biggl(\ind\{i \neq i' \text{ or } s \neq s' \text{ or } a \neq a'\} \\ 
& \qquad \cdot \ind\{\pi_{i,i'}^{m,m'} > 0\} \bigl(\pi_{i,s,i',s'}^{m,a,m',a'} - \pi_{i,s}^{m,a} \pi_{i',s'}^{m',a'}\bigr) \Bigl(- \frac{\pi_{s}}{1-\pi_{s}}\Bigr)^{\ind\{a=c\}} \Bigl(- \frac{\pi_{s'}}{1-\pi_{s'}}\Bigr)^{\ind\{a'=c\}} \\
& \qquad \cdot \frac{w_{i}(s,d) w_{i'}(s',d)}{\bar{n}(d)^{2}} (Y_{i}(m) - \mu_{a}(d)) (Y_{i'}(m') - \mu_{a'}(d)) \Biggr)\\
\end{aligned}
\end{equation*}
\begin{equation*}
\begin{aligned}
\tilde{V}_{aa}(d) & \equiv
\frac{2}{\abs{\mathbb{S}}} \sum_{a\in\{c,t\}} \sum_{s\in\mathbb{S}} \sum_{s'\in\mathbb{S}} \sum_{i\in\mathbb{I}} \sum_{m\in\mathbb{M}_{i}} \sum_{i'\in\mathbb{I}} \sum_{m'\in\mathbb{M}_{i}} \ind\{\pi_{i,i'}^{m,m'} = 0\} \pi_{i,s}^{m,a} \pi_{i',s'}^{m',a} \\
& \qquad \qquad \cdot \Bigl(\frac{\pi_{s}}{1-\pi_{s}} \frac{\pi_{s'}}{1-\pi_{s'}}\Bigr)^{\ind\{a=c\}}
\frac{w_{i}(s,d) w_{i'}(s',d)}{\bar{n}(d)^{2}} \Bigl(\frac{Y_{i}(m) + Y_{i'}(m')}{2} - \mu_{a}(d)\Bigr)^{2} \\
\end{aligned}
\end{equation*}
\begin{equation*}
\begin{aligned}
\tilde{V}_{ct}(d) & \equiv 
\frac{1}{\abs{\mathbb{S}}} \sum_{s\in\mathbb{S}} \sum_{s'\in\mathbb{S}} \sum_{i\in\mathbb{I}} \sum_{m\in\mathbb{M}_{i}} \sum_{i'\in\mathbb{I}} \sum_{m'\in\mathbb{M}_{i}} \ind\{\pi_{i,i'}^{m,m'} = 0\} \pi_{i,s}^{m,t} \pi_{i',s'}^{m',c} \frac{\pi_{s'}}{1-\pi_{s'}} \\
& \qquad \qquad \cdot 
\frac{w_{i}(s,d) w_{i'}(s',d)}{\bar{n}(d)^{2}} \bigl((Y_{i}(m) - Y_{i'}(m')) - (\mu_{t}(d) - \mu_{c}(d))\bigr)^{2}
\end{aligned}
\end{equation*}
where the fixed, computable, weights \(v_{i,s}^{m,a}(d)\) are
\begin{equation*}
\begin{aligned}
v_{i,s}^{m,a}(d) \equiv & \Bigl(\frac{\pi_{s}}{1-\pi_{s}}\Bigr)^{\ind\{a=c\}} \Bigl((1-\pi_{i,s}^{m,a}) \Bigl(\frac{\pi_{s}}{1-\pi_{s}}\Bigr)^{\ind\{a=c\}} \frac{w_{i}(s,d)}{\bar{n}(d)} \\
& \qquad
+ \sum_{i'\in\mathbb{I}} \sum_{m'\in\mathbb{M}_{i}} \sum_{s'\in\mathbb{S}} \sum_{a'\in\{c,t\}} \ind\{\pi_{i,i'}^{m,m'} = 0\} \pi_{i',s'}^{m',a'} \Bigl(\frac{\pi_{s'}}{1-\pi_{s'}}\Bigr)^{\ind\{a'=c\}} \frac{w_{i'}(s',d)}{\bar{n}(d)}
\Bigr) .
\end{aligned}
\end{equation*}

\paragraph*{Proof}

For part (i), apply the mean value theorem to the function 
\(\tilde{\Delta}(\hat{p}_{t},\hat{p}_{c},\tilde{\mu}_{t},\tilde{\mu}_{c}) \equiv 
\frac{p}{\hat{p}_{t}} \tilde{\mu}_{t} - \frac{p}{\hat{p}_{c}} \tilde{\mu}_{c} 
- (\mu_{t} - \mu_{c} + \tilde{\mu}_{t} - \frac{\hat{p}_{t}}{p} \mu_{t} - \tilde{\mu}_{c} + \frac{\hat{p}_{c}}{p} \mu_{c})\)
with endpoints \((\hat{p}_{t},\hat{p}_{c},\tilde{\mu}_{t},\tilde{\mu}_{c})\) and \((p,p,\mu_{t},\mu_{c})\) to obtain
\(\hat{\tau} - \tilde{\tau} = 
(\tilde{\mu}_{t} - \mu_{t}) \Bigl(\frac{1}{\dot{p}_{t}/p} - 1 \Bigr)
- (\tilde{\mu}_{c} - \mu_{c}) \Bigl(\frac{1}{\dot{p}_{c}/p} - 1 \Bigr)
\quad + (\frac{\hat{p}_{c}}{p} - 1)\Bigl(\frac{p^{2}}{\dot{p}_{c}^{2}} \dot{\mu}_{c} - \mu_{c} \Bigr)
- (\frac{\hat{p}_{t}}{p} - 1)\Bigl(\frac{p^{2}}{\dot{p}_{t}^{2}} \dot{\mu}_{t} - \mu_{t} \Bigr)\),
where variables \(\dot{a}\) lie between \(\hat{a}\) and \(a\) for \(a=\mu_{t},\mu_{c},p_{t},p_{c}\).
Then by Theorem~1 of \citet{Li2017} and the arguments given for Part~(5) below, and using Slutsky's Theorem and the Delta Method, each of the factors of the four products is \(\sqrt{\abs{\mathbb{S}_{k}}}\)-asymptotically normal, implying part~(i).

For parts (ii) and (iii), unbiasedness for \(\tau_{\text{marginal}}(d)\) follows directly by taking expectations of the numerator of \(\mathcal{D}\).
Under Assumption~\ref{as:additive-separability}, the expected value simplifies to \(\tau(d)\) because \(\tau_{i}(s\mid S) = \tau_{i}(s)\).

To characterize the variance in part (iv), note that, under Assumption~\ref{as:no-effect-dist}(a), \(\Y_{i} = \sum_{m \in \mathbb{M}_{i}} \M_{i}^{m} Y_{i}(m)\).
One can rewrite \(\mathcal{D}\) in terms of fixed potential outcomes by using the exposure mappings, specifically
\begin{equation*}
\begin{aligned}
& \sum_{s\in\mathbb{S}} \ind\{\S \ni s\} \sum_{i\in\mathbb{I}} w_{i}(s,d) (\Y_{i}-\mu_{t}(d)) - \sum_{s\in\mathbb{S}} \ind\{\S \niton s\} \frac{\pi_{s}}{1-\pi_{s}} \sum_{i\in\mathbb{I}} w_{i}(s,d) (\Y_{i}-\mu_{c}(d)) \\
= &
\sum_{i\in\mathbb{I}} \sum_{m \in \mathbb{M}_{i}} \sum_{s\in\mathbb{S}} \sum_{a\in\{c,t\}} \M_{i}^{m} \T_{s}^{a} \tilde{Y}_{i}^{s,a}(d,m) \\
\end{aligned}
\end{equation*}
with \(\tilde{Y}_{i}^{s,a}(d,m) \equiv \Bigl(-\frac{\pi_{s}}{1-\pi_{s}}\Bigr)^{\ind\{a=c\}} w_{i}(s,d) (Y_{i}(m) - \mu_{a}(d))\).
Importantly, only \(\M_{i}^{m} \T_{s,a}\) is stochastic in the expression above.
Hence, the variance depends on covariances\\
\(\cov(\M_{i}^{m} \T_{s}^{a},\M_{i'}^{m'} \T_{s'}^{a'}) = \pi_{i,s,i',s'}^{m,a,m',a'} - \pi_{i,s}^{m,a} \pi_{i',s'}^{m',a'}\).

Where \(\pi_{i,s,i',s'}^{m,a,m',a'} = 0\) such that \(m\) and \(m'\) cannot be observed simultaneously, rewrite terms \(\sum_{s}\sum_{s'}(Y_{i}(m)-\mu_{a}(d))(Y_{i}(m')-\mu_{a'}(d))\) with \(a,a' \in \{t,c\}\) depending on whether the current locations \(s,s'\) are treated or control under exposures \(m,m'\).
When \(a \neq a'\), rewrite \((Y_{i}(m)-\mu_{t}(d))(Y_{i'}(m')-\mu_{c}(d)) = \frac{1}{2} \Bigl((Y_{i}(m)-\mu_{t}(d))^{2} + (Y_{i'}(m')-\mu_{c}(d))^{2} - (Y_{i}(m) - Y_{i'}(m') - (\mu_{t}(d)-\mu_{c}(d)))^{2} \Bigr)\) as a kind of variance of ``treatment effects.''
When \(a=a'\), these terms are multiplied by a factor of opposite sign.
To obtain a formula suggesting a conservative estimator of the variance, when \(a=a'\) instead rewrite 
\begin{equation*}
\begin{aligned}
&(Y_{i}(m)-\mu_{a}(d))(Y_{i'}(m')-\mu_{a}(d)) \\
= &\frac{1}{2} \Bigl(
(Y_{i}(m) + Y_{i'}(m') - 2\mu_{a}(d)))^{2}
- (Y_{i}(m)-\mu_{a}(d))^{2} - (Y_{i'}(m')-\mu_{a}(d))^{2} \Bigr)
.
\end{aligned}
\end{equation*}
The remaining steps simplify the summations over such terms.
I show step-by-step derivations in Online~Appendix~5.

Part~(v) claims that \(\frac{\tilde{\tau}_{k} - \tau_{k,\text{marginal}}}{\sigma_{k}} \stackrel{d}{\to}\mathcal{N}(0,1)\) as \(k \to \infty\).
First, note that without Assumption~\ref{as:no-effect-dist}(a) \(\tilde{\tau}_{k}(d)-\tau_{k,\text{marginal}}(d) = \mathcal{D}_{k} = \sum_{i\in\mathbb{I}_{k}} \mathcal{Z}_{k,i} + \epsilon_{k}\) where \\
\(\mathcal{Z}_{k,i} \equiv \sum_{m \in \mathbb{M}_{k,i}} \sum_{s\in\mathbb{S}_{k}} \sum_{a\in\{c,t\}} \M_{k,i}^{m} \T_{s}^{a} \frac{\tilde{Y}_{k,i}^{s,a}(d,m)}{\abs{\mathbb{S}_{k}} \bar{n}_{k}(d)}\) and \(\epsilon\) as in Assumption~\ref{as:no-effect-dist}(b).
Because \(\epsilon_{k}/\sigma_{k} = o_{p}(1)\) by Assumption~\ref{as:no-effect-dist}(b), it suffices to show \(\sum_{i\in\mathbb{I}_{k}} \mathcal{Z}_{k,i} / \sigma_{k} \stackrel{d}{\to} \mathcal{N}(0,1)\).
The result follows from Theorem~1 of \citet{Jenish2009}, the conditions of which I verify below.
Note that \(\sum_{i\in\mathbb{I}_{k}} \mathcal{Z}_{k,i} / \sigma_{k} = \sum_{i\in\mathbb{I}_{k}} \tilde{\mathcal{Z}}_{k,i} / \tilde{\sigma}_{k}\) with \(\tilde{\mathcal{Z}}_{k,i} \equiv \abs{\mathbb{I}_{k}} \mathcal{Z}_{k,i}\) and \(\tilde{\sigma}_{k} \equiv \abs{\mathbb{I}_{k}} \sigma_{k}\).
By part~(iv), \(\var(\sum_{i\in\mathbb{I}_{k}} \mathcal{Z}_{k,i}) = \sigma_{k}^{2}\), hence \(\var(\sum_{i\in\mathbb{I}_{k}} \tilde{\mathcal{Z}}_{k,i}) = \abs{\mathbb{I}_{k}}^{2} \sigma_{k}^{2} = \tilde{\sigma}_{k}^{2}\).
Assumption~1 of \citet{Jenish2009} is satisfied given the minimum distance between individuals.
Their Assumption~2 is satisfied with their \(c_{i,n} = 1\) given bounded potential outcomes and weights, and the minimum distance between locations.
Furthermore, given the definition of exposures depending only on locations up to a fixed distance, and \(0\) weight at sufficiently large distances, \(\tilde{\mathcal{Z}}_{k,i}\) and \(\tilde{\mathcal{Z}}_{k,j}\) are independent if \(d(r_{k,i},r_{k,j}) > D\) for some \(D\), such that the random field is \(\phi\)-mixing satisfying their Assumption~4.
Finally, the asymptotic negligibility condition implies their Assumption~5.
Hence, by Theorem~1 of \citet{Jenish2009}, \(\sum_{i\in\mathbb{I}_{k}} \mathcal{Z}_{k,i} / \sigma_{k} = \sum_{i\in\mathbb{I}_{k}} \tilde{\mathcal{Z}}_{k,i} / \tilde{\sigma}_{k} \stackrel{d}{\to} \mathcal{N}(0,1)\).

\subsection{Proof of Theorem~\ref{thm:only-nearest-identification}}\label{app:proof-thm-only-nearest-identification}

Independent assignment implies that for each individual \(i\) and candidate treatment location \(s\), there is a positive probability the location is the nearest realized treatment location.
The assumption that only the nearest realized location matters implies that in this case \(Y_{i}(s)\) is observed, rather than \(Y_{i}(s \cup S)\) for some set of other locations \(S\) farther from \(i\) than \(s\).
The control potential outcome \(Y_{i}(0)\) is observed when no treatment location within distance \(\dnoeffect{}\) is treated, which occurs with positive probability under independent assignment.

\subsection{Proof of Theorem~\ref{thm:dml}}\label{app:proof-thm-dml}

For ease of notation, I suppress dependence on the distance \(d\) throughout the proof.

The proof establishes that \((\hat{\tau}(\hat{\pi},\hat{\mu}) - \hat{\tau}(\pi,\mu))/\sigma = o_{p}(1)\).
The conclusion of the theorem follows because \((\hat{\tau}(\pi,\mu) - \tau_{\text{marginal}}) / \sigma \stackrel{d}{\to} \mathcal{N}(0, 1)\).

First, note that \((\frac{\hat{\pi}_{s}}{1-\hat{\pi}_{s}} - \frac{\pi_{s}}{1-\pi_{s}})^{2} = \frac{(\hat{\pi}_{s}-\pi_{s})^{2}}{(1-\hat{\pi}_{s})^{2}(1-\pi_{s})^{2}} \leq C_{0} (\hat{\pi}_{s}-\pi_{s})^{2}\) for some fixed real number \(C_{0}\) because \(\hat{\pi}_{s}\) and \(\pi_{s}\) are bounded away from one.
Hence, \(\sum_{s\in\mathbb{S}} (\frac{\hat{\pi}_{s}}{1-\hat{\pi}_{s}} - \frac{\pi_{s}}{1-\pi_{s}})^{2} \leq C_{0} \sum_{s\in\mathbb{S}} (\hat{\pi}_{s} - \pi_{s})^{2} = o_{p}(\sqrt{\abs{\mathbb{S}}})\).

Note that \(1/(\sum_{s\in\mathbb{S}} \sum_{i\in\mathbb{I}} w_{i}(s) \T_{s}^{t}) = O_{p}(\abs{\mathbb{S}}^{-1})\) by the regularity conditions and \(1/\sigma = O(\sqrt{\abs{\mathbb{S}}})\) by assumption, so it suffices to show that the difference in numerators of \(\hat{\tau}(\hat{\pi},\hat{\mu})\) and \(\hat{\tau}(\pi,\mu)\) is \(o_{p}(\sqrt{\abs{\mathbb{S}}})\).
One can write the difference in numerators as \(\sum_{f=1}^{F} (\R_{1,f} + \R_{2,f} + \R_{3,f})\) where
\begin{equation*}
\begin{aligned}
\R_{1,f} & = 
\sum_{s \in \mathbb{S}_{f}} \Bigl(\frac{\T_{s}^{t} - \pi_{s}}{1-\pi_{s}}\Bigr) \sum_{i \in \mathbb{I}} w_{i}(s)  (\mu_{s,i} - \hat{\mu}_{s,i}) \\
\R_{2,f} & = 
\sum_{s \in \mathbb{S}_{f}} \sum_{i \in \mathbb{I}} w_{i}(s) \T_{s}^{c}\Bigl(\frac{\pi_{s}}{1-\pi_{s}} - \frac{\hat{\pi}_{s}}{1-\hat{\pi}_{s}} \Bigr) (\Y_{i} - \mu_{s,i})
\\
\R_{3,f} & = 
\sum_{s \in \mathbb{S}_{f}} \sum_{i \in \mathbb{I}} w_{i}(s) \T_{s}^{c} \Bigl(\frac{\pi_{s}}{1-\pi_{s}} - \frac{\hat{\pi}_{s}}{1-\hat{\pi}_{s}}\Bigr) (\mu_{s,i} - \hat{\mu}_{s,i})
\end{aligned}
\end{equation*}

For \(\R_{1,f}\), consider \(E(\R_{1,f}^{2} \mid \mathbb{S}_{f}, (\hat{\mu}_{s})_{s\in\mathbb{S}_{f}})\) which conditions on the identity of observations in fold \(f\) and the out-of-fold estimator \(\hat{\mu}_{s}\) for all \(s \in \mathbb{S}_{f}\), where \(\hat{\mu}_{s} = (\hat{\mu}_{s,i})_{i\in\mathbb{I}: w_{i}(s) \neq 0}\).
Because, by independent assignment and sample splitting, \(E\bigl( (\frac{\T_{s}^{t} - \pi_{s}}{1-\pi_{s}}) (\frac{\T_{s'}^{t} - \pi_{s'}}{1-\pi_{s'}}) \mid \mathbb{S}_{f}, (\hat{\mu}_{s})_{s\in\mathbb{S}_{f}} \bigr) = 0\) for any \(s \neq s'\) in fold \(f\), and all \(\hat{\mu}_{s,i}\) with \(f(s)=f\) are fixed by conditioning,
for fixed real numbers \(C_{1}\) and \(C_{2}\),
\(
E(\R_{1,f}^{2} \mid \mathbb{S}_{f}, (\hat{\mu}_{s,i})_{s\in\mathbb{S}_{f}})
\leq 
C_{1} \sum_{s\in\mathbb{S}_{f}} \Bigl(\sum_{i\in\mathbb{I}} w_{i}(s)(\mu_{s,i} - \hat{\mu}_{s,i}) \Bigr)^{2}
\\\leq
C_{1} C_{2} \sum_{s\in\mathbb{S}_{f}} \sum_{i\in\mathbb{I}_{s}} (\mu_{s,i} - \hat{\mu}_{s,i})^{2},
\)
where the first inequality bounds \(E((\frac{\T_{s}^{t}-\pi_{s}}{1-\pi_{s}})^{2})\),
and the second inequality uses the Cauchy-Schwarz inequality and the bound on \(\abs{w_{i}(s,d)}\).
Hence, \(\sum_{s\in\mathbb{S}_{f}} \sum_{i\in\mathbb{I}_{s}} (\mu_{s,i} - \hat{\mu}_{s,i})^{2} = o_{p}(\abs{\mathbb{S}})\) implies \(\R_{1,f} = o_{p}(\sqrt{\abs{\mathbb{S}}})\) by Markov's inequality as required.

For \(\R_{2,f}\),
define \(\delta_{s,i} = w_{i}(s) \T_{s}^{c}(\frac{\pi_{s}}{1-\pi_{s}} - \frac{\hat{\pi}_{s}}{1-\hat{\pi}_{s}}) (\Y_{i} - \mu_{s,i})\) and \(\bar{\delta}_{s,i} = E(\delta_{s,i} \mid \mathbb{S}_{f}, (\hat{p}_{s})_{s\in\mathbb{S}_{f}})\) and \(\delta_{s} = \sum_{i\in\mathbb{I}} \delta_{s,i}\), \(\bar{\delta}_{s} = \sum_{i\in\mathbb{I}} \bar{\delta}_{s,i}\).
Then \(\R_{2,f} = \sum_{s\in\mathbb{S}_{f}} (\delta_{s} - \bar{\delta}_{s}) + \sum_{s\in\mathbb{S}_{f}} \bar{\delta}_{s}\).

For the first term, similar to \(\R_{1,f}\), consider \(E\bigl((\sum_{s\in\mathbb{S}_{f}} \delta_{s} - \bar{\delta}_{s})^{2} \mid \mathbb{S}_{f}, (\hat{\pi}_{s})_{s\in\mathbb{S}_{f}}\bigr)\).
By definition, \(E(\delta_{s} - \bar{\delta}_{s} \mid \mathbb{S}_{f}, (\hat{\pi}_{s})_{s\in\mathbb{S}_{f}}) = 0\), and \(\delta_{s} - \bar{\delta}_{s}\) and \(\delta_{s'} - \bar{\delta}_{s'}\) are (conditionally) independent unless there exists either an individual \(i'\) with \(w_{i'}(s') \neq 0\) such that treatment at location \(s\) can affect the outcome of \(i\), or individuals \(i\) with \(w_{i}(s) \neq 0\) and \(i'\) with \(w_{i'}(s') \neq 0\) with a common treatment location \(s''\) that can affect the outcome of both \(i\) and \(i'\).
Under the assumptions of a minimum distance between candidate treatment locations, a maximum distance after which weights are zero, and a maximum distance after which the treatment has no effect, for any \(s'\), the number of \(s\) for which the above occurs is bounded.
For these \(s\), \(s'\), bound \((\delta_{s}-\bar{\delta}_{s})(\delta_{s'}-\bar{\delta}_{s'}) \leq (\delta_{s}-\bar{\delta}_{s})^{2} + (\delta_{s'}-\bar{\delta}_{s'})^{2}\).
Steps analogous to \(\R_{1,f}\) then yield, for a fixed real number \(C_{3}\),
\(
E\bigl((\sum_{s\in\mathbb{S}_{f}} \delta_{s} - \bar{\delta}_{s})^{2} \mid \mathbb{S}_{f}, (\hat{p}_{s})_{s\in\mathbb{S}_{f}}\bigr)
\leq
C_{3}
\sum_{s\in\mathbb{S}_{f}} (\frac{\pi_{s}}{1-\pi_{s}} - \frac{\hat{\pi}_{s}}{1-\hat{\pi}_{s}})^{2}
\).
So \(\sum_{s\in\mathbb{S}_{f}} (\frac{\pi_{s}}{1-\pi_{s}} - \frac{\hat{\pi}_{s}}{1-\hat{\pi}_{s}})^{2} = o_{p}(\abs{\mathbb{S}})\) implies that the first term is \(o_{p}(\sqrt{\abs{\mathbb{S}}})\).

For the second term, \(\sum_{s\in\mathbb{S}_{f}} \bar{\delta}_{s}\), use that \(\pi_{s} = p(Z_{s})\) and \(\hat{\pi}_{s} = \hat{\pi}_{s'}\) if \(Z_{s} = Z_{s'}\) and \(f(s) = f(s')\), to write
\(
\sum_{s\in\mathbb{S}_{f}} \bar{\delta}_{s} = 
\sum_{s\in\mathbb{S}_{f}}\sum_{i\in\mathbb{I}} w_{i}(s) (1-\pi_{s})(\frac{\pi_{s}}{1-\pi_{s}} - \frac{\hat{p}_{s}}{1-\hat{p}_{s}}) (\tilde{\mu}_{s,i} - \mu_{s,i})
\).
Then, using the Cauchy-Schwarz inequality and the bound on \(\lvert w_{i}(s)\rvert\),
\\\(
(\sum_{s\in\mathbb{S}_{f}} \bar{\delta}_{s})^{2} \leq C_{4} 
\sum_{s\in\mathbb{S}_{f}} (\frac{\pi_{s}}{1-\pi_{s}} - \frac{\hat{\pi}_{s}}{1-\hat{\pi}_{s}}))^{2} \sum_{s\in\mathbb{S}_{f}}\sum_{i\in\mathbb{I}_{s}} (\tilde{\mu}_{s,i} - \mu_{s,i})^{2} 
\).
So, \(\sum_{s\in\mathbb{S}_{f}} (\frac{\pi_{s}}{1-\pi_{s}} - \frac{\hat{\pi}_{s}}{1-\hat{\pi}_{s}}))^{2} = o_{p}(\sqrt{\abs{\mathbb{S}}})\) and \(\sum_{s\in\mathbb{S}_{f}}\sum_{i\in\mathbb{I}_{s}} (\tilde{\mu}_{s,i} - \mu_{s,i})^{2} = O_{p}(\sqrt{\abs{\mathbb{S}}})\) ensure \(\sum_{s\in\mathbb{S}_{f}} \bar{\delta}_{s} = o_{p}(\sqrt{\abs{\mathbb{S}}})\).
Hence, overall \(\R_{2,f} = o_{p}(\sqrt{\abs{\mathbb{S}}})\) by Markov's inequality.

For \(\R_{3,f}\), applying the Cauchy-Schwarz inequality and bounding \(\lvert w_{i}(s) \rvert\) yields
\\\(
E(\R_{3,f}^{2} \mid \mathbb{S}_{f}, (\hat{\pi}_{s},\hat{\mu}_{s,i})_{s\in\mathbb{S}_{f}})
\leq
C_{5} 
\sum_{s \in \mathbb{S}_{f}} \Bigl(\frac{\pi_{s}}{1-\pi_{s}} - \frac{\hat{\pi}_{s}}{1-\hat{\pi}_{s}}\Bigr)^{2}
\sum_{s \in \mathbb{S}_{f}} \sum_{i \in \mathbb{I}_{s}} (\mu_{s,i} - \hat{\mu}_{s,i})^{2}
\).
Hence, \(\sum_{s \in \mathbb{S}_{f}} \Bigl(\frac{\pi_{s}}{1-\pi_{s}} - \frac{\hat{\pi}_{s}}{1-\hat{\pi}_{s}}\Bigr)^{2} = o_{p}(\sqrt{\abs{\mathbb{S}}})\) and \(\sum_{s \in \mathbb{S}_{f}} \sum_{i \in \mathbb{I}_{s}} (\mu_{s,i} - \hat{\mu}_{s,i})^{2} = o_{p}(\sqrt{\abs{\mathbb{S}}})\) jointly imply that \(\R_{3,f} = o_{p}(\sqrt{\abs{\mathbb{S}}})\) by Markov's inequality.

\subsection{Proof of Theorem~\ref{thm:prop-score-hd}}\label{app:proof-prop-score-hd}

For \(s \in \tilde{\mathbb{S}} \setminus \mathbb{S}\), \(\pi_{s} = 0\) by the definition of \(\mathbb{S}\), and also \(\hat{\pi}_{s} = 0\) with probability one by Assumption~\ref{as:spatial-unconfoundedness}.
Hence, it suffices to focus on \(s\in\mathbb{S}\).
Under Assumption~\ref{as:spatial-unconfoundedness}, one can equivalently define \(n_{z} = \sum_{s\in\mathbb{S}} \ind\{Z_{s} = z\}\) for \(z\in\mathbb{Z}\) because for any \(s\in\tilde{\mathbb{S}}\), if \(Z_{s}\in\mathbb{Z}\) then \(s\in\mathbb{S}\) as \(p(Z_{s}) > 0\) by the definition of \(\mathbb{Z}\).
Note that \(\sum_{s\in\mathbb{S}} (\hat{\pi}_{s}-\pi_{s})^{2} = \sum_{z\in\mathbb{Z}} n_{z} (\hat{p}(z)-p(z))^{2}\) where \(\hat{p}(z) = \sum_{s\in\mathbb{S}} \ind\{Z_{s}=z\} \T_{s}^{t} / n_{z}\).
Let \(\mathcal{E}_{z} \equiv (\hat{p}(z)-p(z))^{2} - \frac{p(z)(1-p(z))}{n_{z}}\).
Then, for \(\epsilon > 0\),
\(\Pr(\sum_{s\in\mathbb{S}} (\hat{\pi}_{s}-\pi_{s})^{2} / a > \epsilon)
\leq
\Pr(\sum_{z\in\mathbb{Z}} n_{z} \mathcal{E}_{z} > a \epsilon - \abs{\mathbb{Z}}/4)\) using \(\sum_{z\in\mathbb{Z}} p(z)(1-p(z)) \leq \abs{\mathbb{Z}}/4\).

Next, by Bernstein's inequality for sums of independent mean-zero sub-Exponential random variables, 
\(\Pr(\sum_{z\in\mathbb{Z}} n_{z} \mathcal{E}_{z} > u) \leq 2 \exp(-c \min(\frac{u^2}{\abs{\mathbb{Z}} C^{2}}, \frac{u}{C}))\) for universal constants \(c\) and \(C\):
Because \(\hat{p}(z)\) is the average of \(n_{z}\) independent bounded (binary) random variables with mean \(p(z)\), there exists a fixed \(C\) such that \(\hat{p}(z)-p(z)\) is sub-Gaussian with norm at most \(C/\sqrt{n_{z}}\), and \(\mathcal{E}_{z}\) is mean zero and sub-Exponential with norm at most \(C/n_{z}\) \citep{Vershynin2018} for all \(z\in\mathcal{Z}\).
So \(n_{z} \mathcal{E}_{z}\) is sub-Exponential with norm at most \(C\).
By Assumption~\ref{as:assign-independent}, \(\mathcal{E}_{z}\) are independent across \(z\).
Hence, Bernstein's inequality applies as claimed.

Substitute \(u = \sqrt{\abs{\mathbb{S}}} \epsilon - \abs{\mathbb{Z}}/4\).
Then, \(\abs{\mathbb{Z}} = o(\sqrt{\abs{\mathbb{S}}})\) implies that \(\min(\frac{u^2}{\abs{\mathbb{Z}} C^{2}}, \frac{u}{C}) \to \infty\) as \(\abs{\mathbb{S}} \to \infty\).
Hence, \(\lim_{\abs{\mathbb{S}}\to\infty} \Pr(\sum_{s\in\mathbb{S}} (\hat{\pi}_{s}-\pi_{s})^{2} / \sqrt{\abs{\mathbb{S}}} > \epsilon) = 0\) such that \(\sqrt{\abs{\mathbb{S}}^{-1} \sum_{s\in\mathbb{S}} (\hat{\pi}_{s}-\pi_{s})^{2}} = o_{p}(\abs{\mathbb{S}}^{-1/4})\) as stated in the theorem.

\subsection{Proof of Theorem~\ref{thm:spatial-lasso}}

Following \citet[Theorem~7.13]{Wainwright2019}, if \(\lambda \geq 2 \max_{l} \{\abs{\boldsymbol{\Z}_{\cdot,l}'(\boldsymbol{\Y}-\boldsymbol{\Z}\beta)}\}/\dot{n}\) and the restricted eigenvalue condition holds, then \(\sqrt{\sum_{l} (\hat{\beta}_{l} - \beta_{l})^{2}} \leq \frac{3}{\kappa} \sqrt{\norm{\beta}_{0}} \lambda\).
The restricted eigenvalue condition is assumed to hold with probability approaching one.
The proof establishes that the first condition holds with high probability for the choice of \(\lambda\) given in the theorem, and establishes the resulting rate.

To show that the choice of \(\lambda\) specified in the theorem satisfies the condition above with probability approaching 1,
use the approximate clustering notation and the triangle inequality to get
\(
\max_{l} \{\abs{\boldsymbol{\Z}_{\cdot,l}'(\boldsymbol{\Y}-\boldsymbol{\Z}\beta)}\}
\leq 
\max_{l} \{\abs{\boldsymbol{\Z}_{\cdot,l}'(\boldsymbol{\Y}-\boldsymbol{Y}^{\boldsymbol{\M}})}\}
+
\max_{l} \{\abs{\boldsymbol{\Z}_{\cdot,l}'(\boldsymbol{Y}^{\boldsymbol{\M}}-\boldsymbol{\Z}\beta)}\}
\).
By the approximate clustering assumption, \(\max_{l} \{\abs{\boldsymbol{\Z}_{\cdot,l}'(\boldsymbol{\Y}-\boldsymbol{Y}^{\boldsymbol{\M}})}\}/\dot{n} = o_{p}(\sqrt{\ln(L)/\abs{\mathbb{C}}})\).
Below, I show that also \(\max_{l} \{\abs{\boldsymbol{\Z}_{\cdot,l}'(\boldsymbol{Y}^{\boldsymbol{\M}}-\boldsymbol{\Z}\beta)}\}/\dot{n} \leq \sqrt{\frac{C}{\abs{\mathbb{C}}} \ln(L)}\) with probability approaching \(1\) for some fixed \(C\).
Hence, there exists a fixed \(\tilde{C}\) such that \(\lambda = \tilde{C} \sqrt{\ln(L)/\abs{\mathbb{C}}} \geq 2 \max_{l}\{\abs{\boldsymbol{\Z}_{\cdot,l}'(\boldsymbol{\Y}-\boldsymbol{\Z}\beta)}\}/\dot{n}\) with probability approaching 1.
Because \(\abs{\mathbb{S}}/\abs{\mathbb{C}}\) is bounded by Assumption~\ref{as:approximate-clusters}(i), one can replace \(\abs{\mathbb{C}}\) by \(\abs{\mathbb{S}}\) in the definition of \(\lambda\) with an updated choice of \(\tilde{C}\) without affecting the validity of the claims.
Substituting the choice \(\lambda = \tilde{C} \sqrt{\ln(L)/\abs{\mathbb{S}}}\) into \(\sqrt{\sum_{l} (\hat{\beta}_{l} - \beta_{l})^{2}} \leq \frac{3}{\kappa} \sqrt{\norm{\beta}_{0}} \lambda\) yields the claim of the theorem.

To bound \(\abs{\boldsymbol{\Z}_{\cdot,l}'(\boldsymbol{Y}^{\boldsymbol{\M}}-\boldsymbol{\Z}\beta)}/\dot{n}\) for fixed \(l\), apply Hoeffding's inequality.
Then, apply the union bound to bound the maximum over \(l\).
Define
\(\rho_{c,l} \equiv \frac{1}{\dot{n}/\abs{\mathbb{C}}} \sum_{s\in\mathbb{S}} \sum_{i\in\mathbb{I}} \ind\{c(s,i)=c\} w_{i}(s,d) \ind\{\S \niton s\} Z_{(s,i),l} (Y_{i}(\M_{s,i}) - Z_{s,i}\beta)\)
such that
\(\boldsymbol{\Z}_{\cdot,l}(\boldsymbol{Y}^{\boldsymbol{\M}}-\boldsymbol{\Z}\beta)/\dot{n} = \frac{1}{\abs{\mathbb{C}}} \sum_{c \in \mathbb{C}} (\rho_{c,l} - E(\rho_{c,l}))\)
because \(\sum_{c\in\mathbb{C}} E(\rho_{c,l}) = 0\) by linearity.
Note that the \(\rho_{c,l}\) terms are independent across \(c\) because \(Y_{i}(\M_{s,i})\) varies only with assignments within the cluster of \((s,i)\) by the definition of the approximate exposures, and \(\ind\{\S \niton s\}\) is independent across clusters by Assumptions~\ref{as:assign-independent}~and~\ref{as:approximate-clusters}(ii).
Furthermore, because all factors inside the summation constituting \(\rho_{c,l}\) are bounded, as is the total number of observations in the cluster, \(\rho_{c,l}\) is bounded.
Hence, \(\boldsymbol{\Z}_{\cdot,l}'(\boldsymbol{Y}^{\boldsymbol{\M}}-\boldsymbol{\Z}\beta)/\dot{n}\) is the average of \(\abs{\mathbb{C}}\) independent bounded mean zero random variables.
Then, by Hoeffding's inequality, for all \(t>0\), \(\Pr(\abs{\frac{1}{\abs{\mathbb{C}}} \sum_{c\in\mathbb{C}} (\rho_{c,l} - E(\rho_{c,l}))} \geq u) \leq 2 \exp(- \frac{2u^{2} \abs{\mathbb{C}}}{C})\) for some \(C < \infty\).
Taking the union bound,
\(\Pr(\max_{l}\abs{\frac{1}{\abs{\mathbb{C}}} \sum_{c\in\mathbb{C}} (\rho_{c,l} - E(\rho_{c,l}))} \geq u) \leq 2 \exp(- \frac{2u^{2} \abs{\mathbb{C}}}{C} + \ln(L))\).
So,
\(\Pr\Bigl(\max_{l} \{\abs{\boldsymbol{\Z}_{\cdot,l}(\boldsymbol{Y}^{\boldsymbol{\M}}-\boldsymbol{\Z}\beta)}\}/\dot{n} \leq \sqrt{\frac{C}{\abs{\mathbb{C}}} \ln(L)}\Bigr) \geq 1 - 2 \exp(- \ln(L)) \to 1\) as \(L\to\infty\).

\end{appendix}

\printbibliography[title={References}]

\end{refsection}

\clearpage

\setcounter{table}{0}
\setcounter{figure}{0}
\setcounter{equation}{0}
\setcounter{section}{0}
\renewcommand\thetable{OA\arabic{table}}
\renewcommand\thefigure{OA\arabic{figure}}
\renewcommand{\theequation}{OA\arabic{equation}}
\renewcommand{\thesection}{\arabic{section}}
\renewcommand{\theHtable}{OA\arabic{table}}
\renewcommand{\theHfigure}{OA\arabic{figure}}
\renewcommand{\theHequation}{OA\arabic{equation}}
\renewcommand{\theHsection}{OA\arabic{section}}

\begin{refsection}

\begin{center}
\LARGE \textbf{Supplement to ``Causal Inference for Spatial Treatments''}
\end{center}

\section{Additional Tables and Figures}

Table~\ref{tab:examples} lists examples of papers studying spatial treatments.
The outcomes or outcome units mentioned in the table are either directly studied in each paper or are closely related to the question studied.
The list is meant to help the reader map empirical objects into the framework of this paper and to illustrate the breadth of topics involving spatial treatments.
Not all of these papers had precise location data on treatments and/or outcome units, but such data could, in principle, be collected in all instances.
\citet{dell2020development} is the only example on this list explicitly considering counterfactual treatment locations.\footnote{There are other empirical studies considering counterfactual treatment locations, but to the best of my knowledge, none include statistical theory allowing design-based inference.}
The theory in the present paper derives standard errors complementing the \(p\)-values of randomization tests of the sharp null hypothesis reported in the original paper.

\begin{sidewaystable}[ht]
\caption{\label{tab:examples}Examples of papers studying spatial treatments, and outcomes or outcomes units that are either directly studied in each paper or are closely related to the question studied.}
\begin{tabular}{l l l}
\hline
paper & spatial treatment & outcome / outcome units \\
\hline
\citet{aliprantis2015blowing} & public housing demolition & crime in local neighborhoods \\
\citet{athey2018estimating} & restaurant opening & utility of consumers \\
\citet{buchmueller2006far} & hospital closure & mortality of residents \\
\citet{cohen2010bednets} & subsidized bed nets sold at hospitals & adoption of bed nets in local communities \\
\citet{currie2015environmental} & toxic plant opening and closing & house prices, infant health \\
\citet{dell2020development} & site of historic sugar mill & economic development of nearby towns \\
\citet{diamond2019wants} & low income housing projects & house prices \\
\citet{DiTella2004} & police presence in city blocks & number of car thefts \\
\citet{duflo2001schooling} & school construction & educational attainment in nearby villages \\
\citet{ellickson2013wal} & Walmart entry & entry, exit of competitors \\
\citet{feyrer2017shocks} & fracking site & income of local residents \\
\citet{greenstone2003bidding} & large manufacturing plant entry & property values, labor earnings of residents \\
\citet{greenstone2010identifying} & large manufacturing plant entry & TFP of other plants \\
\citet{jia2008happens} & Walmart entry & profit/exit of small discount stores \\
\citet{keiser2019consequences} & wastewater treatment plants & commercial \& recreational value of rivers \\
\citet{linden2008estimates} & sex offenders moving in & house prices \\
\citet{miguel2004worms} & deworming administered at schools & worm prevalence in local population \\
\citet{oates1969effects} & (spending on) local public goods & property values \\
\citet{seim2006empirical} & video store entry & effect on local competitors \\
\citet{siegfried2000economics} & sport stadiums & local businesses, property values \\
\citet{stock1991nonparametric} & toxic waste cleanup & property values \\
\hline
\end{tabular}
\end{sidewaystable}

\begin{figure}
\begin{tabular}{@{} K{1.75in} @{} K{1.4in} @{} K{1.4in} @{} K{1.4in} @{}}
{\tiny restaurants dist. \(0.1\)mi--\(0.125\)mi} & {\tiny \(0.125\)mi--\(0.15\)mi to grocery} & {\tiny \(0.15\)mi--\(0.175\)mi to grocery} & {\tiny \(0.175\)mi--\(0.2\)mi to grocery} \\
\end{tabular}
\rotatebox{90}{\tiny \hspace{2em} restaurant}
\includegraphics{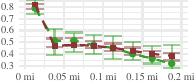} \hfill
\includegraphics{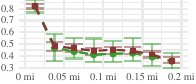} \hfill
\includegraphics{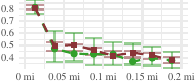} \hfill
\includegraphics{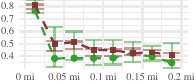} \hfill \\
\rotatebox{90}{\tiny \hspace{1em} amusement}
\includegraphics{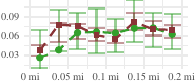} \hfill
\includegraphics{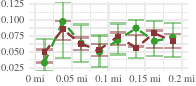} \hfill
\includegraphics{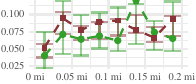} \hfill
\includegraphics{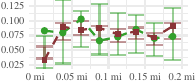} \hfill \\
\rotatebox{90}{\tiny \hspace{1em} museum}
\includegraphics{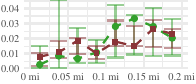} \hfill
\includegraphics{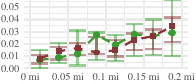} \hfill
\includegraphics{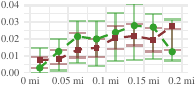} \hfill
\includegraphics{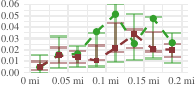} \hfill \\
\rotatebox{90}{\tiny \hspace{1em} religious org}
\includegraphics{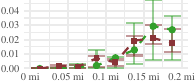} \hfill
\includegraphics{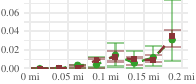} \hfill
\includegraphics{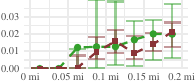} \hfill
\includegraphics{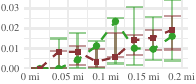} \hfill \\
\rotatebox{90}{\tiny \hspace{2em} dentist}
\includegraphics{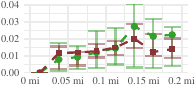} \hfill
\includegraphics{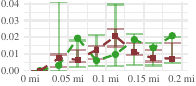} \hfill
\includegraphics{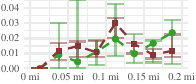} \hfill
\includegraphics{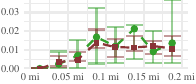} \hfill \\
\rotatebox{90}{\tiny \hspace{1em} auto repair}
\includegraphics{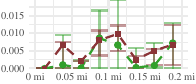} \hfill
\includegraphics{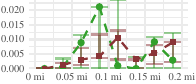} \hfill
\includegraphics{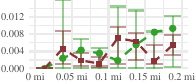} \hfill
\includegraphics{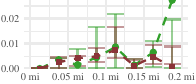} \hfill
\caption{\label{fig:balance-long-dist}Balance in neighborhood characteristics.
Each plot shows the fraction of businesses with, in different rows, 4-digit NAICS 7225, 7139, 7121, 8131, 6212, or 8111 at multiple distances (horizontal axis) from treated (red squares) and untreated (green dots) restaurants.
Columns show restaurants in different distance bins around candidate grocery store locations.
To give a sense of statistical uncertainty in characteristics balance, I take 10,000 draws from the treatment assignment distribution (given by Assumptions~1~and~5) and display error bars covering the balance in characteristics realized in 95\% of these draws.}
\end{figure}

\FloatBarrier

\section{Conley-Type Standard Errors in Design-Based Analysis}

\subsection{Undercoverage of design-based implementations}\label{app:conley-design}

Two simple examples show that natural design-based implementations of \citet{conley1999gmm}-type standard errors, similar to one studied by \citet{Wang2025}, can be invalid for design-based inference.

In both examples, the treatment effect estimator studied in Theorem~1 is numerically identical to the slope coefficient in a linear regression of each individual's observed outcome on an intercept and the treatment status of their nearest candidate treatment location.
I consider the variance estimator \(\hat{\sigma}^{2} = ((X'X)^{-1} \Omega (X'X)^{-1})_{(2,2)}\) where \(\Omega = \sum_{i\in\mathbb{I}} \sum_{j\in\mathbb{I}} K_{i,j} X_{i}'X_{j} e_{i} e_{j}\) and \(X\) is the regressor matrix with row \(i\) given by \(X_{i} = [1, W_{s(i)}]\).
\(W_{s(i)}\) is the treatment status of the nearest candidate location, \(e_{i}\) is the \(i\)th regression residual, and \(K_{i,j} \in [0,1]\) is a kernel weight.
Design-based implementations set \(K_{i,j} = 0\) when \(i\) and \(j\) are design-based independent and hold \(K_{i,j}\) fixed along the asymptotic sequence of populations for fixed individuals \(i\) and \(j\).

The first example behaves identically to an experiment with clustered assignment.
Locations are in \(\mathbb{R}\).
The location of each ``town'' begins at a different multiple of \(10\), and each town contains two individuals and one candidate treatment location.
The individuals differ in their location offset \(1,2\) relative to the beginning of their town, and outcomes differ by town.
Hence, \(r_{i} \in \{1,2,11,12,\dots\}\) for \(i\in\mathbb{I}\) and \(\mathbb{S} = \{1.5,11.5,\dots\}\).
In towns located at even multiples of 10, outcomes in the absence of treatment in the town equal \(0\); in ``odd'' towns such outcomes equal \(1\).
The single candidate treatment location is at location offset \(1.5\).
If the treatment is realized in a town, all outcomes in the town increase by \(1\) unit.
The probability of treatment is \(\pi_{s} = 0.5\) for \(s \in \mathbb{S}\), and assignment is independent across locations.
The researcher estimates the effect of the treatment on individuals who are distance \(0.5\) away.
In this example, \(n \var(\hat{\tau}) \to^{p} 2\) where \(n\) is the number of individuals and \(\hat{\tau}\) the treatment effect estimator.

For the design-based \citet{conley1999gmm} standard errors, the kernel weight satisfies \(K_{i,j} = k(d(r_{i},r_{j}))\) with \(k(0) = 1\) and \(k(d) = 0\) for \(d>1\).
The choice of \(k\) determines the weight \(k(1) \in [0,1]\) applied to cross-products of distinct individuals in the same town.
Simple derivations show \(n \hat{\sigma}^{2} \to^{p} 1 + k(1)\).
Hence, any kernel other than the rectangular kernel, which sets \(k(1)=1\), yields standard errors that are systematically too small.

However, it is known that the rectangular kernel can lead to negative variance estimates with finite bandwidth \citep[p. 11]{conley1999gmm}.
Even when the variance estimate is non-negative, design-based inference based on the rectangular kernel can be anti-conservative.

The second example shows such issues for the rectangular kernel.
Locations are in \(\mathbb{R}\).
The location of each ``town'' begins at a different multiple of 10, and each town contains one individual of each of four types and one candidate location of each of two types.
The four individual types differ in their location offset \(1,2,3,4\) relative to the beginning of their town, as well as their treatment effect scale \(a\).
The two candidate location types also differ in their location offsets \(1.5\) and \(3.5\).
Hence, \(r_{i} \in \{1,2,3,4,11,12,13,14,\dots\}\) for \(i\in\mathbb{I}\) and \(\mathbb{S} = \{1.5,3.5,11.5,13.5,\dots\}\).
The probability of treatment is \(\pi_{s} = 0.5\) for \(s \in \mathbb{S}\), and assignment is independent across locations.
Potential outcomes are \(Y_{i}(S) = \sum_{s\in S} \ind\{d(r_{i},s) = 0.5\} a_{i} + \ind\{d(r_{i},s) = 1.5\} a_{i}/10\) where \(a_{i} = 0\) when \(i\) is of type 1 or 4 (\(r_{i}=1,4,11,14,\dots\)) and \(a_{i}=1\) when \(i\) is of type \(3\) (\(r_{i}=3,13,\dots\)).
I vary \(a_{2}\), the treatment effect scale parameter of type \(2\), below.
This definition of potential outcomes ensures that the treatment does not affect individuals beyond a distance of \(1.5\).
The researcher estimates the effect of the treatment on individuals who are distance \(0.5\) away.

\begin{figure}\centering
\includegraphics{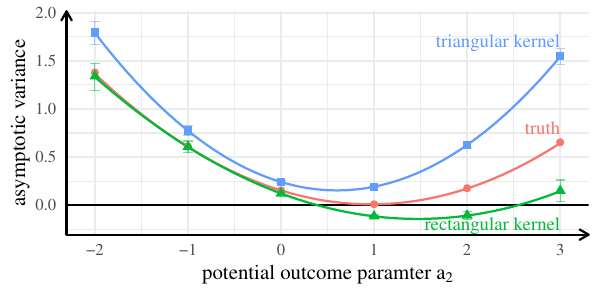}
\caption{\label{fig:app-conley-var}
Asymptotic variance (red solid line) of the estimator and probability limits of Conley variance estimators using rectangular (green) and triangular (blue) kernels in the setup described in Online~Appendix~\ref{app:conley-design}.
The horizontal axis varies the scale of one of the treatment effects.
Red circles show the variance of estimates using 100 ``towns'' (groups of 4 individuals and 2 candidate treatment locations) across 10,000 simulated assignments.
Green triangles and blue squares show the average estimated variances across the same simulations, with bars indicating the interquartile range of estimated variances.}
\end{figure}

In a design-based implementation, the kernel weight satisfies \(K_{i,j} = k(d(r_{i},r_{j}))\) with \(k(0) = 1\) and \(k(d) = 0\) for \(d\geq3\).
This variance estimator enforces zero correlation when \(i\) and \(j\) are design-based independent.
With \(k(1) = k(2) = 1\) as for the rectangular kernel, the variance estimator allows arbitrary correlation whenever the outcomes of observations \(i\) and \(j\) are design-based correlated.
By considering the four possible configurations of \(X_{i}'X_{j}\), their probabilities, and associated residuals, tedious but uninteresting algebra yields
\begin{equation*}
\begin{aligned}
n \hat{\sigma}^{2} \to^{p} \frac{667 - 442 a_{2} + 667 a_{2}^{2} - 371 k(1) + 66 a_{2} k(1) - 371 a_{2}^{2} k(1) - 100 (1 + a_{2})^2 k(2)}{1600}
\end{aligned}
\end{equation*}
Consider two examples:
(i) For the rectangular kernel that sets \(k(1)=k(2)=1\), \(n \hat{\sigma}^{2} \to^{p} \sigma_{\text{rect}}^{2} \equiv (49 - 144 a_{2} + 49 a_{2}^2)/400\).
(ii) For the triangular kernel with bandwidth \(3\) that sets \(k(1) = 2/3\) and \(k(2)=1/3\), \(n \hat{\sigma}^{2} \to^{p} \sigma_{\text{tri}}^{2} \equiv (1159 - 1394 a_{2} + 1159 a_{2}^2)/4800\).

To approximate the scaled asymptotic variance of the treatment effect estimator, I derive the exact finite sample variance of the estimator \(\tilde{\beta}_{1} = \beta_{1} + \frac{\sum_{i\in\mathbb{I}} W_{s(i)} (Y_{i} - \beta_{0} - \beta_{1})}{\sum_{i\in\mathbb{I}} E(W_{s(i)})} - \frac{\sum_{i\in\mathbb{I}} (1-W_{s(i)}) (Y_{i} - \beta_{0})}{\sum_{i\in\mathbb{I}} E(1-W_{s(i)})}\), where \((\beta_{0},\beta_{1})\) are the probability limits of the regression coefficients.
The estimator \(\tilde{\beta}_{1}\) is based on the same approximations used in the proofs of the main text.
Note that this estimator is the simple average over separate estimates of the same form for each town, and each town is design-based independent.
Hence, it suffices to consider the variance of the estimator given a single town, scaled appropriately.
In this way, one can find that \(n \var(\tilde{\beta}_{1}) = (123 - 238 a_{2} + 123 a_{2}^2)/800\).

Figure~\ref{fig:app-conley-var} shows \(\sigma_{\text{rect}}^{2}\), \(\sigma_{\text{tri}}^{2}\), and \(n \var(\tilde{\beta}_{1})\) for different values of \(a_{2}\) in this example.
The rectangular kernel yields an underestimate of the asymptotic variance, \(\sigma_{\text{rect}}^{2} < n \var(\tilde{\beta}_{1})\), except when \(a_{2}=1\) and the two are equal.
Furthermore, the rectangular kernel can yield a negative probability limit, \(\sigma_{\text{rect}}^{2} < 0\) when \(a_{2} \in ((72 - 11 \sqrt{23})/49, (72 + 11 \sqrt{23})/49)\).
The figure also shows that, in this example, other kernels can yield (conservatively) valid inference when the rectangular kernel does not.
For instance, the triangular kernel yields an overestimate of the asymptotic variance, \(\sigma_{\text{tri}}^{2} > n \var(\tilde{\beta}_{1})\) for all \(a_{2}\).

To compare this example to the setting of \citet{Wang2025}, note that I use disaggregate data whereas \citet{Wang2025} average outcomes by treatment location, affecting their inference procedure.
When aggregating individuals at the level of candidate treatment locations, the variance estimator of \citet{Wang2025} includes products of ``residuals'' of type 1 and type 4 individuals.
Because type 1 and type 4 individuals are never affected by the same treatment location, such products are included in neither the design-based variance formulas derived in the present paper nor the variance estimator described in this appendix.
Including these extraneous products through aggregation, even if disaggregated data are available, does not appear generally advisable.
\citet{Wang2025} require an additional treatment effect homophily assumption to ensure the variance estimator is not anti-conservative.
However, the inclusion of extraneous terms may similarly lead to needlessly conservative estimates of the variance.
The variance estimator in the present paper includes only products of outcomes that appear in the variance formulas derived in this paper.
The validity of this variance estimator does not require restricting treatment effect heterogeneity.
The variance estimator is based on worst-case bounds on products of potential outcomes that are not simultaneously observable.
Under restrictions of treatment effect heterogeneity, such as the treatment effect homophily assumption of \citet{Wang2025}, it may be possible to derive tighter estimable bounds on the variance.

\subsection{Overcoverage of a sampling-based implementation}\label{app:conley-sampling}

Another example shows that the original sampling-based implementation of \citet{conley1999gmm} can be arbitrarily too large for design-based inference.
In the example, the ratio of the sampling-based \citet{conley1999gmm} variance estimator to the design-based variance estimator (which itself is slightly conservative due to treatment effect heterogeneity) is approximately \(1 + 2h + 2h^{2}\) in large samples, where \(h\) is a tuning parameter for which \citet{conley1999gmm} requires \(h\to \infty\).
Even in finite samples, the \citet{conley1999gmm} variance estimator can be noticeably too large:
The ratio of variance estimators tends to be around \(3\) for a sample size of \(32\) and around \(11\) for a sample size of \(512\) in simulations for this example (with \(h = (n/2)^{1/8}\) where \citet{conley1999gmm} requires \(h = o(n^{1/6})\) for sample size \(n\)).

Consider a sequence of finite populations indexed by \(k\).
In population \(k\), there is one individual \(i\) located at each point \((r_{i,1},r_{i,2})\) such that \(r_{i,1} \in \{1,\dots,k\}\) and \(r_{i,2} \in \{1,\dots,k\}\), as well as one individual located at each point \((r_{i,1},r_{i,2})\) such that \(r_{i,1} \in \{-1,\dots,-k\}\) and \(r_{i,2} \in \{-1,\dots,-k\}\).
In other words, there is one individual at each point of the first and third quadrant of the two-dimensional bounded integer space excluding zero, \(\{-k,\dots,-1\}^{2} \cup \{1,\dots,k\}^{2}\).
Hence, the total number of individuals is \(n_{k} = 2 k^{2}\).
There are candidate treatment locations at an offset of \((0,-0.1)\) from each individual, and the treatment is known not to affect individuals beyond a distance of \(0.5\).
Hence, each individual is potentially affected by exactly one treatment.
Potential outcomes in the absence of treatment equal \(0\) for all individuals.
Potential outcomes in the presence of treatment at the associated candidate location equal \(1\) for individuals in the first quadrant and \(-1\) for individuals in the third quadrant.
Treatment is assigned independently to candidate locations with a constant probability of treatment \(p=1/2\).

Suppose the researcher is interested in the effect of the treatment at a distance of \(0.1\) and estimates the effect by taking the difference in means between ``treated'' and ``control'' individuals:
Here, treated individuals have a treated location at distance \(0.1\), and control individuals have no treated location at distance \(0.1\).

The setting described above effectively corresponds to a standard randomized experiment with individual-level treatment.
While there is spatial correlation in treatment effects (through treated potential outcomes), independent assignment of the treatment implies that the correct design-based variance need not specifically take this spatial correlation into account.
If treatment were assigned randomly to a fixed number of candidate locations (rather than independently), the exact finite sample variance of the estimator would be \(\frac{1}{2k^{2}-1}\) by \citet[Theorem~6.2]{imbens2015causal}.
The common design-based estimator of this variance is
\begin{equation*}
\hat{V}_{\text{db},k} = \frac{\frac{1}{n_{k,0}} \sum_{i=1}^{n_{k}} (1-W_{s(i)})(Y_{i} - \bar{Y}_{0})^{2}}{n_{k,0}} + \frac{\frac{1}{n_{k,1}} \sum_{i=1}^{n_{k}} W_{s(i)}(Y_{i} - \bar{Y}_{1})^{2}}{n_{k,1}}
\end{equation*}
where \(W_{s(i)}\) is the treatment indicator for the candidate location within \(0.1\) of \(i\), \(n_{k,1} \equiv \sum_{i=1}^{n_{k}} W_{s(i)}\) and \(n_{k,0} \equiv n_{k} - n_{k,1}\) are the number of treated and control in the sample, and \(\bar{Y}_{0} \equiv \sum_{i=1}^{n_{k}} (1-W_{s(i)})Y_{i}/n_{k,0}\) and \(\bar{Y}_{1} \equiv \sum_{i=1}^{n_{k}} W_{s(i)} Y_{i}/n_{k,1}\) are the mean outcomes by treatment group in the sample.
Then \(E(\hat{V}_{\text{db},k}) \approx \frac{1}{n_{k,1}} = \frac{1}{k^{2}}\) because the first term is identically zero and the numerator of the second term is close to \(1\) as \(\bar{Y}_{1} \approx 0\) and \(Y_{i}^{2} = 1\) with probability \(1\) when \(W_{s(i)} = 1\).

\citet{conley1999gmm} proposes an estimator of the sampling variance that, when applied to the setting above, takes the form
\begin{equation*}
\hat{V}_{\text{Conley},k} \equiv \frac{1}{n_{k}} \Bigl(\frac{1}{n_{k}} \sum_{i=1}^{n_{k}} X_{i}'X_{i}\Bigr)^{-1} \Sigma_{k} \Bigl(\frac{1}{n_{k}} \sum_{i=1}^{n_{k}} X_{i}'X_{i}\Bigr)^{-1} \quad \text{with} \quad \Sigma_{k} \equiv \frac{1}{n_{k}} \sum_{i=1}^{n_{k}} \sum_{j=1}^{n_{k}} K_{i,j}(h_{k}) X_{i}'X_{j} e_{i} e_{j},
\end{equation*}
where \(X_{i} = [1, W_{s(i)}]\) is the row vector with second element \(W_{s(i)}\) equaling the treatment indicator of the unique candidate treatment location at distance \(0.1\) of individual \(i\), and \(e_{i}\) is the residual of observation \(i\) in the regression of the observed outcome \(Y_{i}\) on the regressors \(X_{i}\).
For simplicity of the analytical derivations, I study weights \(K_{i,j}\) of the form
\begin{equation*}
K_{i,j}(h_{k}) = \ind\{\abs{r_{i,1}-r_{j,1}} \leq h_{k} \text{ and } \abs{r_{i,2}-r_{j,2}} \leq h_{k}\}
\end{equation*}
which equal \(1\) if \(i\) and \(j\) are within a square of side length \(h_{k}\) and equal \(0\) otherwise, where \(h_{k}\) is a bandwidth satisfying \(h_{k} \to \infty\) and \(h_{k} = o(k^{1/3})\) to satisfy the conditions of \citet{conley1999gmm}.
Similar derivations to the below are possible for the coordinate-wise Bartlett-window weights of \citet[Section~3.1.1]{conley1999gmm}, yielding identical conclusions.

For \((\frac{1}{n_{k}} \sum_{i=1}^{n_{k}} X_{i}'X_{i})^{-1}\), note that
\begin{equation*}
E\Bigl(\frac{1}{n_{k}} \sum_{i=1}^{n_{k}} X_{i}'X_{i}\Bigr) =
\begin{bmatrix}
1 & p \\
p & p
\end{bmatrix}
,
\qquad
E\Bigl(\frac{1}{n_{k}} \sum_{i=1}^{n_{k}} X_{i}'X_{i}\Bigr)^{-1} =
\begin{bmatrix}
\frac{1}{1-p} & \frac{-1}{1-p} \\
\frac{-1}{1-p} & \frac{1}{p(1-p)}
\end{bmatrix}
.
\end{equation*}

For \(\Sigma_{k}\), use that, in this example, \(e_{i} \to Y_{i}\), which equals zero except when \(i\) is treated, in which case it equals \(\pm 1\) depending on the quadrant \(i\) is located in.
Hence,
\begin{equation*}
E(\Sigma_{k}) \approx \Bigl(p + p^{2} \underbrace{\frac{1}{n_{k}} \sum_{i=1}^{n_{k}} \sum_{j \neq i} K_{i,j}(h_{k}) \frac{r_{i,1}r_{j,1}}{\abs{r_{i,1}r_{j,1}}}}_{\equiv \tilde{K}(h_{k})} \Bigr) \begin{bmatrix} 1 & 1 \\ 1 & 1 \end{bmatrix}.
\end{equation*}
The initial \(p\) is due to the \(i=j\) terms of the double summation using that \(Y_{i}=0\) unless \(W_{s(i)} = 1\) which happens with probability \(p\), in which case \(Y_{i}^{2} = 1\).
For \(i \neq j\), the sign of \(Y_{i}Y_{j}\) when both \(i\) and \(j\) are treated (which occurs with probability \(p^{2}\)) depends on whether \(i\) and \(j\) are in the same or different quadrant, which the ratio of coordinate products, \(\tilde{K}(h_{k})\), determines.
Hence,
\begin{equation*}
E\Bigl(\frac{1}{n_{k}} \sum_{i=1}^{n_{k}} X_{i}'X_{i}\Bigr)^{-1} E(\Sigma_{k}) E\Bigl(\frac{1}{n_{k}} \sum_{i=1}^{n_{k}} X_{i}'X_{i}\Bigr)^{-1} 
\approx
(p + p^{2} \tilde{K}(h_{k}))
\begin{bmatrix}
\frac{1}{1-p} & \frac{-1}{1-p} \\
\frac{-1}{1-p} & \frac{1}{p(1-p)}
\end{bmatrix}
\begin{bmatrix} 1 & 1 \\ 1 & 1 \end{bmatrix}
\begin{bmatrix}
\frac{1}{1-p} & \frac{-1}{1-p} \\
\frac{-1}{1-p} & \frac{1}{p(1-p)}
\end{bmatrix}
\end{equation*}
which has \((2,2)\) element \(\frac{1}{p} + \tilde{K}(h_{k})\) such that \(E(\hat{V}_{\text{Conley},k})\rvert_{(2,2)} \approx \frac{1}{k^{2}} + \frac{\tilde{K}(h_{k})}{2k^{2}}\).

Hence,
\begin{equation*}
\frac{E(\hat{V}_{\text{Conley},k})\rvert_{(2,2)}}{E(V_{\text{db},k})} \approx 1 + \tilde{K}(h_{k})/2.
\end{equation*}
Tedious but uninteresting counting and application of the geometric series show that
\begin{equation*}
\tilde{K} = \frac{3}{4} \frac{\floor{h_{k}}^2}{k^{2}} + \frac{5}{2} \frac{\floor{h_{k}}^3}{k^{2}} + \frac{3}{4} \frac{\floor{h_{k}}^4}{k^{2}} - 2 \frac{\floor{h_{k}}}{k} - 6 \frac{\floor{h_{k}}^2}{k} - 4 \frac{\floor{h_{k}}^3}{k} + 4 \floor{h_{k}} + 4 \floor{h_{k}}^2.
\end{equation*}
when \(k \geq 2 \floor{h_{k}} + 1\).
Recall that \(h_{k} \to \infty\) and \(h_{k} = o(k^{1/3})\) such that in large samples the final term, \(4 \floor{h_{k}}^2\), dominates.
Hence, \(\frac{E(\hat{V}_{\text{Conley},k})\rvert_{(2,2)}}{E(V_{\text{db},k})} \to \infty\).

\section{Setup and training of neural networks for finding counterfactual locations}

The proposed implementation of inverse probability weighting estimators for observational data relies on estimates of treatment probabilities at any location as a function of the spatial distribution of characteristics in the neighborhood of the location.
The probability of treatment across space resembles the spatial distribution of treatment locations \(\S_{j} \sim G(Z_{j})\), where \(Z_{j}\) are the characteristics of region \(j\), potentially relative locations of all individuals in the region as well as moments of their covariates.
One could then use the estimated \(\hat{G}\) to inform the treatment probabilities at each point in the region as inputs in the estimators proposed in this paper.

In practice, it is typically sufficient to find a finite number of candidate treatment locations that offer a plausible counterfactual to the realized treatment locations.
With a continuous distribution across space, a simple approximation of the assignment process, such as independent assignment, can lead to unrealistic assignments that are considered in computing standard errors.
More complex assignment processes for continuous distributions may instead be analytically intractable.
In addition, computationally, it is often impractical to use a continuous distribution \(G\) because the weight of individual \(i\) when estimating effects at distance \(d\) would depend on the integral of the noisy \(\hat{G}\) along a ring with radius \(d\) around her location, \(r_{i}\), for each of the typically many individuals \(i \in \mathbb{I}\).
Instead, I recommend finding a finite number of candidate locations.
The average across these finitely many candidate locations approximates the strategy based on the complete distribution \(G\), setting \(\hat{G}\) to exactly zero for many of the implausible locations.

I propose taking draws \(\S_{j} \sim G(Z_{j})\) to obtain candidate treatment locations, where \(G(Z_{j})\) is estimated implicitly.
Perhaps surprisingly, recent machine learning methods achieve good results at this task, despite the difficulty of estimating \(G\) itself.
Specifically, I recommend a formulation similar to generative adversarial networks \citep{Goodfellow2014gan}; see \citet{liang2018well} and \citet{singh2018nonparametric} on the relationship between generative adversarial networks and density estimation.
Most closely related to this paper, \citet{Athey2024} use generative adversarial networks to draw artificial observations from the distribution that generated the (real) sample, for use in Monte Carlo simulations.

Generative adversarial methods for drawing \(\S_{j} \sim G(Z_{j})\) are based on iteration between two steps.
First, a generator generates draws \(\tilde{\S}_{j} \sim \tilde{G}(Z_{j})\), where \(\tilde{G}\) is an implicit estimate of the density maintained by the generator in the current iteration.
Second, a discriminator receives as input either counterfactual locations proposed by the generator, \(\tilde{\S}_{j} \mid Z_{j}\), or real treatment locations, \(\S_{j} \mid Z_{j}\), and guesses whether its input is real.
Both the generator and the discriminator are highly flexible models (typically neural networks) designed for their given tasks.
The discriminator is trained by taking (stochastic) gradient descent steps in the direction that improves discrimination between real and counterfactual locations.
The generator is trained by taking (stochastic) gradient descent steps in the direction that leads to fooling the discriminator into classifying counterfactual locations as real.

Effectively, the output of such models is a set of counterfactual candidate treatment locations \(\tilde{\S}_{j} \mid Z_{j}\) that are indistinguishable (to the discriminator) from real treatment locations \(\S_{j} \mid Z_{j}\).
With a sufficiently flexible discriminator, the process is similar to matching.\footnote{Standard matching methods, however, are unlikely to perform well due to high-dimensional covariates that describe spatial data, such as relative spatial locations between many individuals as well as their characteristics.}
If a proposed candidate location \(\tilde{\S}_{j}\) is noticeably different from all real treatment locations \(\S\), a flexible discriminator will learn to reject \(\tilde{\S}_{j}\).
In contrast, synthetic control-type methods \citep[cf.][]{abadie2010synthetic} would average multiple candidate locations, for instance, \(\tilde{\S}_{a}\) and \(\tilde{\S}_{b}\), to create a synthetic counterfactual for a real treatment location \(\S_{j}\).
If \(\tilde{\S}_{a}\) and \(\tilde{\S}_{b}\) individually differ from all real treatment locations \(\S\), the discriminator will reject them despite their average resembling \(\S_{j}\).

Intuitively, the goal is to find ``false positives:''
Occasions when the discriminator fails to reject a counterfactual location suggested by the generator.
Discriminator networks do not necessarily make binary predictions but may give a continuous activation score that indicates how likely a location is to be real.
In practice, I recommend matching on the activation score, rather than taking all locations with high activation scores because some \emph{real} treatment locations may have low activation scores.
Matching on the activation score helps find suitable counterfactual locations resembling each real location.
Such locations are likely to be decent matches for the real treatment locations because they must share features of realized locations to achieve these comparable activation scores.

I discuss how to tune generic machine learning methods to find suitable candidate treatment locations in social science applications.
I recommend four high-level implementation choices in adapting these methods.
First, the discretization of geographic space into a fine grid for tractability.
Second, \emph{convolutional} neural networks capture the idea that spatial neighborhoods matter in a parsimonious way.
Third, incorporating the adversarial task of the discriminator into a classification task for the generator substantially simplifies training. 
Fourth, data augmentation (rotation, mirroring, shifting) for settings where absolute locations and orientation are irrelevant.

\paragraph*{Discretization}
To tractably summarize the relative spatial locations of individuals and treatment locations, I recommend discretizing geographic space into a fine grid.
Discretization provides an approximation that is particularly tractable for the convolutional neural networks recommended below.
In principle, future improvements to, for instance, Capsule Neural Networks \citep{hinton2011transforming} or other novel methods, may replace convolutional neural networks as the preferred architecture and eliminate the need for discretization.

For each grid cell, one can include a count of individuals with residence in the cell, potentially separately for individuals with different values of covariates, as well as average covariate values of the individuals in the cell or other moments of their covariates.
Based on the architecture of convolutional neural networks, suggested below, it is typically not necessary to also pre-compute covariates describing the neighborhood of each cell.
The convolutional neural network can compute such neighborhood averages if they help predict the outcome (here, whether a location is likely to be treated).
If the grid is very fine, discretization retains almost all meaningful information about relative locations.
For instance, in the application of this paper, each grid cell has size \(0.025 \mathrm{mi} \times 0.025 \mathrm{mi}\) (approximately \(40 \mathrm{m} \times 40 \mathrm{m}\)).
The discretized grid creates a three-dimensional array:
The first two dimensions determine spatial location, and the third dimension enumerates the different covariates that are summarized.
Rather than taking the spatial dimensions to be entire regions, I recommend using smaller (square) areas within a region such that the probability of treatment in the approximate center of the area is plausibly only affected by individuals and covariates within the area.

\paragraph*{Convolutional neural networks}
Convolutional neural networks have been particularly successful at image recognition \citep{Krizhevsky2012convolution}.
In image recognition, the input is a 3D array: a 2D grid of pixels, with a third dimension given by multiple RGB color channels.
For spatial treatments, the input also is a 3D array: the 2D spatial grid, with a third dimension given by the covariates as described above.

Convolutional steps in neural networks generally retain the shape of the 2D grid, but the value of each neuron is a function of the covariates (or neurons) of the previous step, not just at the same grid cell, but also the covariates (or neurons) at neighboring grid cells.
Importantly, convolutional layers average the neighborhoods of grid cells at any point in the grid with the same weights.
Reusing parameters across points in space makes convolutional layers substantially more parsimonious than fully connected layers, and allows the neural network to capture neighborhood patterns appearing in different parts of a region in a unified way.

In particular, I recommend using at least two convolutions with reasonably large spatial reach.
Consider the application in this paper, where grocery stores are spatial treatments and restaurants are outcome units with foot traffic as the outcome variable.
The first convolution allows each grid cell to see the covariates of grid cells around it.
In the application of this paper, the output of the first convolution for a particular grid cell may be:
``There are 3 grocery stores nearby, 4 competing restaurants very close, and 10 restaurants within walking distance.''
The second convolution then uses the information on such neighborhoods to determine whether treatment is likely in a grid cell:
``If there are many grid cells nearby (in all directions) containing restaurants or grocery stores facing much competition, this location is probably in the center of a shopping area and reasonably likely to contain another grocery store.''
Intuitively, the first convolution may measure what is important to the restaurants, while the second convolution translates how that is important for the treatment location choice.

\paragraph*{Adversarial Classification}
Generative adversarial networks \citep{Goodfellow2014gan} are oftentimes difficult to train despite recent advances such as networks with Wasserstein-type criterion function \citep{arjovsky2017principled,arjovsky2017wasserstein}.
The difficulty arises because the training of the generator and discriminator networks needs to be sufficiently balanced such that both improve.
For instance, if the discriminator early on becomes (close to) perfect at discriminating between the proposals of the generator and the real treatment locations, the gradient for the generator is relatively flat (little improvement in any direction), and hence the generator fails to improve.
Similarly, if the discriminator is insufficiently flexible, even poor proposals by the generator may pass, such that the false positives are not necessarily similar to the real treatment locations.

In contrast, convolutional neural networks for image classification are much easier to train, and, in this case, can be adapted to the same task.
Hence, I recommend setting up the problem of finding candidate treatment locations as a classification task.
Specifically, the convolutional neural network takes a 3D input array and ``classifies'' it into, say, 101 categories, where categories correspond either to the \(10 \times 10 = 100\) grid cells in the center of the input area or an additional ``no missing treatment location'' category.
The distinction from other generation tasks is that here the set of possible outputs is relatively small, for instance, the 101 categories described above.
In contrast, in image generation, there are infinitely many possible images that could be generated.

To retain the adversarial nature of the task, I propose simultaneously training the classification on three sets of data and adding a final fully connected layer.
The three sets of data are as follows:
First, areas with at least one real treatment location, but with one treatment location removed.
The correct classification of such input data is into the category corresponding to the grid cell where the treatment location was removed.
Second, areas with at least one real treatment location, but without any treatment location removed.
The correct classification of such input data is into the no-missing-treatment-location category.
Third, areas without treatment locations.
These areas are also correctly classified as not missing any treatment location.
The output of the convolutional layers is a prediction for each grid cell of whether it is missing a treatment location.
A final fully connected layer combines the location-specific predictions into the categories mentioned above:
one category for each of the central grid cells, plus one category for no-missing-treatment-location.

This neural network architecture balances two tasks:
a generative task of picking the correct location if a treatment location is missing, predominantly performed by the convolutional layers; and a discriminatory task of deciding whether a treatment location is missing at all, predominantly performed by the final fully connected layer.
This structure retains the attractive interpretation of generative adversarial networks but is substantially easier to train.
It also resembles denoising autoencoders \citep[cf.][]{vincent2008extracting}, where the removal of a real treatment location represents noise added to the input, with the autoencoder trained to remove the noise, here meaning to add the removed real treatment location.
The idea of using the second and third sets of training examples without missing treatment location has precedents in the literature on adversarial examples and adversarial training \citep[see][]{biggio2013evasion,szegedy2013intriguing}.

The setup as an adversarial task, as well as the prediction of categories, additionally is beneficial because it generates draws near the local \emph{modes} rather than the \emph{mean} of the treatment location distribution \citep[cf.][]{goodfellow2016nips,lotter2016unsupervised}.
The importance of sampling from the mode rather than the mean of the location distribution becomes clear in a simple example.
Suppose all areas contain three possible locations in one-dimensional space: 1, 2, and 3.
For instance, 2 may be the city center, while 1 and 3 are suburbs on either side of the city.
In the data, if a region is treated, treatment always occurs in the suburbs; at either location 1 or location 3, each with probability \(0.5\).
However, estimating the likely location of the treatment with the familiar mean squared error loss function will estimate the mean of the treatment location distribution, predicting treatment at location 2.
In contrast, the adversarial loss function, as well as loss functions used for classification tasks, are minimized by predicting either 1 or 3 because these categories are most likely to correspond to the correct location.\footnote{In general adversarial networks, one input to the network is white noise.
This noise effectively chooses between the different local modes of the distribution.
In the setup as a classification task proposed here, data augmentation, as described below, plays a similar role.}
In contrast, location 2 is rejected as a candidate treatment location because treatment is never observed at such a location.

\paragraph*{Data Augmentation}
Data augmentation serves two closely related purposes.
First, rotating, mirroring, and shifting of input areas produce additional, albeit dependent, observations but preserve all relative distances.
Additional observations are helpful because training neural networks requires many training samples.
Second, these transformations effectively regularize the parameters of the estimated model.
One can choose transformations that induce equivariance to rotation, mirroring, and shifts as appropriate for the particular setting.
For instance, in many applications in the social sciences, North-South and East-West orientation are irrelevant on a small scale; only the relative distances matter.\footnote{Applications in environmental economics are notable exceptions if, for instance, wind direction is relevant.
In such cases, rotation hinders the ability of the model to capture patterns due to, for instance, wind consistently blowing from one direction, and may require the inclusion of wind direction in estimation.
The choice of appropriate data augmentation is, therefore, application-specific.}
Suppose there is an individual who visits a business to the North of her home because it is on the way to work in the North.
If the whole space was rotated counterclockwise by 90 degrees, the individual would equally visit the same business now to the West as it is still on the way to work, now also rotated to be to the West of her home.
In image classification, the use of data augmentation is common and associated with a reduction in overfitting and greater generalizability of the learned models \citep{yaeger1996effective,simard2003best,Krizhevsky2012convolution}.

Shifting the entire grid has two additional desirable effects:
First, imposing a continuous shift of the grid relative to covariates renders the exact discretization less relevant.
The \emph{average} (across draws from the shift distribution) distance in grid cells between two observations becomes directly proportional to their actual distance.
Second, the location of an observation \emph{within} a grid cell is no longer fixed.
Shifting within-cell location is attractive because the classification is not informative of whether the candidate treatment location is at the center or towards the edge of a grid cell.
With a continuous shift of the observations, the center of the grid cell points to different absolute locations depending on the shift.
One can then average over several realizations of the shift to reduce the influence of the particular translation of grid cells to absolute locations.

\section{Implementation details for the empirical application}

\subsection{Data Processing}\label{app:data-processing}

I use the July 2021 release of \citet{safegraph2021}'s data for the year 2020.
In this release of the data, SafeGraph applied its then-current algorithm to the data it collected in 2020 and updated its data sets attributing smartphone pings to businesses.
In this paper, I focus on businesses in the San Francisco Bay Area, specifically in the Peninsula and South Bay between South San Francisco and Sunnyvale, see Figure~2 in the main text.
To create the initial sample of all possibly relevant businesses for which SafeGraph has recorded data, I keep all businesses that either lie within six miles of several points throughout the Bay Area or have a SafeGraph-determined ZIP code falling within a list of relevant ZIP codes, see Table~\ref{tab:app-sg-geocodes-included}.

\begin{table}\centering
\caption{\label{tab:app-sg-geocodes-included}The initial sample of all possibly relevant businesses consists of all businesses in the SafeGraph ``point of interest'' data with location within six miles of one of the five cities or with a zip code given in the table.}
\begin{tabular}{l @{\hskip 0.5in} c @{\hskip 0.5in} c c}
\hline
city & latitude & longitude \\
\hline
South San Francisco & 37.653540     & -122.416866 \\
Burlingame          & 37.584103     & -122.366083 \\
Belmont             & 37.516493     & -122.294191 \\
Menlo Park          & 37.451967     & -122.177993 \\
Mountain View       & 37.389389     & -122.083210 \\
\hline
ZIP codes: \\
\hline
\multicolumn{4}{l}{\hspace{1em} 94002, 94005, 94010, 94014, 94015, 94016, 94019,} \\
\multicolumn{4}{l}{\hspace{1em} 94020, 94022, 94024, 94025, 94027, 94028,} \\
\multicolumn{4}{l}{\hspace{1em} 94030, 94032, 94035, 94037, 94040, 94041, 94042, 94043, 94044,} \\
\multicolumn{4}{l}{\hspace{1em} 94061, 94062, 94063, 94064, 94065, 94066, 94070,} \\
\multicolumn{4}{l}{\hspace{1em} 94080, 94083, 94085, 94086, 94087, 94089,} \\
\multicolumn{4}{l}{\hspace{1em} 94101, 94102, 94104, 94105, 94110, 94112, 94114, 94117,} \\
\multicolumn{4}{l}{\hspace{1em} 94121, 94124, 94127, 94128, 94129,} \\
\multicolumn{4}{l}{\hspace{1em} 94130, 94131, 94132, 94133, 94134, 94169, 94192,} \\
\multicolumn{4}{l}{\hspace{1em} 94301, 94303, 94304, 94305, 94306, 94309,} \\
\multicolumn{4}{l}{\hspace{1em} 94401, 94402, 94403, 94404, 94497, 94530, 94538, 94555, 94603,} \\
\multicolumn{4}{l}{\hspace{1em} 95014, 95015, 95051, 95054, 95101, 95112} \\
\hline
\end{tabular}
\end{table}

To define the units of interest and ensure high-quality data for this application, I take three additional steps in processing the data.
First, I determine the grocery and convenience stores that I consider ``treatments'' in this paper.
Second, I manually set the location of each of these treatments to correspond to the main entrance of the store.
Third, I check and de-duplicate restaurant location data to restrict the sample to real restaurants that were likely to be open in early 2020.

Based on SafeGraph's ``point of interest'' data, I find 167 unique grocery and convenience store (treatment) locations that were open in 2020 in the interior of the study area.
Starting from the sample defined above, I define the possible businesses of interest as those within 3 miles of Burlingame, 5 miles of Belmont, 5.5 miles of Menlo Park, or 2.95 miles of Mountain View, with the city locations as in Table~\ref{tab:app-sg-geocodes-included}.
Focusing on grocery stores in the interior of the study area guarantees that the full sample includes data on all businesses that are within different distances of interest from the grocery stores.
To find locations consumers typically visit to purchase groceries, I start with all businesses with 4-digit NAICS code 4451 (grocery and convenience stores) assigned by SafeGraph, and then add all Costco, Target, and Walmart stores (which SafeGraph classifies as general merchandise stores, 4523), for a total of 313 stores.
Of these stores, I exclude 28 stores that SafeGraph determines to have closed permanently before the COVID-19 pandemic (in or before February 2020; there were no further grocery store closures until July as recorded by SafeGraph), as well as 1 store that SafeGraph determines to have opened only in November 2020.
For the remaining 284 stores, I verify manually that they fit my definition of grocery or convenience store.
I exclude 100 stores, primarily convenience stores that are part of gas stations, delis, and food producers and importers/exporters that are incorrectly classified as grocery stores by SafeGraph's algorithm.
I confirm, based on newspaper articles, Yelp entries, and Google Street View imagery, that another 17 grocery stores were either not open in 2020 (closed before or opened after) or were duplicate entries in the data set.
Overall, I consider 167 treatment locations; 139 locations are labeled as grocery (or general merchandise) stores by SafeGraph, with the remaining 28 labeled as convenience stores by SafeGraph.

For the 167 grocery and convenience stores in the sample, I manually determine the latitude and longitude of the main entrance, which serves two related purposes.
First, the main entrance (and exit) is the relevant location to measure distances to or from for trip sequencing:
If a consumer considers visiting a coffee shop before or after a grocery store, the additional distance she has to travel is based on the front door of the grocery store, not a location in the interior.
Second, placing the location of grocery stores at their main entrances typically reduces the differences between taking straight-line distance (as in this paper) and walking distance (likely the economically relevant distance metric) between grocery stores and restaurants.
When the grocery store location is instead placed in the interior of the store, restaurants that are \emph{behind} the grocery store can appear closer than restaurants that are next door.
Hence, placing the location of the grocery store at its front entrance improves the interpretability of estimates by distance.
The latitude and longitude given in the SafeGraph data instead reflect ``the general center of the business,''\footnote{SafeGraph documentation, \url{https://docs.safegraph.com/docs/core-places\#section-latitude-longitude} accessed on July 29, 2021.} typically in the interior of the store.
I use Google Maps satellite as well as Street View imagery to locate the main entrances of all grocery stores.
For about three-quarters of the grocery and convenience stores, the difference in locations is less than 20 meters.
The largest differences in locations (of around 70 meters) occur for a handful of particularly large Costco, Safeway, Target, and Walmart stores.

I audit the data on restaurant (outcome unit) locations in three steps.
First, I de-duplicate observations by checking the similarity of business names between any two businesses with locations within 50 meters of each other according to SafeGraph data.
To detect duplicates based on name similarity, I focus my attention on businesses with high relative Levenshtein distance.
This distance measures the minimum number of character edits needed to make the names of the two businesses equal, relative to the length of the longer business name.
Most duplicates I detect are clear typos in the name of one of the observations, and some are abbreviations of business names that I verify to indeed describe the same business using Google Maps and Street View data.
Second, I audit the SafeGraph location data by comparing the latitude and longitude in the SafeGraph ``point of interest'' data to the latitude and longitude obtained by searching for the business name and street address (also from the ``point of interest'' data) on Google Maps.
This analysis confirms the high quality of the SafeGraph location data.
Randomly inspecting the locations of a few dozen restaurants in more detail, I find that neither the SafeGraph nor the Google Maps locations are systematically closer to the entrance of the restaurants.
Given the much smaller size (area) of restaurants compared to grocery stores, as well as the much greater number of restaurants, I do not manually record the latitudes and longitudes of their entrances.
Third, I focus on businesses that were reliably assigned visits by SafeGraph.
I restrict the non-grocery store sample to businesses for which SafeGraph reported at least 7 visits in each of the four weeks starting in January 2020.
This step excludes businesses that were not open at the time, not properly assigned visits by SafeGraph's algorithm, or are too small to reliably measure visits for, but retains 95-97.5\% of all \emph{visits} (depending on the week) in the SafeGraph data.
Importantly, I take each of the three steps without knowledge of which businesses are, in the later analysis, considered treated or control.

\subsection{Convolutional Neural Network}\label{app:CNN}

I use a convolutional neural network (CNN) to identify plausible counterfactual locations.
First, I specify the input for the training of the CNN.
Second, I describe the architecture of the CNN.
Third, I use the trained CNN to predict many plausible counterfactual locations, followed by additional matching steps, to select the final counterfactual locations used in the analysis.

I project the latitude and longitude of all businesses into two-dimensional Cartesian space using the NAD83 (2011) projection, EPSG:6419 California zone 3.
This projection gives the location in meters East and North relative to a point near the San Francisco Bay Area.
In applications where the data come from different regions, the researcher should choose the appropriate projection for each region to ensure the accuracy of relative distances within regions.

The CNN learns to predict treatment locations in the areas around prespecified locations: real grocery store locations and semi-randomly chosen locations.
The semi-randomly chosen locations, together with the real grocery store locations, are meant to cover the areas in which counterfactual locations could plausibly occur.
I start with the locations of all businesses for which the nearest grocery store is between 0.2 miles and 2 miles away.
The areas around businesses even closer to a grocery store are already included in the consideration set by including the area of that grocery store.
Next, I jitter these locations by adding independent shocks from a normal distribution with mean \(\pm\)0.0004 and standard deviation 0.0001 to their latitudes and longitudes, where the sign of the mean is independently drawn to be \(+1\) or \(-1\) for each location and coordinate.
This step ensures that the \emph{center} of each area does not fall exactly onto a business because real grocery store locations never exactly coincide with the locations of other businesses.
Finally, to avoid including an area multiple times, I detect all pairs of jittered locations that are within 100 meters of one another.
I drop locations that are listed ``first'' (in the arbitrary order based on the row numbers of the businesses the location is based on) in any such pair.
The areas around both the resulting \(1,900\) semi-random locations and the 167 real grocery store locations are used as input to the CNN.

The CNN predictions are based on observable characteristics describing small 2D grid cells around the prespecified locations.
Each grid cell covers an area of \(0.025\mathrm{mi} \times 0.025\mathrm{mi}\) (approximately \(40\mathrm{m} \times 40\mathrm{m}\)).
I use the count of businesses by 4-digit NAICS code for the codes given in Table~\ref{tab:app-sg-naics} as observable characteristics of each grid cell.
That is, a cell covering two gasoline stations, one car dealership, and no other businesses, will have ``covariate value'' 2 for the covariate indicating industry group 4471 (gasoline stations) and 3 for the covariate indicating ``any'' industry, with the remaining covariates at 0 because there is no separate covariate for the relatively rare car dealerships (NAICS code 4411, less than 100 in the study area).

\begin{table}\centering
\caption{\label{tab:app-sg-naics}Number of businesses by 4-digit NAICS code that are in the larger neighborhoods forming the input into the convolutional neural network. The number of grocery stores exceeds 167 here because additional grocery stores that are not in the interior of the main study area are included in these larger neighborhoods.}
\begin{tabular}{l l r}
\hline
NAICS code & description & \# unique businesses \\
\hline
7225 & Restaurants and Other Eating Places & 1975 \\
7139 & Other Amusement and Recreation Industries &  606 \\
7121 & Museums, Historical Sites, and Similar Institutions &  409 \\
8131 & Religious Organizations &  324 \\
6111 & Elementary and Secondary Schools &  265 \\
6244 & Child Day Care Services &  264 \\
4451 & Grocery Stores &  244 \\
4471 & Gasoline Stations &  182 \\
\hline
any & -- & 7845 \\
\hline
\end{tabular}
\end{table}

Each input observation to the CNN consists of one of the 2,067 areas described above.
The covariates of the 2D grid are separate ``channels'' constituting a 3D tensor for each such observation.
Each area consists of \(50 \times 50\) grid cells.
All coordinates within an area are jointly shifted, rotated, and mirrored randomly using independent uniform distributions for each of the three operations.
The maximum absolute shift is such that the original center is placed within one of the central \(10 \times 10\) grid cells.
Hence, there are at least another 20 grid cells (\(0.5\mathrm{mi}\)) of ``padding'' on all sites of the original center until the edge of the area.

The CNN consists of 4 sequential 2D convolutions and a final linear (fully connected) layer yielding \(10 \times 10 + 1 = 101\) outputs.
I use 2D instance normalization and leaky rectified linear activation for all neurons in the CNN, and replication padding to ensure the output of each convolution has the same spatial dimension as the input.
The first convolution takes the 9 input channels (eight specific industries and one for any industry) and convolves them with a kernel size of 5 (considering the \(5 \times 5\) grid cells centered around a given grid cell) into 18 channels.
This layer can ``smooth'' the input such that the hard borders between grid cells due to discretization become less relevant.
The increase in the number of channels allows the neural network to learn a larger number of nonlinearities.
The second convolution takes the 18 channels of the previous layer and convolves them with a kernel size of 21 with a stride of 2 into 36 channels, such that each grid cell can view grid cells up to 20 cells away in any direction, but skipping every other cell for parsimony.
This layer allows each grid cell to learn about its neighborhood up to even relatively large distances (approximately \(20 \times 0.025\textrm{mi} = 0.5\textrm{mi}\)).
The third convolution takes the 36 channels of the previous layer and convolves them with a kernel size of 5 into 36 channels, again allowing some smoothing across grid cells to counteract the skipping of every other grid cell of the previous layer.
The fourth convolution takes the 36 channels of the previous layer and convolves them with a kernel size of 21 with a stride of 2 into a single channel.
Intuitively, this layer forces a single prediction for each grid cell based on the large neighborhood (up to 20 cells away in any direction).
The final layer linearly combines the \(50 \times 50\) grid cells of the single channel of the previous layer into \(101\) ``categories'' that constitute the predictions of whether and where an additional grocery store may be located.

The \(101\) categories correspond to the central \(10 \times 10\) grid, as well as one category indicating a prediction of no additional grocery store.
I train the CNN on batches consisting of 64 observations (areas).
Half (32) of the observations are areas around a real grocery store, but with that grocery store removed from the input channel count of grocery stores per grid cell.
The random shift and rotation of the input are such that this removed grocery store could have been in any of the central \(10 \times 10\) grid cells.
For these observations, the prediction maximizing the cross-entropy loss is the category corresponding to the cell that the grocery store has been removed from.
All other categories are equal in terms of loss and worse than the correct category, which trains the CNN to identify the mode, rather than the average, location.
A quarter (16) of the observations are areas around real grocery stores with no grocery store removed.
The correct classification of such observations is into the category corresponding to ``no missing grocery store'' instead of the \(10 \times 10\) grid cells.
The last quarter (16) of the observations of each batch are areas around the semi-random prespecified locations.
Their correct classification is also the category corresponding to ``no missing grocery store.''

After training, I evaluate the areas of the prespecified locations for possible grocery store locations according to the CNN.
In this step, I input batches consisting of 32 observations into the trained CNN.
In each batch, 4 observations are areas around real grocery stores: 2 have the grocery store removed from the input, while 2 do not have the grocery store removed.
An additional 28 observations are areas around the semi-random prespecified locations.
The trained neural network calculates predictions for \(5,000\) batches.
Predictions for observations with removed grocery stores allow me to learn the activation scores of real grocery store locations.
The remaining observations yield possible counterfactual locations.

I find good matches for real grocery store locations among the possible counterfactual locations in two steps.
In the first step, I find for each real grocery store location possible counterfactual locations with similar CNN activation.
Specifically, I take each prediction for a removed real grocery store separately (there are multiple such predictions for each real grocery store under different random shifts, rotation, and mirroring), and match in descending order of activation, with replacement, within the possible counterfactual locations (excluding the prediction category for ``no missing grocery store'').
I repeat the same matching process (matching with replacement using the complete set of possible counterfactual locations) using relative activation within neighborhood-observation, corresponding to the cross-entropy loss function.
Taking the union of these matches, I obtain \(19,857\) possible locations that the CNN evaluated as similar to a real grocery store location under at least one shift, rotation, and mirroring.
I drop \(43\) of these locations that are closer to the nearest real grocery store than two-thirds of the minimum distance between any two real grocery stores.
In the second step, I use ``propensity score'' matching to pick the final counterfactual locations among the \(19,814\) remaining locations.
I estimate a propensity score model using the real and possible counterfactual locations as observations in a logistic regression.
There are three sets of regressors:
1) the number of restaurants in each distance bin of width \(0.025\) miles from the location, up to a distance of \(0.2\) miles;
2) the average number of grocery stores near the restaurants in each bin, broken out for each bin into similar bins of distance from the restaurant;
3) the total number of businesses (of any industry) in distance bins of width \(0.25\) miles, up to a distance of 1 mile.
I match, with replacement, each grocery store location to the possible counterfactual location with the closest estimated propensity score.
The final sample consists of 162 counterfactual locations and all 167 real grocery store locations.

For the final sample of real grocery stores and most plausible counterfactual locations, I estimate treatment probabilities to analyze the sample as a quasi-experiment conditional on these locations and treatment probabilities.
The treatment probability estimation uses the same regressors as the propensity scores used for matching.
The inverse probability weighting estimator only uses these probabilities to weight the ``control'' observations (restaurants near counterfactual locations) because the average treatment effect on the treated (ATT) estimator does not require reweighting of the ``treated'' observations (restaurants near real grocery stores).
The primary purpose of estimating the treatment probabilities, rather than re-using the propensity scores, is to balance exposure to grocery stores appropriately between treated and control restaurants.
When estimating the average effect of one marginal grocery store on restaurants at a distance \(d\), the treated and control restaurants at that distance indeed differ on average by one grocery store at distance \(d\) and have similar average exposure to grocery stores at other distances as Figure~3 in the main text illustrates.
By selecting the counterfactual locations from the CNN predictions based on the relative locations of other businesses in the area, these locations and treatment probability weights also balance exposure to other businesses in the neighborhood of restaurants as shown in Figure~4 of the main text.

For the double machine learning estimator, the outcome model is based on a post-LASSO regression at the restaurant-level on the number of grocery stores in each distance bin (same as above) with the distance of interest always included.
To predict the outcome in the absence of the marginal grocery store, I make predictions for each restaurant-grocery store pair with augmented data that removes the grocery store from the distance bin counts.
For double machine learning, both the outcome model and the second step estimate of the treatment probabilities are based on cross-fitting as described in the main text.

\section{Variance in Single Region Settings}

Write the infeasible estimator as:
\begin{equation*}
\begin{aligned}
\tilde{\tau} & = \mu_{t} - \mu_{c}
+ \frac{\sum_{s\in\mathbb{S}} \ind\{\S \ni s\} \sum_{i\in\mathbb{I}} w_{i}(s,d) (\Y_{i}-\mu_{t})}{\sum_{s\in\mathbb{S}} \pi_{s} \sum_{i\in\mathbb{I}} w_{i}(s,d)} 
- \frac{\sum_{s\in\mathbb{S}} \ind\{\S \niton s\} \frac{\pi_{s}}{1-\pi_{s}} \sum_{i\in\mathbb{I}} w_{i}(s,d) (\Y_{i}-\mu_{c})}{\sum_{s\in\mathbb{S}} \pi_{s} \sum_{i\in\mathbb{I}} w_{i}(s,d)} \\
& = \mu_{t} - \mu_{c} 
+ \frac{\sum_{s\in\mathbb{S}} \ind\{\S \ni s\} \sum_{i\in\mathbb{I}} w_{i}(s,d) (\Y_{i}-\mu_{t}) - \sum_{s\in\mathbb{S}} \ind\{\S \niton s\} \frac{\pi_{s}}{1-\pi_{s}} \sum_{i\in\mathbb{I}} w_{i}(s,d) (\Y_{i}-\mu_{c})}{\sum_{s\in\mathbb{S}} \pi_{s} \sum_{i\in\mathbb{I}} w_{i}(s,d)}
\end{aligned}
\end{equation*}
where, for brevity, I suppress the dependence of \(\mu\) on \(d\) throughout.

Define exposure mappings \citep{aronow2017estimating} based on Assumption~3 as follows.
\(\mathbb{M}_{i} \equiv 2^{\{s \in \mathbb{S}: \; d(s,r_{i}) \leq \dnoeffect{}\}}\) is the set of all possible ways in which treatment can be assigned to those locations that possibly affect \(i\).
With slight abuse of notation, denote \(i\)'s potential outcome under exposure \(m \in \mathbb{M}_{i}\) by \(Y_{i}(m)\).
Let the random variable \(\M_{i}^{m}\) be the indicator for whether exposure \(m\) of individual \(i\) is realized.
Then \(\Y_{i} = \sum_{m \in \mathbb{M}_{i}} \M_{i}^{m} Y_{i}(m)\).
Denote the marginal and joint probabilities of exposures by \(\pi_{m}^{i} \equiv \Pr(\M_{i}^{m}=1)\) and \(\pi_{m,m'}^{i,i'} \equiv \Pr(\M_{i}^{m}=1 \text{ and } \M_{i'}^{m'}=1)\).
Let
\begin{equation*}
\T_{s}^{a}
\equiv
\begin{cases}
1 & \text{if } a=t \text{ and } \S \ni s \\
1 & \text{if } a=c \text{ and } \S \niton s \\
0 & \text{otherwise}
\end{cases}
\end{equation*}
be an indicator for the events \(\S \ni s\)~(\(a=t\)) and \(\S\niton s\)~(\(a=c\)).

For the variance of the estimator, note that only the numerator of the ratio in the definition of \(\tilde{\tau}\) is stochastic.
Using the definitions above, rewrite the numerator:
\begin{equation*}
\begin{aligned}
& \sum_{s\in\mathbb{S}} \ind\{\S \ni s\} \sum_{i\in\mathbb{I}} w_{i}(s,d) (\Y_{i}-\mu_{t}) - \sum_{s\in\mathbb{S}} \ind\{\S \niton s\} \frac{\pi_{s}}{1-\pi_{s}} \sum_{i\in\mathbb{I}} w_{i}(s,d) (\Y_{i}-\mu_{c}) \\
= & 
\sum_{s\in\mathbb{S}} \sum_{a\in\{c,t\}} \T_{s}^{a} \Bigl(\ind\{a=t\} \sum_{i\in\mathbb{I}} w_{i}(s,d) (\Y_{i}-\mu_{t}) - \ind\{a=c\} \frac{\pi_{s}}{1-\pi_{s}} \sum_{i\in\mathbb{I}} w_{i}(s,d) (\Y_{i}-\mu_{c})\Bigr) \\
= &
\sum_{i\in\mathbb{I}} \sum_{m \in \mathbb{M}_{i}} \sum_{s\in\mathbb{S}} \sum_{a\in\{c,t\}} \M_{i}^{m} \T_{s}^{a} \Bigl(\ind\{a=t\} w_{i}(s,d) (Y_{i}(m)-\mu_{t}) - \ind\{a=c\} \frac{\pi_{s}}{1-\pi_{s}} w_{i}(s,d) (Y_{i}(m)-\mu_{c})\Bigr) \\
\end{aligned}
\end{equation*}
where, importantly, only \(\M_{i}^{m} \T_{s,a}\) is stochastic.
For ease of notation, define
\begin{equation*}
\begin{aligned}
\tilde{Y}_{i}^{s,a}(m) & \equiv \ind\{a=t\} w_{i}(s,d) (Y_{i}(m)-\mu_{t}) - \ind\{a=c\} \frac{\pi_{s}}{1-\pi_{s}} w_{i}(s,d) (Y_{i}(m)-\mu_{c}) \\
& =
\Bigl(-\frac{\pi_{s}}{1-\pi_{s}}\Bigr)^{\ind\{a=c\}} w_{i}(s,d) (Y_{i}(m) - \mu_{a})
\end{aligned}
\end{equation*}
where, for brevity, I suppress the dependence of \(\tilde{Y}\) on \(d\) throughout.

Then
\begin{equation}\label{eq:single-region-expanded-cov-sums}
\begin{aligned}
& \var\Big(\sum_{i\in\mathbb{I}} \sum_{m \in \mathbb{M}_{i}} \sum_{s\in\mathbb{S}} \sum_{a\in\{c,t\}} \M_{i}^{m} \T_{s}^{a} \tilde{Y}_{i}^{s,a}(m)\Bigr) \\
= &
\sum_{i\in\mathbb{I}} \sum_{m \in \mathbb{M}_{i}} \sum_{s\in\mathbb{S}} \sum_{a\in\{c,t\}}
\sum_{i'\in\mathbb{I}} \sum_{m' \in \mathbb{M}_{i'}} \sum_{s'\in\mathbb{S}} \sum_{a'\in\{c,t\}}
    \cov(\M_{i}^{m}\T_{s}^{a},\M_{i'}^{m'}\T_{s'}^{a'})
    \tilde{Y}_{i}^{s,a}(m) \tilde{Y}_{i'}^{s',a'}(m') \\
= & 
\sum_{i\in\mathbb{I}} \sum_{m\in\mathbb{M}_{i}} \sum_{s\in\mathbb{S}} \sum_{a\in\{c,t\}} \var(\M_{i}^{m}\T_{s}^{a}) \tilde{Y}_{i}^{s,a}(m)^{2} \\
&
+ \sum_{i\in\mathbb{I}} \sum_{m\in\mathbb{M}_{i}} \sum_{s\in\mathbb{S}} \sum_{a\in\{c,t\}} \sum_{s'\in\mathbb{S}} \sum_{a'\in\{c,t\}} \ind\{s \neq s' \text{ or } a \neq a'\} \cov(\M_{i}^{m}\T_{s}^{a},\M_{i}^{m}\T_{s'}^{a'}) \tilde{Y}_{i}^{s,a}(m) \tilde{Y}_{i}^{s',a'}(m) \\
&
+ \sum_{i\in\mathbb{I}} \sum_{m\in\mathbb{M}_{i}} \sum_{s\in\mathbb{S}} \sum_{a\in\{c,t\}} \sum_{m'\in\mathbb{M}_{i}} \sum_{s'\in\mathbb{S}} \sum_{a'\in\{c,t\}} \ind\{m \neq m'\} \cov(\M_{i}^{m}\T_{s}^{a},\M_{i}^{m'}\T_{s'}^{a'}) \tilde{Y}_{i}^{s,a}(m) \tilde{Y}_{i}^{s',a'}(m') \\
&
+ \sum_{i\in\mathbb{I}} \sum_{m\in\mathbb{M}_{i}} \sum_{s\in\mathbb{S}} \sum_{a\in\{c,t\}} \sum_{i'\in\mathbb{I}} \sum_{m'\in\mathbb{M}_{i}} \sum_{s'\in\mathbb{S}} \sum_{a'\in\{c,t\}} \ind\{i \neq i'\} \cov(\M_{i}^{m}\T_{s}^{a},\M_{i'}^{m'}\T_{s'}^{a'}) \tilde{Y}_{i}^{s,a}(m) \tilde{Y}_{i'}^{s',a'}(m')
.
\end{aligned}
\end{equation}

Define
\begin{equation*}
\pi_{i,s}^{m,a} \equiv \Pr(\M_{i}^{m}\T_{s}^{a}=1) \qquad \pi_{i,s,i',s'}^{m,a,m',a'} \equiv \Pr(\M^{i}_{m}\T_{s}^{a}=1 \text{ and } \M^{i'}_{m'}\T_{s'}^{a'}=1)
\end{equation*}
such that \(\cov(\M_{i}^{m}\T_{s}^{a},\M_{i'}^{m'}\T_{s'}^{a'}) = \pi_{i,s,i',s'}^{m,a,m',a'} - \pi_{i,s}^{m,a} \pi_{i',s'}^{m',a'}\) and \(\var(\M_{i}^{m}\T_{s}^{a}) = \pi_{i,s}^{m,a}(1-\pi_{i,s}^{m,a})\).

Initially, consider the first two (lines of) summations in the final expression in Equation~(\ref{eq:single-region-expanded-cov-sums}), which each have a single summation over \(i\) and \(m\).
Substituting the (co-) variances and then the definitions of \(\tilde{Y}_{i}^{s,a}(m)\) yields
\begin{equation}\label{eq:single-region-observable-within-i}
\begin{aligned}
& 
\sum_{i\in\mathbb{I}} \sum_{m\in\mathbb{M}_{i}} \sum_{s\in\mathbb{S}} \sum_{a\in\{c,t\}} \var(\M_{i}^{m}\T_{s}^{a}) \tilde{Y}_{i}^{s,a}(m)\\
&
+ \sum_{i\in\mathbb{I}} \sum_{m\in\mathbb{M}_{i}} \sum_{s\in\mathbb{S}} \sum_{a\in\{c,t\}} \sum_{s'\in\mathbb{S}} \sum_{a'\in\{c,t\}} \ind\{s \neq s' \text{ or } a \neq a'\} \cov(\M_{i}^{m}\T_{s}^{a},\M_{i}^{m}\T_{s'}^{a'}) \tilde{Y}_{i}^{s,a}(m) \tilde{Y}_{i}^{s',a'}(m) \\
= &
\sum_{i\in\mathbb{I}} \sum_{m\in\mathbb{M}_{i}} \sum_{s\in\mathbb{S}} \sum_{a\in\{c,t\}} \pi_{i,s}^{m,a} (1-\pi_{i,s}^{m,a}) \tilde{Y}_{i}^{s,a}(m)^{2} \\
&
+ \sum_{i\in\mathbb{I}} \sum_{m\in\mathbb{M}_{i}} \sum_{s\in\mathbb{S}} \sum_{a\in\{c,t\}} \sum_{s'\in\mathbb{S}} \sum_{a'\in\{c,t\}} \ind\{s \neq s' \text{ or } a \neq a'\} \bigl(\pi_{i,s,i,s'}^{m,a,m,a'} - \pi_{i,s}^{m,a} \pi_{i,s'}^{m,a'}\bigr) \tilde{Y}_{i}^{s,a}(m) \tilde{Y}_{i}^{s',a'}(m) \\
= &
\sum_{i\in\mathbb{I}} \sum_{m\in\mathbb{M}_{i}} \sum_{s\in\mathbb{S}} \sum_{a\in\{c,t\}} \pi_{i,s}^{m,a} w_{i}(s,d) (Y_{i}(m) - \mu_{a})^{2} \cdot \Bigl((1-\pi_{i,s}^{m,a}) \Bigl(\frac{\pi_{s}}{1-\pi_{s}}\Bigr)^{2 \cdot \ind\{a=c\}} w_{i}(s,d) \Bigr)\\
&
+ \sum_{i\in\mathbb{I}} \sum_{m\in\mathbb{M}_{i}} \sum_{s\in\mathbb{S}} \sum_{a\in\{c,t\}} \sum_{s'\in\mathbb{S}} \sum_{a'\in\{c,t\}} \ind\{s \neq s' \text{ or } a \neq a'\} \bigl(\pi_{i,s,i,s'}^{m,a,m,a'} - \pi_{i,s}^{m,a} \pi_{i,s'}^{m,a'}\bigr) (-1)^{\ind\{a \neq a'\}}\\
& \qquad \qquad \cdot 
\Bigl(\frac{\pi_{s}}{1-\pi_{s}}\Bigr)^{\ind\{a=c\}} \Bigl(\frac{\pi_{s'}}{1-\pi_{s'}}\Bigr)^{\ind\{a'=c\}} w_{i}(s,d) w_{i}(s',d) (Y_{i}(m) - \mu_{a}) (Y_{i}(m) - \mu_{a'})
.
\end{aligned}
\end{equation}

Next, consider the summations in the third~(\(m \neq m'\)) and fourth~(\(i \neq i'\)) lines of the final expression in Equation~(\ref{eq:single-region-expanded-cov-sums}).
Separate these summations based on whether \(\M_{i}^{m} \M_{i'}^{m'} = 0\) with probability 1, such that \(\pi_{i,i'}^{m,m'} = 0\).
For any given treatment assignment, only the potential outcome corresponding to a single exposure of each individual is observed.
Hence, for \(m \neq m'\), \(\M_{i}^{m} \M_{i}^{m'} = 0\) with probability 1, and, by definition, \(\pi_{i,s,i,s'}^{m,a,m',a'} = 0\) irrespective of \(s,s',a,a'\).
Similarly, even when \(i \neq i\), \(\M_{i}^{m} \M_{i'}^{m'} = 0\) with probability 1 for some \(i,m\), \(i',m'\) if there is at least one candidate treatment location that can affect both \(i\) and \(i'\) and \(m\) and \(m'\) correspond to different assignments for such a location.
Then, by definition, \(\pi_{i,i'}^{m,m'} = 0\) and also \(\pi_{i,s,i',s'}^{m,a,m',a'} = 0\).
Hence,
\begin{equation*}
\begin{aligned}
&
\sum_{i\in\mathbb{I}} \sum_{m\in\mathbb{M}_{i}} \sum_{s\in\mathbb{S}} \sum_{a\in\{c,t\}} \sum_{m'\in\mathbb{M}_{i}} \sum_{s'\in\mathbb{S}} \sum_{a'\in\{c,t\}} \ind\{m \neq m'\} \cov(\M_{i}^{m}\T_{s}^{a},\M_{i}^{m'}\T_{s'}^{a'}) \tilde{Y}_{i}^{s,a}(m) \tilde{Y}_{i}^{s',a'}(m') \\
&
+ \sum_{i\in\mathbb{I}} \sum_{m\in\mathbb{M}_{i}} \sum_{s\in\mathbb{S}} \sum_{a\in\{c,t\}} \sum_{i'\in\mathbb{I}} \sum_{m'\in\mathbb{M}_{i}} \sum_{s'\in\mathbb{S}} \sum_{a'\in\{c,t\}} \ind\{i \neq i'\} \cov(\M_{i}^{m}\T_{s}^{a},\M_{i'}^{m'}\T_{s'}^{a'}) \tilde{Y}_{i}^{s,a}(m) \tilde{Y}_{i'}^{s',a'}(m') \\
= & 
- \sum_{i\in\mathbb{I}} \sum_{m\in\mathbb{M}_{i}} \sum_{s\in\mathbb{S}} \sum_{a\in\{c,t\}} \sum_{m'\in\mathbb{M}_{i}} \sum_{s'\in\mathbb{S}} \sum_{a'\in\{c,t\}} \ind\{m \neq m'\} \pi_{i,s}^{m,a} \pi_{i,s'}^{m',a'} \tilde{Y}_{i}^{s,a}(m) \tilde{Y}_{i}^{s',a'}(m') \\
&
- \sum_{i\in\mathbb{I}} \sum_{m\in\mathbb{M}_{i}} \sum_{s\in\mathbb{S}} \sum_{a\in\{c,t\}} \sum_{i'\in\mathbb{I}} \sum_{m'\in\mathbb{M}_{i}} \sum_{s'\in\mathbb{S}} \sum_{a'\in\{c,t\}} \ind\{i \neq i'\} \ind\{\pi_{i,i'}^{m,m'} = 0\} \pi_{i,s}^{m,a} \pi_{i',s'}^{m',a'} \tilde{Y}_{i}^{s,a}(m) \tilde{Y}_{i'}^{s',a'}(m') \\
&
+ \sum_{i\in\mathbb{I}} \sum_{m\in\mathbb{M}_{i}} \sum_{s\in\mathbb{S}} \sum_{a\in\{c,t\}} \sum_{i'\in\mathbb{I}} \sum_{m'\in\mathbb{M}_{i}} \sum_{s'\in\mathbb{S}} \sum_{a'\in\{c,t\}} \ind\{i \neq i'\} \ind\{\pi_{i,i'}^{m,m'} > 0\} \bigl(\pi_{i,s,i',s'}^{m,a,m',a'} - \pi_{i,s}^{m,a} \pi_{i',s'}^{m',a'}\bigr) \\
& \qquad \qquad \qquad \qquad \qquad \qquad \qquad \qquad \qquad \qquad \cdot \tilde{Y}_{i}^{s,a}(m) \tilde{Y}_{i'}^{s',a'}(m')
.
\end{aligned}
\end{equation*}
The first line equals exactly the ``missing'' \(i=i'\) terms of the second line because \(\pi_{i,i}^{m,m'} = 0\) if and only if \(m \neq m'\).
Combining these lines, it is then convenient to treat cases \(a = a'\) and \(a \neq a'\) separately because the sign of the terms multiplying potential outcomes \(Y_{i}(m) Y_{i'}(m')\) differs across the two cases such that they need to be bounded differently (in estimation because the potential outcomes cannot be observed simultaneously for conflicting exposures).
The expression above, therefore, equals
\begin{equation}\label{eq:single-region-across-i-base}
\begin{aligned}
= & 
- \sum_{i\in\mathbb{I}} \sum_{m\in\mathbb{M}_{i}} \sum_{s\in\mathbb{S}} \sum_{a\in\{c,t\}} \sum_{i'\in\mathbb{I}} \sum_{m'\in\mathbb{M}_{i}} \sum_{s'\in\mathbb{S}} \ind\{\pi_{i,i'}^{m,m'} = 0\} \pi_{i,s}^{m,a} \pi_{i',s'}^{m',a} \tilde{Y}_{i}^{s,a}(m) \tilde{Y}_{i'}^{s',a}(m') \\
&
- 2 \sum_{i\in\mathbb{I}} \sum_{m\in\mathbb{M}_{i}} \sum_{s\in\mathbb{S}}\sum_{i'\in\mathbb{I}} \sum_{m'\in\mathbb{M}_{i}} \sum_{s'\in\mathbb{S}} 
\ind\{\pi_{i,i'}^{m,m'} = 0\} \pi_{i,s}^{m,t} \pi_{i',s'}^{m',c} \tilde{Y}_{i}^{s,t}(m) \tilde{Y}_{i'}^{s',c}(m') \\
&
+ \sum_{i\in\mathbb{I}} \sum_{m\in\mathbb{M}_{i}} \sum_{s\in\mathbb{S}} \sum_{a\in\{c,t\}} \sum_{i'\in\mathbb{I}} \sum_{m'\in\mathbb{M}_{i}} \sum_{s'\in\mathbb{S}} \sum_{a'\in\{c,t\}} \ind\{i \neq i'\} \ind\{\pi_{i,i'}^{m,m'} > 0\} \bigl(\pi_{i,s,i',s'}^{m,a,m',a'} - \pi_{i,s}^{m,a} \pi_{i',s'}^{m',a'}\bigr) \\
& \qquad \qquad \qquad \qquad \qquad \qquad \qquad \qquad \qquad \qquad \cdot \tilde{Y}_{i}^{s,a}(m) \tilde{Y}_{i'}^{s',a'}(m')
.
\end{aligned}
\end{equation}

Substituting for \(\tilde{Y}_{i}^{s,a}(m)\), the products \(\tilde{Y}_{i}^{s,a}(m) \tilde{Y}_{i'}^{s',a'}(m')\) are
\begin{equation*}
\begin{aligned}
\tilde{Y}_{i}^{s,a}(m) \tilde{Y}_{i'}^{s',a}(m') & =
\Bigl(\frac{\pi_{s}}{1-\pi_{s}} \frac{\pi_{s'}}{1-\pi_{s'}}\Bigr)^{\ind\{a=c\}}
w_{i}(s,d) w_{i'}(s',d)
(Y_{i}(m) - \mu_{a}) (Y_{i'}(m') - \mu_{a}) \\
\tilde{Y}_{i}^{s,t}(m) \tilde{Y}_{i'}^{s',c}(m') & =
- \frac{\pi_{s'}}{1-\pi_{s'}}
w_{i}(s,d) w_{i'}(s',d)
(Y_{i}(m) - \mu_{t}) (Y_{i'}(m') - \mu_{c})
,
\end{aligned}
\end{equation*}
and using the first and second binomial formulas:
\begin{equation*}
- (Y_{i}(m) - \mu_{a}) (Y_{i'}(m') - \mu_{a})
=
\frac{1}{2} (Y_{i}(m) - \mu_{a})^{2} + \frac{1}{2} (Y_{i'}(m') - \mu_{a})^{2} - 2 \big(\frac{Y_{i}(m) + Y_{i'}(m')}{2} - \mu_{a}\bigr)^{2}
\end{equation*}
\begin{equation*}
2 (Y_{i}(m) - \mu_{t}) (Y_{i'}(m') - \mu_{c})
=
(Y_{i}(m) - \mu_{t})^{2} + (Y_{i'}(m') - \mu_{c})^{2} - \bigl((Y_{i}(m) - Y_{i'}(m')) - (\mu_{t} - \mu_{c})\bigr)^{2}
\end{equation*}

Substituting these equations sequentially into the first and second lines of Equation~(\ref{eq:single-region-across-i-base}):
\begin{equation*}
\begin{aligned}
= & 
- \sum_{i\in\mathbb{I}} \sum_{m\in\mathbb{M}_{i}} \sum_{s\in\mathbb{S}} \sum_{a\in\{c,t\}} \sum_{i'\in\mathbb{I}} \sum_{m'\in\mathbb{M}_{i}} \sum_{s'\in\mathbb{S}} \ind\{\pi_{i,i'}^{m,m'} = 0\} \pi_{i,s}^{m,a} \pi_{i',s'}^{m',a} \Bigl(\frac{\pi_{s}}{1-\pi_{s}} \frac{\pi_{s'}}{1-\pi_{s'}}\Bigr)^{\ind\{a=c\}} \\
& \qquad \qquad \cdot
 w_{i}(s,d) w_{i'}(s',d) (Y_{i}(m) - \mu_{a}) (Y_{i'}(m') - \mu_{a}) \\
&
+ 2 \sum_{i\in\mathbb{I}} \sum_{m\in\mathbb{M}_{i}} \sum_{s\in\mathbb{S}} \sum_{i'\in\mathbb{I}} \sum_{m'\in\mathbb{M}_{i}} \sum_{s'\in\mathbb{S}} \ind\{\pi_{i,i'}^{m,m'} = 0\} \pi_{i,s}^{m,t} \pi_{i',s'}^{m',c} \frac{\pi_{s'}}{1-\pi_{s'}} \\
& \qquad \qquad \cdot 
 w_{i}(s,d) w_{i'}(s',d) (Y_{i}(m) - \mu_{t}) (Y_{i'}(m') - \mu_{c}) \\
&
+ \sum_{i\in\mathbb{I}} \sum_{m\in\mathbb{M}_{i}} \sum_{s\in\mathbb{S}} \sum_{a\in\{c,t\}} \sum_{i'\in\mathbb{I}} \sum_{m'\in\mathbb{M}_{i}} \sum_{s'\in\mathbb{S}} \sum_{a'\in\{c,t\}} \ind\{i \neq i'\} \ind\{\pi_{i,i'}^{m,m'} > 0\} \bigl(\pi_{i,s,i',s'}^{m,a,m',a'} - \pi_{i,s}^{m,a} \pi_{i',s'}^{m',a'}\bigr) \\
& \qquad \qquad \qquad \qquad \qquad \qquad \qquad \qquad \qquad \qquad \cdot \tilde{Y}_{i}^{s,a}(m) \tilde{Y}_{i'}^{s',a'}(m') \\
\end{aligned}
\end{equation*}
\begin{equation*}
\begin{aligned}
= &
\sum_{i\in\mathbb{I}} \sum_{m\in\mathbb{M}_{i}} \sum_{s\in\mathbb{S}} \sum_{a\in\{c,t\}} \sum_{i'\in\mathbb{I}} \sum_{m'\in\mathbb{M}_{i}} \sum_{s'\in\mathbb{S}} \ind\{\pi_{i,i'}^{m,m'} = 0\} \pi_{i,s}^{m,a} \pi_{i',s'}^{m',a} \Bigl(\frac{\pi_{s}}{1-\pi_{s}} \frac{\pi_{s'}}{1-\pi_{s'}}\Bigr)^{\ind\{a=c\}} \\
& \qquad \qquad \cdot 
w_{i}(s,d) w_{i'}(s',d) \Bigl( 
\frac{1}{2} (Y_{i}(m) - \mu_{a})^{2} + \frac{1}{2} (Y_{i'}(m') - \mu_{a})^{2} - 2 \bigl(\frac{Y_{i}(m) + Y_{i'}(m')}{2} - \mu_{a}\bigr)^{2} \Bigr) \\
& 
+ \sum_{i\in\mathbb{I}} \sum_{m\in\mathbb{M}_{i}} \sum_{s\in\mathbb{S}} \sum_{i'\in\mathbb{I}} \sum_{m'\in\mathbb{M}_{i}} \sum_{s'\in\mathbb{S}} \ind\{\pi_{i,i'}^{m,m'} = 0\} \pi_{i,s}^{m,t} \pi_{i',s'}^{m',c} \frac{\pi_{s'}}{1-\pi_{s'}} \\
& \qquad \qquad \cdot 
w_{i}(s,d) w_{i'}(s',d) \Bigl(
(Y_{i}(m) - \mu_{t})^{2} + (Y_{i'}(m') - \mu_{c})^{2} - \bigl((Y_{i}(m) - Y_{i'}(m')) - (\mu_{t} - \mu_{c})\bigr)^{2} \Bigr) \\
&
+ \sum_{i\in\mathbb{I}} \sum_{m\in\mathbb{M}_{i}} \sum_{s\in\mathbb{S}} \sum_{a\in\{c,t\}} \sum_{i'\in\mathbb{I}} \sum_{m'\in\mathbb{M}_{i}} \sum_{s'\in\mathbb{S}} \sum_{a'\in\{c,t\}} \ind\{i \neq i'\} \ind\{\pi_{i,i'}^{m,m'} > 0\} \bigl(\pi_{i,s,i',s'}^{m,a,m',a'} - \pi_{i,s}^{m,a} \pi_{i',s'}^{m',a'}\bigr) \\
& \qquad \qquad \qquad \qquad \qquad \qquad \qquad \qquad \qquad \qquad \cdot  \tilde{Y}_{i}^{s,a}(m) \tilde{Y}_{i'}^{s',a'}(m') \\
\end{aligned}
\end{equation*}

Splitting the summations into some that square single potential outcomes and others that square averages or differences of potential outcomes:
\begin{equation}\label{eq:single-region-across-i-unobservable}
\begin{aligned}
= &
\sum_{i\in\mathbb{I}} \sum_{m\in\mathbb{M}_{i}} \sum_{s\in\mathbb{S}} \sum_{a\in\{c,t\}} 
\pi_{i,s}^{m,a} \Bigl(\frac{\pi_{s}}{1-\pi_{s}}\Bigr)^{\ind\{a=c\}} w_{i}(s,d) (Y_{i}(m) - \mu_{a})^{2} \\
& \qquad \qquad \cdot
\sum_{i'\in\mathbb{I}} \sum_{m'\in\mathbb{M}_{i}} \sum_{s'\in\mathbb{S}} \ind\{\pi_{i,i'}^{m,m'} = 0\} \pi_{i',s'}^{m',a} \Bigl(\frac{\pi_{s'}}{1-\pi_{s'}}\Bigr)^{\ind\{a=c\}} w_{i'}(s',d) \\
&
- 2 \sum_{i\in\mathbb{I}} \sum_{m\in\mathbb{M}_{i}} \sum_{s\in\mathbb{S}} \sum_{a\in\{c,t\}} \sum_{i'\in\mathbb{I}} \sum_{m'\in\mathbb{M}_{i}} \sum_{s'\in\mathbb{S}} \ind\{\pi_{i,i'}^{m,m'} = 0\} \pi_{i,s}^{m,a} \pi_{i',s'}^{m',a} \Bigl(\frac{\pi_{s}}{1-\pi_{s}} \frac{\pi_{s'}}{1-\pi_{s'}}\Bigr)^{\ind\{a=c\}} \\
& \qquad \qquad \cdot 
w_{i}(s,d) w_{i'}(s',d) \Bigl(\frac{Y_{i}(m) + Y_{i'}(m')}{2} - \mu_{a}\Bigr)^{2} \\
&
+ \sum_{i\in\mathbb{I}} \sum_{m\in\mathbb{M}_{i}} \sum_{s\in\mathbb{S}} \sum_{a\in\{c,t\}}
\pi_{i,s}^{m,a} \Bigl(\frac{\pi_{s}}{1-\pi_{s}}\Bigr)^{\ind\{a=c\}} w_{i}(s,d) (Y_{i}(m) - \mu_{a})^{2} \\
& \qquad \qquad \cdot 
\sum_{i'\in\mathbb{I}} \sum_{m'\in\mathbb{M}_{i}} \sum_{s'\in\mathbb{S}} \sum_{a'\in\{c,t\}\setminus\{a\}} \ind\{\pi_{i,i'}^{m,m'} = 0\} \pi_{i',s'}^{m',a'} \Bigl(\frac{\pi_{s'}}{1-\pi_{s'}}\Bigr)^{\ind\{a'=c\}} w_{i'}(s',d) \\
& 
- \sum_{i\in\mathbb{I}} \sum_{m\in\mathbb{M}_{i}} \sum_{s\in\mathbb{S}} \sum_{i'\in\mathbb{I}} \sum_{m'\in\mathbb{M}_{i}} \sum_{s'\in\mathbb{S}} \ind\{\pi_{i,i'}^{m,m'} = 0\} \pi_{i,s}^{m,t} \pi_{i',s'}^{m',c} \frac{\pi_{s'}}{1-\pi_{s'}} \\
& \qquad \qquad \cdot 
w_{i}(s,d) w_{i'}(s',d) \bigl((Y_{i}(m) - Y_{i'}(m')) - (\mu_{t} - \mu_{c})\bigr)^{2} \\
&
+ \sum_{i\in\mathbb{I}} \sum_{m\in\mathbb{M}_{i}} \sum_{s\in\mathbb{S}} \sum_{a\in\{c,t\}} \sum_{i'\in\mathbb{I}} \sum_{m'\in\mathbb{M}_{i}} \sum_{s'\in\mathbb{S}} \sum_{a'\in\{c,t\}} \ind\{i \neq i'\} \ind\{\pi_{i,i'}^{m,m'} > 0\} \bigl(\pi_{i,s,i',s'}^{m,a,m',a'} - \pi_{i,s}^{m,a} \pi_{i',s'}^{m',a'}\bigr) \\
& \qquad \qquad \qquad \qquad \qquad \qquad \qquad \qquad \qquad \qquad \cdot  \tilde{Y}_{i}^{s,a}(m) \tilde{Y}_{i'}^{s',a'}(m') \\
\end{aligned}
\end{equation}

Finally, combine the results in Equations~(\ref{eq:single-region-expanded-cov-sums}),~(\ref{eq:single-region-observable-within-i}),~and~(\ref{eq:single-region-across-i-unobservable}),
and substitute \(\tilde{Y}_{i}^{s,a}(m)\).
Then
\begin{equation*}
\begin{aligned}
& \var\Big(\sum_{i\in\mathbb{I}} \sum_{m \in \mathbb{M}_{i}} \sum_{s\in\mathbb{S}} \sum_{a\in\{c,t\}} \M_{i}^{m} \T_{s}^{a} \tilde{Y}_{i}^{s,a}(m)\Bigr) \\
= &
\sum_{i\in\mathbb{I}} \sum_{m\in\mathbb{M}_{i}} \sum_{s\in\mathbb{S}} \sum_{a\in\{c,t\}} \pi_{i,s}^{m,a} w_{i}(s,d) (Y_{i}(m) - \mu_{a})^{2} \cdot \Bigl((1-\pi_{i,s}^{m,a}) \Bigl(\frac{\pi_{s}}{1-\pi_{s}}\Bigr)^{2 \cdot \ind\{a=c\}} w_{i}(s,d) \Bigr)\\
&
+ \sum_{i\in\mathbb{I}} \sum_{m\in\mathbb{M}_{i}} \sum_{s\in\mathbb{S}} \sum_{a\in\{c,t\}} \sum_{s'\in\mathbb{S}} \sum_{a'\in\{c,t\}} \ind\{s \neq s' \text{ or } a \neq a'\} \bigl(\pi_{i,s,i,s'}^{m,a,m,a'} - \pi_{i,s}^{m,a} \pi_{i,s'}^{m,a'}\bigr) (-1)^{\ind\{a \neq a'\}}\\
& \qquad \qquad \cdot 
\Bigl(\frac{\pi_{s}}{1-\pi_{s}}\Bigr)^{\ind\{a=c\}} \Bigl(\frac{\pi_{s'}}{1-\pi_{s'}}\Bigr)^{\ind\{a'=c\}} w_{i}(s,d) w_{i}(s',d) (Y_{i}(m) - \mu_{a}) (Y_{i}(m) - \mu_{a'}) \\
&
+ \sum_{i\in\mathbb{I}} \sum_{m\in\mathbb{M}_{i}} \sum_{s\in\mathbb{S}} \sum_{a\in\{c,t\}} \pi_{i,s}^{m,a} w_{i}(s,d) (Y_{i}(m) - \mu_{a})^{2} \\
& \qquad \qquad \cdot 
\Bigl(\frac{\pi_{s}}{1-\pi_{s}}\Bigr)^{\ind\{a=c\}} \sum_{i'\in\mathbb{I}} \sum_{m'\in\mathbb{M}_{i}} \sum_{s'\in\mathbb{S}} \ind\{\pi_{i,i'}^{m,m'} = 0\} \pi_{i',s'}^{m',a} \Bigl(\frac{\pi_{s'}}{1-\pi_{s'}}\Bigr)^{\ind\{a=c\}} w_{i'}(s',d) \\
& 
- 2 \sum_{i\in\mathbb{I}} \sum_{m\in\mathbb{M}_{i}} \sum_{s\in\mathbb{S}} \sum_{a\in\{c,t\}} \sum_{i'\in\mathbb{I}} \sum_{m'\in\mathbb{M}_{i}} \sum_{s'\in\mathbb{S}} \ind\{\pi_{i,i'}^{m,m'} = 0\} \pi_{i,s}^{m,a} \pi_{i',s'}^{m',a} \Bigl(\frac{\pi_{s}}{1-\pi_{s}} \frac{\pi_{s'}}{1-\pi_{s'}}\Bigr)^{\ind\{a=c\}} \\
& \qquad \qquad \cdot 
w_{i}(s,d) w_{i'}(s',d) \Bigl(\frac{Y_{i}(m) + Y_{i'}(m')}{2} - \mu_{a}\Bigr)^{2} \\
& 
+ \sum_{i\in\mathbb{I}} \sum_{m\in\mathbb{M}_{i}} \sum_{s\in\mathbb{S}} \sum_{a\in\{c,t\}} \pi_{i,s}^{m,a} w_{i}(s,d) (Y_{i}(m) - \mu_{a})^{2} \Bigl(\frac{\pi_{s}}{1-\pi_{s}}\Bigr)^{\ind\{a=c\}} \\
& \qquad \qquad
\cdot \sum_{i'\in\mathbb{I}} \sum_{m'\in\mathbb{M}_{i}} \sum_{s'\in\mathbb{S}} \sum_{a'\in\{c,t\}\setminus\{a\}} \ind\{\pi_{i,i'}^{m,m'} = 0\} \pi_{i',s'}^{m',a'} \Bigl(\frac{\pi_{s'}}{1-\pi_{s'}}\Bigr)^{\ind\{a'=c\}} w_{i'}(s',d) \\
& 
- \sum_{i\in\mathbb{I}} \sum_{m\in\mathbb{M}_{i}} \sum_{s\in\mathbb{S}} \sum_{i'\in\mathbb{I}} \sum_{m'\in\mathbb{M}_{i}} \sum_{s'\in\mathbb{S}} \ind\{\pi_{i,i'}^{m,m'} = 0\} \pi_{i,s}^{m,t} \pi_{i',s'}^{m',c} \frac{\pi_{s'}}{1-\pi_{s'}} \\
& \qquad \qquad \cdot 
w_{i}(s,d) w_{i'}(s',d) \bigl((Y_{i}(m) - Y_{i'}(m')) - (\mu_{t} - \mu_{c})\bigr)^{2} \\
&
+ \sum_{i\in\mathbb{I}} \sum_{m\in\mathbb{M}_{i}} \sum_{s\in\mathbb{S}} \sum_{a\in\{c,t\}} \sum_{i'\in\mathbb{I}} \sum_{m'\in\mathbb{M}_{i}} \sum_{s'\in\mathbb{S}} \sum_{a'\in\{c,t\}} \ind\{i \neq i'\} \ind\{\pi_{i,i'}^{m,m'} > 0\} \bigl(\pi_{i,s,i',s'}^{m,a,m',a'} - \pi_{i,s}^{m,a} \pi_{i',s'}^{m',a'}\bigr) \\
& \qquad \qquad \cdot \Bigl(- \frac{\pi_{s}}{1-\pi_{s}}\Bigr)^{\ind\{a=c\}} \Bigl(- \frac{\pi_{s'}}{1-\pi_{s'}}\Bigr)^{\ind\{a'=c\}} w_{i}(s,d) w_{i'}(s',d) (Y_{i}(m) - \mu_{a}) (Y_{i'}(m') - \mu_{a'})
.
\end{aligned}
\end{equation*}
The first, third, and fifth summations all contain \(\pi_{i,s}^{m,a} w_{i}(s,d) (Y_{i}(m) - \mu_{a})^{2}\) post-multiplied by different factors.
Hence, they can be combined.

Recall that the denominator used in \(\tilde{\tau}\) equals \(\sum_{s\in\mathbb{S}} \pi_{s} \sum_{i\in\mathbb{I}} w_{i}(s,d)\).
Define
\begin{equation*}
\bar{n}(d) \equiv \frac{1}{\mathbb{S}} \sum_{s\in\mathbb{S}} \pi_{s} \sum_{i\in\mathbb{I}} w_{i}(s,d).
\end{equation*}

Then
\begin{equation}
\begin{aligned}
& \var(\tilde{\tau}) = \frac{1}{\abs{\mathbb{S}}^{2}} \var\Big(\sum_{i\in\mathbb{I}} \sum_{m \in \mathbb{M}_{i}} \sum_{s\in\mathbb{S}} \sum_{a\in\{c,t\}} \M_{i}^{m} \T_{s}^{a} \tilde{Y}_{i}^{s,a}(m)\Bigr)/\bar{n}(d)^{2} \\
= & \frac{1}{\abs{\mathbb{S}}} \Biggl(
\frac{1}{\abs{\mathbb{S}}} \sum_{a\in\{c,t\}} \sum_{s\in\mathbb{S}} \sum_{i\in\mathbb{I}} \sum_{m\in\mathbb{M}_{i}} \pi_{i,s}^{m,a} \frac{w_{i}(s,d)}{\bar{n}(d)} v_{i,s}^{m,a}(d) (Y_{i}(m) - \mu_{a})^{2} \\
& \quad
+ \frac{1}{\abs{\mathbb{S}}} \sum_{s\in\mathbb{S}} \sum_{s'\in\mathbb{S}} \sum_{i\in\mathbb{I}} \sum_{m\in\mathbb{M}_{i}} \sum_{i'\in\mathbb{I}} \sum_{m'\in\mathbb{M}_{i}} \sum_{a\in\{c,t\}} \sum_{a'\in\{c,t\}} \ind\{i \neq i'\} \ind\{\pi_{i,i'}^{m,m'} > 0\} \bigl(\pi_{i,s,i',s'}^{m,a,m',a'} - \pi_{i,s}^{m,a} \pi_{i',s'}^{m',a'}\bigr) \\
& \qquad \qquad \cdot \Bigl(- \frac{\pi_{s}}{1-\pi_{s}}\Bigr)^{\ind\{a=c\}} \Bigl(- \frac{\pi_{s'}}{1-\pi_{s'}}\Bigr)^{\ind\{a'=c\}} \frac{w_{i}(s,d) w_{i'}(s',d)}{\bar{n}(d)^{2}} (Y_{i}(m) - \mu_{a}) (Y_{i'}(m') - \mu_{a'}) \\
& \quad
+ \frac{1}{\abs{\mathbb{S}}} \sum_{s\in\mathbb{S}} \sum_{s'\in\mathbb{S}} \sum_{i\in\mathbb{I}} \sum_{m\in\mathbb{M}_{i}} \sum_{a\in\{c,t\}} \sum_{a'\in\{c,t\}} \ind\{s \neq s' \text{ or } a \neq a'\} \bigl(\pi_{i,s,i,s'}^{m,a,m,a'} - \pi_{i,s}^{m,a} \pi_{i,s'}^{m,a'}\bigr) \\
& \qquad \qquad \cdot 
\Bigl(-\frac{\pi_{s}}{1-\pi_{s}}\Bigr)^{\ind\{a=c\}} \Bigl(-\frac{\pi_{s'}}{1-\pi_{s'}}\Bigr)^{\ind\{a'=c\}} \frac{w_{i}(s,d) w_{i}(s',d)}{\bar{n}(d)^{2}} (Y_{i}(m) - \mu_{a}) (Y_{i}(m) - \mu_{a'}) \\
& \quad
- \frac{2}{\abs{\mathbb{S}}} \sum_{a\in\{c,t\}} \sum_{s\in\mathbb{S}} \sum_{s'\in\mathbb{S}} \sum_{i\in\mathbb{I}} \sum_{m\in\mathbb{M}_{i}} \sum_{i'\in\mathbb{I}} \sum_{m'\in\mathbb{M}_{i}} \ind\{\pi_{i,i'}^{m,m'} = 0\} \pi_{i,s}^{m,a} \pi_{i',s'}^{m',a} \Bigl(\frac{\pi_{s}}{1-\pi_{s}} \frac{\pi_{s'}}{1-\pi_{s'}}\Bigr)^{\ind\{a=c\}} \\
& \qquad \qquad \cdot 
\frac{w_{i}(s,d) w_{i'}(s',d)}{\bar{n}(d)^{2}} \Bigl(\frac{Y_{i}(m) + Y_{i'}(m')}{2} - \mu_{a}\Bigr)^{2} \\
& \quad
- \frac{1}{\abs{\mathbb{S}}} \sum_{s\in\mathbb{S}} \sum_{s'\in\mathbb{S}} \sum_{i\in\mathbb{I}} \sum_{m\in\mathbb{M}_{i}} \sum_{i'\in\mathbb{I}} \sum_{m'\in\mathbb{M}_{i}} \ind\{\pi_{i,i'}^{m,m'} = 0\} \pi_{i,s}^{m,t} \pi_{i',s'}^{m',c} \frac{\pi_{s'}}{1-\pi_{s'}} \\
& \qquad \qquad \cdot 
\frac{w_{i}(s,d) w_{i'}(s',d)}{\bar{n}(d)^{2}} \bigl((Y_{i}(m) - Y_{i'}(m')) - (\mu_{t} - \mu_{c})\bigr)^{2}
\Biggr)
\end{aligned}
\end{equation}
where
\begin{equation*}
\begin{aligned}
v_{i,s}^{m,a}(d) \equiv & \Bigl(\frac{\pi_{s}}{1-\pi_{s}}\Bigr)^{\ind\{a=c\}} \Bigl((1-\pi_{i,s}^{m,a}) \Bigl(\frac{\pi_{s}}{1-\pi_{s}}\Bigr)^{\ind\{a=c\}} \frac{w_{i}(s,d)}{\bar{n}(d)} \\
& \qquad 
+ \sum_{i'\in\mathbb{I}} \sum_{m'\in\mathbb{M}_{i}} \sum_{s'\in\mathbb{S}} \sum_{a'\in\{c,t\}} \ind\{\pi_{i,i'}^{m,m'} = 0\} \pi_{i',s'}^{m',a'} \Bigl(\frac{\pi_{s'}}{1-\pi_{s'}}\Bigr)^{\ind\{a'=c\}} \frac{w_{i'}(s',d)}{\bar{n}(d)}
\Bigr) .
\end{aligned}
\end{equation*}

Define
\begin{equation*}
\begin{aligned}
\tilde{V}_{a}(d) & \equiv 
\frac{1}{\abs{\mathbb{S}}} \sum_{s\in\mathbb{S}} \sum_{i\in\mathbb{I}} \sum_{m\in\mathbb{M}_{i}} \pi_{i,s}^{m,a} \frac{w_{i}(s,d)}{\bar{n}(d)} v_{i,s}^{m,a}(d) (Y_{i}(m) - \mu_{a})^{2} \\
\end{aligned}
\end{equation*}
\begin{equation*}
\begin{aligned}
\tilde{V}_{\times}(d) & \equiv 
\frac{1}{\abs{\mathbb{S}}} \sum_{s\in\mathbb{S}} \sum_{s'\in\mathbb{S}} \sum_{i\in\mathbb{I}} \sum_{m\in\mathbb{M}_{i}} \sum_{i'\in\mathbb{I}} \sum_{m'\in\mathbb{M}_{i}} \sum_{a\in\{c,t\}} \sum_{a'\in\{c,t\}} \Biggl(\ind\{i \neq i' \text{ or } s \neq s' \text{ or } a \neq a'\} \\ 
& \qquad \cdot \ind\{\pi_{i,i'}^{m,m'} > 0\} \bigl(\pi_{i,s,i',s'}^{m,a,m',a'} - \pi_{i,s}^{m,a} \pi_{i',s'}^{m',a'}\bigr) \Bigl(- \frac{\pi_{s}}{1-\pi_{s}}\Bigr)^{\ind\{a=c\}} \Bigl(- \frac{\pi_{s'}}{1-\pi_{s'}}\Bigr)^{\ind\{a'=c\}} \\
& \qquad \cdot \frac{w_{i}(s,d) w_{i'}(s',d)}{\bar{n}(d)^{2}} (Y_{i}(m) - \mu_{a}(d)) (Y_{i'}(m') - \mu_{a'}(d)) \Biggr)\\
\end{aligned}
\end{equation*}
\begin{equation*}
\begin{aligned}
\tilde{V}_{aa}(d) & \equiv
\frac{2}{\abs{\mathbb{S}}} \sum_{a\in\{c,t\}} \sum_{s\in\mathbb{S}} \sum_{s'\in\mathbb{S}} \sum_{i\in\mathbb{I}} \sum_{m\in\mathbb{M}_{i}} \sum_{i'\in\mathbb{I}} \sum_{m'\in\mathbb{M}_{i}} \ind\{\pi_{i,i'}^{m,m'} = 0\} \pi_{i,s}^{m,a} \pi_{i',s'}^{m',a} \\
& \qquad \qquad \cdot \Bigl(\frac{\pi_{s}}{1-\pi_{s}} \frac{\pi_{s'}}{1-\pi_{s'}}\Bigr)^{\ind\{a=c\}}
\frac{w_{i}(s,d) w_{i'}(s',d)}{\bar{n}(d)^{2}} \Bigl(\frac{Y_{i}(m) + Y_{i'}(m')}{2} - \mu_{a}\Bigr)^{2} \\
\end{aligned}
\end{equation*}
\begin{equation*}
\begin{aligned}
\tilde{V}_{ct}(d) & \equiv 
\frac{1}{\abs{\mathbb{S}}} \sum_{s\in\mathbb{S}} \sum_{s'\in\mathbb{S}} \sum_{i\in\mathbb{I}} \sum_{m\in\mathbb{M}_{i}} \sum_{i'\in\mathbb{I}} \sum_{m'\in\mathbb{M}_{i}} \ind\{\pi_{i,i'}^{m,m'} = 0\} \pi_{i,s}^{m,t} \pi_{i',s'}^{m',c} \frac{\pi_{s'}}{1-\pi_{s'}} \\
& \qquad \qquad \cdot 
\frac{w_{i}(s,d) w_{i'}(s',d)}{\bar{n}(d)^{2}} \bigl((Y_{i}(m) - Y_{i'}(m')) - (\mu_{t} - \mu_{c})\bigr)^{2}
\end{aligned}
\end{equation*}

Then \(\var(\tilde{\tau}(d)) = \frac{1}{\abs{\mathbb{S}}} (\tilde{V}_{t}(d) + \tilde{V}_{c}(d) + \tilde{V}_{\times}(d) - \tilde{V}_{tt}(d) - \tilde{V}_{cc}(d) - \tilde{V}_{ct}(d))\) as stated in the theorem.

\section{Estimator when only the nearest realized location matters}

The identification argument in the proof of Theorem~2 suggests the estimator
\begin{equation*}
\begin{aligned}
\hat{\tau}_{\text{nearest}}(d) \equiv &
\frac{\sum_{s \in \mathbb{S}} \ind\{\S \ni s\} \sum_{i \in \mathbb{I}} \frac{\N_{i}(s)}{\Pr(\N_{i}(s)=1 \mid \S \ni s)} w_{i}(s,d) \Y_{i}}{\sum_{s \in \mathbb{S}} \ind\{\S \ni s\} \sum_{i \in \mathbb{I}} \frac{\N_{i}(s)}{\Pr(\N_{i}(s)=1 \mid \S \ni s)} w_{i}(s,d)} \\
&
-
\frac{\sum_{s \in \mathbb{S}} \frac{\ind\{\S \niton s\}}{1-\pi_{s}} \pi_{s} \sum_{i \in \mathbb{I}} \frac{\N_{i}(0)}{\Pr(\N_{i}(0)=1 \mid \S \niton s)} w_{i}(s,d) \Y_{i}}{\sum_{s \in \mathbb{S}} \frac{\ind\{\S \niton s\}}{1-\pi_{s}} \pi_{s} \sum_{i \in \mathbb{I}} \frac{\N_{i}(0)}{\Pr(\N_{i}(0)=1 \mid \S \niton s)} w_{i}(s,d)}
\end{aligned}
\end{equation*}
where \(\N_{i}(s)\) is an indicator for \(s\) being the nearest realized treatment location to \(i\), and \(\N_{i}(0)\) is an indicator for no treatment location within \(\dnoeffect{}\) of \(i\) being realized:
\begin{equation*}
\begin{aligned}
\N_{i}(s) & = \ind\{\S \ni s\} \prod_{s' \in \mathbb{S} \setminus \{s\}} (1 - \ind\{\S \ni s'\})^{\ind\{d(s',r_{i})<d(s,r_{i})\}} \\
\N_{i}(0) & = \prod_{s \in \mathbb{S}} (1 - \ind\{\S \ni s\})^{\ind\{d(s,r_{i})<\dnoeffect{}\}}
\end{aligned}
\end{equation*}
and the (conditional) probabilities of these events are, under independent assignment,
\begin{equation*}
\begin{aligned}
\Pr(\N_{i}(s)=1 \mid \S \ni s) & = \prod_{s' \in \mathbb{S} \setminus \{s\}} (1 - \pi_{s})^{\ind\{d(s',r_{i})<d(s,r_{i})\}} \\
\Pr(\N_{i}(0)=1 \mid \S \niton s) & = \frac{1}{1-\pi_{s}} \prod_{s' \in \mathbb{S}} (1 - \pi_{s'})^{\ind\{d(s,r_{i})<\dnoeffect{}\}}
.
\end{aligned}
\end{equation*}
It is straightforward to show that \(E(\hat{\tau}_{\text{nearest}}(d)) \approx \tau(d)\) and the approximate variance of the estimator can be derived analogously to the previous results.

If the event \(\N_{i}(0)\) is rare, the variance of the \(\hat{\tau}_{\text{nearest}}(d)\) will likely be large.
The difficulty lies in estimating the weighted mean of \(Y_{i}(0)\).
Additive separability allows identifying this mean from differences in exposure, but Assumption~4 only allows using individuals who are unexposed to the treatment (within distance \(\dnoeffect{}\)).
Under Assumption~4, the estimator, therefore, tends to use drastically fewer observations, increasing the variance.

There are, effectively, two options for addressing this issue.
First, the researcher can impose additional structure.
As discussed, under, for instance, Assumption~2, alternative estimators with likely smaller variance are feasible.
Other assumptions more in the spirit of Assumption~4 may be conceivable.
Second, the researcher can change the target of estimation.
Minor improvements in the variance are possible by choosing weights \(w_{i}(s,d)\) including a factor \(\Pr(\N_{i}(s)=1 \mid \S \ni s)\) or, for interpretation likely less attractively, \(\Pr(\N_{i}(0)=1 \mid \S \niton s)\).
More substantial gains arise by changing the estimand to not rely on treatment effects \(\tau_{i}(s)\) but instead build on \(\tau_{i}(s\mid S_{i}(s))\) for some \(S_{i}(s) \subset \{s' \in \mathbb{S}: \; d(s,r_{i}) \leq d(s',r_{i})\}\).
More research is needed to develop recommendations for the choice of \(S_{i}(s)\) with desirable interpretation and inferential properties.

\section{Assessing the Identifying Assumption}

\begin{figure}
\includegraphics[width=0.49\linewidth]{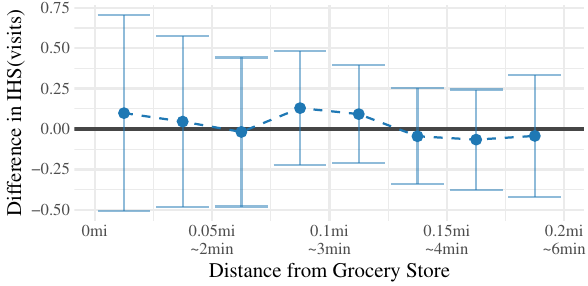}
\hfill
\includegraphics[width=0.49\linewidth]{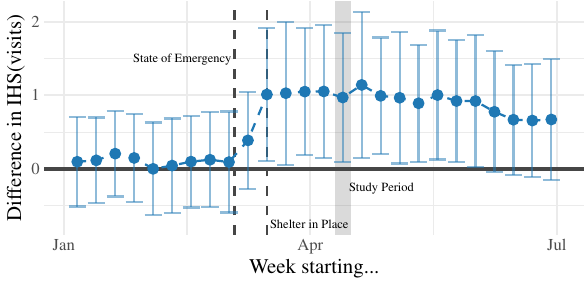}
\caption{\label{fig:effect-by-week}Estimated average effect of grocery stores on restaurants.
The left panel shows estimated effects in the week starting January 6, 2020, analogous to Figure~5 in the main text.
The right panel shows effects in the shortest distance bin (\(0\)mi--\(0.025\)mi) over time.
The Governor of California declared a State of Emergency on March 4.
Shelter-in-place policies were announced on March 16 in the Bay Area.
The week studied in the main text is highlighted.
}
\end{figure}

A popular approach to ``assessing unconfoundedness'' (here: characteristics-determined treatment probabilities) is to estimate effects on ``pseudo outcomes'' that are proxies for potential outcomes but not themselves affected by the treatment \citep[c.f.][ch.~21.3]{imbens2015causal}.
Figure~\ref{fig:effect-by-week} shows estimated effects on the inverse hyperbolic sine of visits in the week starting January 6, 2020, at each distance (left), and at the shortest distance for each week between January and June (right).
The figure is suggestive of there being no effects and no differential selection into neighborhoods near real vs. counterfactual grocery store locations prior to the pandemic.

Note that, in this application, lagged outcomes are not necessarily ideal ``pseudo outcomes'' for two reasons.
First, because grocery stores (the treatment) were in place before January 2020, lagged outcomes may themselves be affected by the treatment.
Second, pre-pandemic outcomes may not be good proxies for pandemic outcomes for the purpose of assessing the identifying assumption.
As discussed in the main text, an unconfoundedness-type assumption may be plausible during shelter-in-place policies even if it was violated before the pandemic:
While restaurants may strategically locate based on (proxies for) pre-pandemic potential outcomes, they are unlikely to have sorted in advance into locations based on the potential outcomes of an unexpected and unprecedented pandemic.

As an alternative, in Figure~4 in the main text, I assess covariate balance.
Because dentists and automotive businesses are not directly used in neural network and treatment probability estimation, the corresponding figures can also be viewed as assessing ``pseudo outcomes'' using a subset of the covariates \citep[c.f.][ch.~21.3]{imbens2015causal}.

\FloatBarrier

\printbibliography[title={References for Online Appendix}]

\end{refsection}

\end{document}